\documentclass{pasj01}

\Received{$\langle$reception date$\rangle$}
\Accepted{$\langle$acception date$\rangle$}
\Published{$\langle$publication date$\rangle$}

\usepackage{color}
\usepackage{url}
\usepackage{natbib}
\usepackage{ulem}

\newcommand{\msun}{\, M_\odot}

\newcommand{\Mbh}{M_{\rm BH}}
\newcommand{\Mpc}{\rm Mpc}
\newcommand{\hMpc}{h^{-1}{\rm Mpc}}
\newcommand{\Dx}{D_{\rm opt-IR}}

\begin{document}

\title{Properties of Galaxies around AGNs with Most Massive Supermassive Black Hole
       Revealed by the Clustering Analysis}

\author{Yuji Shirasaki\altaffilmark{1,}\altaffilmark{2}}
\email{yuji.shirasaki@nao.ac.jp}

\author{Yutaka Komiya\altaffilmark{3}}

\author{Masatoshi Ohishi\altaffilmark{1,}\altaffilmark{2}}

\and
\author{Yoshihiko Mizumoto\altaffilmark{1,}\altaffilmark{2}}

\altaffiltext{1}{National Astronomical Observatory of Japan, 2-21-1 Osawa, Mitaka, Tokyo 181-8588, Japan}
\altaffiltext{2}{Department of Astronomical Science, School of Physical Sciences, SOKENDAI (The Graduate University for Advanced Studies), Okazaki 444-8787, Japan}
\altaffiltext{3}{Research center for the early universe, University of Tokyo, 7-3-1 Hongo, Bunkyo-ku Tokyo, 113-0033, Japan}


\maketitle

\begin{abstract}

We present results of the clustering analysis between active galactic nuclei (AGNs)
and galaxies at redshift 0.1--1.0, which was performed for 
investigating properties of galaxies associated with the AGNs and
revealing the nature of fueling mechanism of supermassive black holes (SMBHs).
We used 8059 SDSS AGNs/QSOs for which virial masses of individual 
SMBHs were measured, and divided them into four mass groups.
Cross-correlation analysis was performed to reconfirm our previous result
that cross-correlation length increases with SMBH mass $\Mbh$; we obtained 
consistent results.
A linear bias of AGN for each mass group was measured as
1.47 for $\Mbh = 10^{7.5}$--$10^{8.2} \msun$
and
3.08 for $\Mbh = 10^{9}$--$10^{10} \msun$.
The averaged color and luminosity distributions of galaxies around the AGNs/QSOs 
were also derived for each mass group.
The galaxy color $\Dx$ was estimated from an SED 
constructed with a merged SDSS and UKIDSS catalog.
The distributions of color and luminosity were derived by a 
subtraction method, which does not require redshift information of galaxies.
The main results of this work are:
(1) a linear bias increases by a factor two from the
  lower mass group to the highest mass group;
(2) the environment around AGNs with the most massive SMBH ($\Mbh > 10^{9} \msun$)
is dominated by red sequence galaxies;
(3) marginal indication of decline in luminosity function at dimmer side of 
$M_{\rm IR} > -19.5$ is found for galaxies around AGNs with $\Mbh = 10^{8.2} - 10^{9} \msun$
and nearest redshift group ($z=$0.1--0.3).
These results indicate that AGNs with the most massive SMBHs reside in haloes
where large fraction of galaxies have been transited to the red sequence. 
The accretion of hot halo gas as well as recycled gas from evolving stars
can be one of the plausible mechanisms to fuel the SMBHs above $\sim 10^{9} \msun$.

\end{abstract}

\section{Introduction}\label{sec:intro}

There are a lot of observational evidences that
a Supermassive Black Hole (hereafter SMBH) is located in the center
of all but the smallest galaxies \citep{Richstone+98}.
Although the evolution mechanism of SMBH is still not well known, a recent
growing evidence suggests that there is a strong link between the growth of SMBH 
and the star formation in the host galaxy.
One of the observational evidence is the similarity between the evolutions of 
black hole accretion rate and the star formation rate of galaxies 
\citep[e.g.,][]{Madau+96,Boyle+98,Ueda+03,Zheng+09}.
Another important evidence is the correlation between mass of the SMBH 
($\Mbh$) and mass ($M_{b}$) / velocity dispersion ($\sigma$) of the bulge component 
of its host galaxy \citep{Magorrian+98,Ferrarese+00,Gebhardt+00,Ho-07}.
These relations indicate that BHs and bulges coevolve regulating 
each others growth \citep[see the recent review by][and references therein]{Kormendy+13}. 

According to the theoretical works \citep[e.g.,][]{King-14}, mass of a
SMBH can be regulated by its own outflow from an active galactic nucleus 
(AGN) in a way that the interstellar gas is expelled far away from the host galaxy 
once $\Mbh$ reaches to the critical 
mass given by the $\Mbh$-$\sigma$ relation.
As a result both the star formation
in the bulge and mass accretion onto the SMBH are terminated.
Therefore understanding the evolution mechanism of SMBHs and AGN phenomenon
is crucial to shed light on the evolution of galaxies.

The most important open questions are the nature of the fueling of SMBHs 
and triggering mechanisms of AGNs.
The following three modes have been proposed for transferring gas onto the
center of galaxy \citep[e.g.,][]{Croton+06,Lagos+08,Keres+09,Fanidakis+13}:

First one is a secular mode which arise through internal dynamical processes
in the disk such as a bar instability or external processes of galaxy 
interactions.
Some authors claim that low luminosity AGNs hosted by late type disk
galaxies are driven by this mode \citep[see review by][and references therein]{Kormendy+04}.

Second one is a merger mode in which gravitational torques
induced by galaxy-galaxy major/minor mergers drive
inflows of cold gas toward the center of galaxies, triggering the central 
starbursts and also accretion on to the SMBH, i.e. AGN \citep[e.g.,][]{Hopkins+08}.
There are observational evidence that some of the luminous AGN, i.e. QSOs,
are likely triggered by a major merger \citep[e.g.,][]{Sanders+88,Treister+12}.

There has been an observational evidence that AGN activity is enhanced
for close companion galaxies which are assumed to be undergoing early
stage of interaction \citep{Silverman+11}.
However, the fraction of such close pairs is only $\sim$18\%, and 
visual inspection of HST image revealed that $\sim$ 85\% of AGN host 
galaxies show no strong distortions on their morphologies \citep{Cisternas+11}.
Thus galaxy interaction including major merger cannot be the only
mechanism for fueling SMBHs, at least for AGNs with moderate luminosity
sampled by \citet{Silverman+11} and \cite{Cisternas+11}.
\citet{Kaviraj+14} analyzed SDSS data to probe the role of minor mergers
in driving stellar mass and BH growth in galaxies, and suggest
that around half of the star formation activity is triggered by 
the minor-merger process.
Third one is a hot halo mode which is characterized by quiescent accretion 
of gas from the hot halo~\citep{Keres+09,Fanidakis+13}.
In this mode the accretion rate is lower than the other two modes. 
However it becomes increasingly effective accretion mode at higher 
dark matter halo masses of $M_{\rm h} > 10^{12.5}\msun$,
as the accretion by
the other two modes becomes
inefficient due to
AGN feedback.
Thus larger fraction of SMBHs 
are expected to evolve through the hot halo mode 
as the host halo mass increases above $M_{\rm h} > 10^{12.5}\msun$,
and the corresponding mass of SMBH is $> 10^{9} \msun$ \citep{Fanidakis+13b}.

Due to the lower accretion rate, the expected AGN luminosity in this mode 
is typically lower than those of the other two modes 
in a dark matter halo with the same mass.
Observationally such accretion is often associated with BHs
that are characterized by very low radiative efficiency.

\citet{Kauffmann+09} observationally found that there are two distinct modes 
in BH growth.
One is associated with galaxy bulges that are undergoing significant 
star formation, and the other is associated with those with
little or no on-going star formation.
The former one could be related with the star formation induced by secular
and/or merger mode, and the latter with the hot halo mode.

In addition to the three modes, a model of accretion of recycled gas from
evolving star has been studied \citep{Ciotti+01,Ciotti+07}.
According to this model, the recycled gas accumulates within a galaxy under the
influence of gas heating from Type Ia supernovae, and a fraction of the gas
intermittently accretes to the central SMBH under the effect of AGN feedback.
With this scenario the central SMBH grows up to $10^{9.5} \msun$ in giant
ellipticals.
The mass of dark matter halo in which SMBH resides can be estimated 
through the large-scale environments.
As an estimator of large-scale environments, cross-correlation length 
and/or a bias parameter are usually used.
The absolute bias of AGNs distribution relative to that of the dark matter
can be derived from the observed two point correlation function, which is 
then used to estimate mass of the dark matter halo.

\citet{Ross+09} performed correlation analysis for SDSS QSOs with
redshift from 0.3 to 2.2, and obtained a result that the QSOs inhabit dark 
matter haloes of constant mass of $\sim 2\times10^{12} h^{-1} \msun$ at any
redshifts.
Combining this observational result with the theoretical work by 
\citet{Fanidakis+13}, the optical AGNs/QSOs are expected to be mostly 
powered by either secular or merger mode.
It is also inferred that they appear only in the dark matter halo of which 
the mass does not exceed a critical mass that is determined by the AGN 
feedback.

\citet{Hickox+09} analyzed the data of radio, X-ray, and IR detected AGN
samples, and found that the corresponding masses of dark matter haloes are 
$\sim 10^{13.5} h^{-1}\msun$, $\sim 10^{13.0} h^{-1}\msun$, and 
$\sim 10^{12.0} h^{-1}\msun$, respectively.
\citet{Krumpe+12} also derived the halo mass for the X-ray selected ROSAT 
AGNs and the optically selected SDSS AGNs, and they obtained $\sim 10^{13.2} h^{-1}\msun$ 
and $\sim 10^{12.7} h^{-1}\msun$, respectively.

\citet{Shen+11} studied the dependence of quasar clustering on luminosity, 
mass of SMBH, quasar color, and radio loudness. 
They did not find significant dependence on either luminosity, mass, or color,
while they marginally found that the most luminous and most massive 
QSOs are more strongly clustered than the remainder of the sample at
$\sim$2 sigma level, and radio-loud QSOs are more strongly clustered than
radio-quiet QSOs.

Many other studies on clustering and/or environment of AGNs/QSOs have been 
reported elsewhere 
\citep[e.g.,][]{Croom+05,Coil+09,Silverman+09,Donoso+10,Allevato+11,Bradshaw+11,Mountrichas+13,
  Zhang+13,Georgakakis+14}.

All these studies indicates that some types of AGNs such as those bright in
the radio and/or X-ray band inhabits relatively larger
dark matter haloes than optically bright AGNs, and at least part of them
may be fueled 
by a different mechanism such as hot halo mode or accretion of recycled
gas from evolved stars.
To investigate the relation between the environment and intrinsic properties
of AGNs with better accuracy, we have developed a clustering analysis method 
which does not require measurements of galaxy redshift and thus can utilize
all the galaxies detected in the imaging data \citep{Shirasaki+11,Komiya+13}.
\citet{Komiya+13} examined dependence of AGN-galaxy clustering on BH mass,
and found an indication of an increasing trend of cross-correlation length
above $10^{8.2} \msun$.

In this paper, following the result of \citet{Komiya+13}, we study the
properties of galaxies, such as color and luminosity function,
around AGNs and its dependence on BH mass by combining optical and near
infrared dataset of SDSS and UKIDSS survey.
We also derived AGN bias parameter to be compared with
other studies.
It is expected that, if the increase of clustering strength for AGNs with
higher BH mass is due to the prominence of hot halo mode, most of the
galaxies around the AGNs are in red sequence since they reside in haloes 
where gas cooling and star formation have been shut off by AGN feedback.
Accretion of recycled gas from evolving stars may overcome in such a
circumstance.
Which accretion occurs may depends on the availability of the gas
accumulated inside the galaxy and the strength of the AGN feedback.
As another possibility, if the AGNs with the most massive SMBH are mostly
fueled by mergers with gas-rich galaxies, the environment of the AGNs may be
dominated with gas-rich blue star-forming galaxies.

Although it is already known that the massive dark matter halos are
dominated by massive red galaxies~\citep[e.g][]{Zehavi+11} and it may be
inferred that those red galaxies are the cause of increase in clustering
found by \citet{Komiya+13}, such speculation has not been proven by any 
observations so far.
This work gives the first observational evidence about the nature of
galaxies around the AGNs as a function of BH mass based on statistically 
significant number of samples, and provides an important clue to identify
the feeding mechanism of the most massive SMBH.

Throughout this paper, we assume a cosmology with $\Omega_{m} = 0.3$, 
$\Omega_{\lambda} = 0.7$, $h = 0.7$ and $\sigma_{8} = 0.8$.
All magnitudes are given in the AB system.
All the distances are measured in comoving coordinates.
The correlation length is presented in unit of $h^{-1}$Mpc.

\section{Datasets}

\subsection{AGNs}

\begin{figure*}
\includegraphics[width=0.8\textwidth]{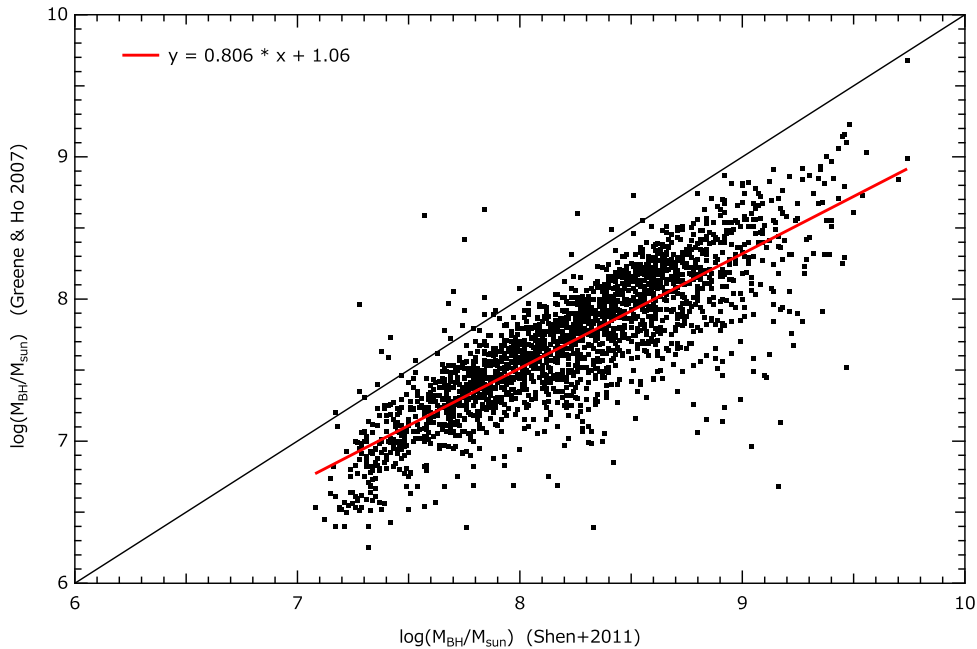}
\caption{Comparison of black hole masses derived by \citet{Shen+11} (horizontal axis) 
and \citet{Greene+07a} (vertical axis). The solid red line represents a liner function
fitted to the data points.}
\label{fig:mass_comparison}
\end{figure*}

The AGN samples were extracted from two AGN properties catalogs 
created by \citet{Shen+11} and \citet{Greene+07a}.
These catalog contains virial mass estimates of SMBHs, which were 
measured from the FWHM of emission lines and continuum flux density.
We examined systematic difference of the mass estimates between 
the two catalogs by extracting AGNs which are contained in both of 
the catalogs.
As shown in Figure~\ref{fig:mass_comparison}, the mass estimates of
\citet{Greene+07a} $M_{\rm G07}$ is relatively lower than the estimates
of \citet{Shen+11} $M_{\rm S11}$ by 0.5 dex.
We therefore calibrated $M_{\rm G07}$ with $M_{\rm S11}$ using the following 
formula:
\begin{equation}
M'_{\rm G07} = (M_{\rm G07} - 1.06) / 0.806.
\end{equation}
We used $M'_{\rm G07}$ as mass estimates of AGNs which are included only
in the catalog of \citet{Greene+07a}.
For the other AGNs, we used $M_{\rm S11}$ as their mass estimates.

\begin{figure*}
\includegraphics[width=0.8\textwidth]{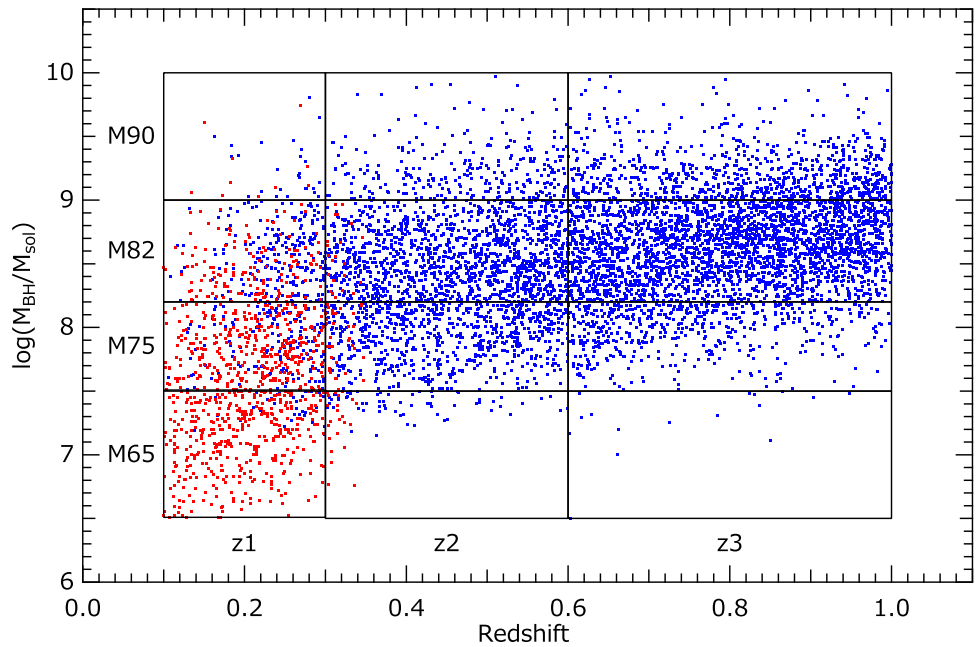}
\caption{Distribution of redshift and BH mass of AGNs. AGNs from \citet{Greene+07a}
and \citet{Shen+11} are plotted in red and blue colors, respectively. The mass correction
is applied to AGNs of \citet{Greene+07a} to compensate the systematic difference between
the two catalogs.}
\label{fig:M-z}
\end{figure*}

\begin{figure*}
\includegraphics[width=0.8\textwidth]{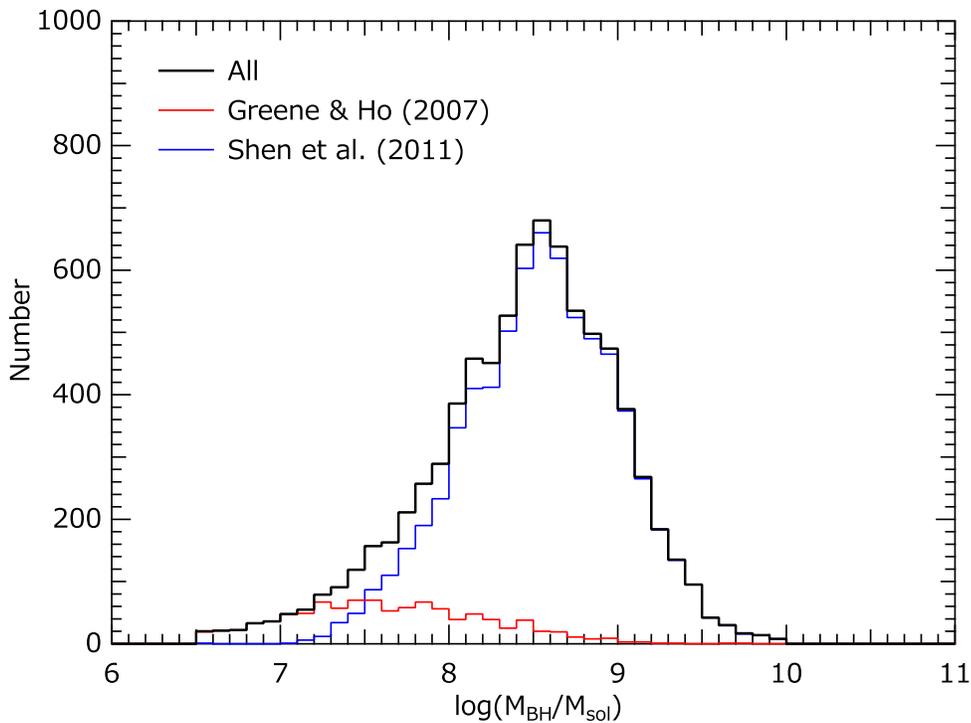}
\caption{Distribution of BH mass for the same sample as shown in Figure~\ref{fig:M-z}.
The red histogram with smaller masses is for \citet{Greene+07a}, the blue one
is for \citet{Shen+11}. The sum of them is shown with the black histogram.
}
\label{fig:M}
\end{figure*}

\begin{figure*}
\includegraphics[width=0.8\textwidth]{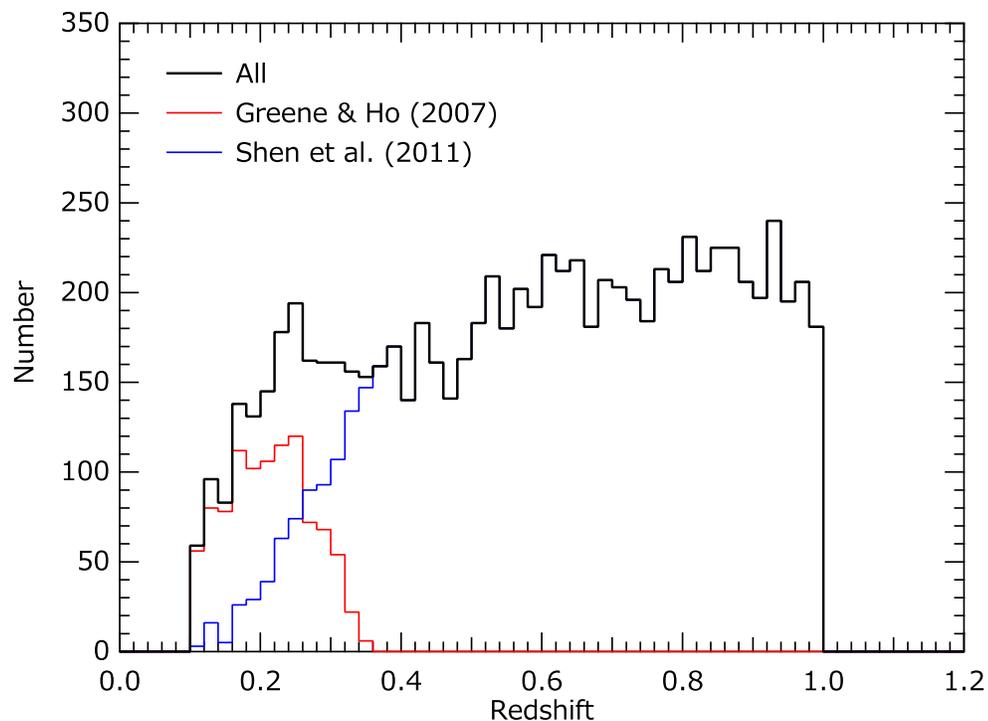}
\caption{Distribution of redshift for the same sample as shown in Figure~\ref{fig:M-z}.
The red histogram with smaller redshift is for \citet{Greene+07a}, the blue one
is for \citet{Shen+11}. The sum of them is shown with the black histogram.}
\label{fig:z}
\end{figure*}

In this work the range of redshift is restricted to 0.1--1.0, and 
mass of SMBH is restricted to the range of $10^{6.5}$--$10^{10} \msun$.
The redshift and mass distributions of AGNs used in this work are shown
in Figures~\ref{fig:M-z}--\ref{fig:z}.
The AGN samples from \citet{Greene+07a} mostly comprise lower redshift
and lower mass samples.
All the samples are divided into 12 groups shown in Figure~\ref{fig:M-z},
and analysis are performed for each group with enough
statistics or a combined group.
Sample selection is applied following the criteria described in 
section~\ref{sec:data_selection}.
The number of analyzed AGNs is 991 from \citet{Greene+07a}, 
7068 from \citet{Shen+11}, and 8059 in total.

\subsection{Galaxies}\label{sec:galaxies}

The galaxy samples were retrieved from Virtual Observatory (VO) services 
of SDSS DR8~\citep{Aihara+11} and UKIDSS DR9 LAS catalog.
The UKIDSS project is defined in \citet{Lowrence+07}.
The data retrieval and creation of the merged catalog were performed for each 
AGN.
The JVO command line tool (jc client~\footnote{\url{http://jvo.nao.ac.jp/jc_clinet/}})
was used to automate the data retrieval.
The following criteria are used to retrieve the galaxy sample:
For SDSS catalog data of which \verb|resolveStatus| attribute equals to 257 
or 258, which corresponds to select unique objects, are retrieved.
For UKIDSS LAS catalog data of which \verb|mergedClass| attribute equals to 1 or $-3$, 
which corresponds to select objects flagged as ``Galaxy'' or ``Probable galaxy'',
are retrieved.
For both catalog data within 1 degree from the AGN coordinates are retrieved.

Photometric magnitude used in this work is an aperture magnitude in 2 arcsec 
diameter for both SDSS and UKIDSS.
Due to the difference of the point spread functions (PSF) between the two surveys 
($\sim$1.3 arcsec for SDSS in FWHM of PSF, $\sim$0.8 arcsec for UKIDSS), color
measurements suffer a systematic error especially for objects whose apparent 
size exceeds the aperture size.
In this work the comparisons of galaxy color are made for two samples of lower 
and higher BH mass, and both of the samples are equally affected by the 
error of the color estimates.
Thus the inaccuracy in the color estimates is not so harmless to derive the
relative difference of the distributions of galaxy color.
The two catalogs were merged into a single one.
In creating the merged catalog, UKIDSS objects detected in the K band data were 
used as a reference, and a nearest neighbor was selected from the SDSS catalog
within 2 arcsec distance for each UKIDSS object.
The UKIDSS objects for which SDSS counterpart was not found were preserved in
the merged catalog, while the SDSS objects which were not selected as a 
counterpart were not included in the merged catalog.
Thus our galaxy samples are K-band selected ones.

\subsection{Data selection}
\label{sec:data_selection}

Data selection was applied to the AGN datasets to ensure their quality as follows:
at first, coverage of the SDSS and UKIDSS galaxy samples were investigated
by using the window function for SDSS and frame metadata
describing the region of observation for UKIDSS.
Regions outside the observed area were identified as a dead region.

We found that some of the UKIDSS galaxy samples were spurious events caused by 
cosmic rays, bright stars, or unknown reason.
The area that contains the spurious events was identified by looking for a
high density region in the count map of the samples detected in the K-band and 
not detected in SDSS.
Those areas were taken as dead regions as well.

The fractions of the dead region were calculated as a function of projected 
distance from the AGN position with 0.2~Mpc bin width.
If the fraction exceeds 0.2 at a projected distance less than 5~Mpc, the AGN dataset
was removed from the samples.
Otherwise the fractions are used to correct the effective area in deriving 
the number density of galaxies.
Next we examined the uniformity of the galaxy number density around AGN.
Since the number of galaxies associated with the AGNs is usually 
much smaller than
the total number of background/foreground galaxies, a flat distribution is 
expected for the number density as a function of distance from AGN.
If there exist a cluster/group of galaxies or stars in front of the AGN
field, it can produce strong fluctuation in the distribution of number density.
Inhomogeneity of the depth of observations can also produce discontinuous
density distribution.

To reduce 
a false signal produced by those cases, we calculated three parameters,
$\chi^{2}$, $\sigma_{\mbox{\scriptsize max}}$, and $B_{\rm QG}$, and they were used to
filter the datasets showing non-uniformity in number density.
The parameter $\chi^{2}$ is a square sum of the deviation from a flat distribution,
and the parameter $\sigma_{\mbox{\scriptsize max}}$ is a maximum deviation from the
average density calculated at a distance range from 2 to 5~Mpc.
The adapted criteria are $\chi^{2}/(n-1) \le 3.0$ and 
$\sigma_{\mbox{\scriptsize max}} \le 5$, where $n$ is the number of data points in
the galaxy number density distribution.
$B_{\rm QG}$ was calculated as \citep{Longair+79},
\begin{equation}
 B_{\rm QG} = \frac{3-\gamma}{2 \pi} \frac{N_{\rm total}-N_{\rm bg}}{\rho_{0}},
\end{equation}
where $N_{\rm total}$ is the total number of galaxies at projected distance
less than 1~Mpc from AGN, $N_{\rm bg}$ is the expected number of 
background/foreground galaxies which are not associated with the AGN and is
estimated from the number density at $r_{p} =$ 3.0--5.0~Mpc, $\rho_{0}$ is the
average number density of observed galaxy at the AGN redshift.
This parameter is usually used to estimate the clustering strength, but here
it is used to remove datasets showing extraordinary large positive/negative
excess density around AGNs.
The criterion adapted is $-10000 \le B_{\rm QG} \le 10000$.
Only two samples were discarded with this selection.
We checked that the samples for which clustering strength is smaller 
than $\sim$80~Mpc in term of correlation length don't give rise a 
prominent fluctuation detected with the above criteria in the galaxy 
density histogram.
As such, these criteria are almost free from selection bias caused by
the clustering strength itself at the source.
To maximize the signal-to-noise ratio, we removed datasets of shallow observations
based on the parameter $\rho_{0}$, which is an average number density of galaxies 
at the AGN redshift;
$\rho_{0}$ is calculated from the luminosity function which is parametrized as 
a function of redshift $z$ and rest-frame wavelength $\lambda$ as discussed in 
\citet{Komiya+13}.
We adapted a criteria of $\rho_{0} \ge 10^{-4}$~Mpc$^{-3}$.

The total number of AGN datasets that have passed above criteria is 8059 out of 
the original number of 9986.
About 20\% of datasets are discarded.
Among the 8059 samples, $\sim$80\% of the samples overlap with the \citet{Komiya+13}
samples.

\section{Analysis Method}

\subsection{Correlation length}

The analysis method used for calculating the AGN-galaxy cross-correlation function
is completely the same as that used in \citet{Komiya+13}, which is briefly described
here.

The cross-correlation function of AGNs and galaxies $\xi(r)$ can be expressed
as an excess in number density of galaxies $\rho(r)$ relative to the average
density $\rho_{0}$ at the AGN redshift,
\begin{equation}
   \xi(r) = \frac{\rho(r)}{\rho_{0}} - 1,
   \label{eq:xi}
\end{equation}
where $r$ represents the distance from an AGN.
We assume the power-law form for the cross-correlation function,
\begin{equation}
   \xi(r) = \left( \frac{r}{r_{0}} \right)^{-\gamma},
\end{equation}
where $r_{0}$ is the correlation length and $\gamma$ is the power-law index
fixed to 1.8, which is a canonical value measured by many other clustering 
study of galaxies and QSOs.
The projected cross-correlation function $\omega(r_{p})$ is calculated by integrating
Equation~(\ref{eq:xi}) as
\begin{eqnarray}
   \omega(r_{p}) & = & 2 \int_{0}^{\infty} \xi(r_{p}, \pi)d\pi
      = 2 \int_{r_{p}}^{\infty} \frac{r \xi(r)}{\sqrt{r^{2} - r_{p}^{2}}} dr \nonumber \\
      & = & r_{p} \left( \frac{r_{0}}{r_{p}} \right)^{\gamma} 
      \frac{\Gamma(\frac{1}{2})\Gamma(\frac{\gamma-1}{2})}{\Gamma(\frac{\gamma}{2})},
      \label{eq:omega_1}
\end{eqnarray}
where $\pi$ and $r_{p}$ are distance along and perpendicular to the line of sight,
respectively, and $\Gamma$ is the gamma function.
$\omega(r_{p})$ can be derived observationally from the surface density of 
galaxies $n(r_{p})$ as
\begin{equation}
   \omega(r_{p}) = \frac{n(r_{p}) - n_{\rm bg}}{\rho_{0}},
   \label{eq:omega_2}
\end{equation}
where $n_{\rm bg}$ represents the surface density of background/foreground galaxies 
which are unassociated with the corresponding AGN.
From Equations~(\ref{eq:omega_1}) and (\ref{eq:omega_2}), the surface density of 
galaxies around an AGN can be modeled as
\begin{equation}
   n(r_{p}) = C(\gamma) \cdot \rho_{0} \cdot r_{p} \left( \frac{r_{0}}{r_{p}} \right) ^{\gamma} 
            + n_{\rm bg},
   \label{eq:model}
\end{equation}
where the term of the gamma function is represented by $C(\gamma)$.
Fitting this model function to the observed surface density, we can obtain
the best estimates of $r_{0}$ and $n_{\rm bg}$.
As already mentioned, $\gamma$ is fixed to 1.8 and $\rho_{0}$ is calculated from
the empirical formula of galaxy luminosity function.
The details of the parametrized luminosity function are described in \citet{Komiya+13}.
Since the clustering signal is too weak to obtain meaningful parameter values 
for each AGN dataset, we applied the fitting to the average of $n(r_{p})$ and
$\rho_{0}$ for a given AGN group.

The uncertainty of $r_{0}$ is calculated as a root sum square of
one sigma statistical error and systematic error derived from the uncertainty 
of $\rho_{0}$ as discussed in \citet{Komiya+13}.
The uncertainty related with an intrinsic variance of clustering 
is also estimated from a difference between the cross correlation lengths calculated
for two independent sub samples which are constructed by dividing the original
samples.

It should be noted that the cross-correlation length obtained by this method is
not a simple average for the AGN group, but an average weighted with $\rho_{0}$.
Thus the result is biased to the low-$z$ AGN samples.

\subsection{AGN absolute bias}
\label{sec:halo_mass}

The linear bias of AGNs relative to the dark matter distribution can be 
derived from the cross-correlation length between AGNs and galaxies as follows:
Assuming linear bias, the autocorrelation function of AGNs $\xi_{\rm AA}$ 
is related with the cross-correlation function between AGNs and galaxies 
$\xi_{\rm AG}$
in the form,
\begin{equation}
  \xi_{\rm AA} = \xi_{\rm AG}^{2} / \xi_{\rm GG},
  \label{eq:xi_AA}
\end{equation}
where $\xi_{\rm GG}$ is the autocorrelation function of galaxies.
The autocorrelation function of galaxies depends on galaxies color and luminosity, 
and measured by \citet{Zehavi+11} as
$\xi_{\rm GG,blue} = (r/3.6~{h^{-1}\rm Mpc})^{-1.7}$ and
$\xi_{\rm GG,red} = (r/6.6~{h^{-1}\rm Mpc})^{-1.9}$
for blue and red galaxies with brightness of $-20 < M_{\rm r} < -19$.
The luminosity dependence is small in the range of $-22 < M_{\rm r} < -17$, while
for galaxies with $M_{\rm r} < -22$ the autocorrelation function significantly
changes to the form of 
$\xi_{\rm GG,red} = (r/10.7~{h^{-1}\rm Mpc})^{-1.9}$ 
In order to take into account the difference of blue/red/brightest galaxy fraction 
in the samples we calculated the autocorrelation function of galaxies with the 
following formulas:
\begin{equation}
  \xi_{\rm GG} = f_{\rm red} \xi_{\rm GG,red} + f_{\rm blue} \xi_{\rm GG,blue},
                                           + f_{\rm bright} \xi_{\rm GG,bright},
  \label{eq:xi_GG}
\end{equation}
where $f_{\rm red}$, $f_{\rm blue}$, $f_{\rm bright}$ is a fraction of red, blue, and brightest
galaxies in the sample which are measured as described in the next section.
The autocorrelation function of AGN is calculated by using
Equation~(\ref{eq:xi_AA}) and (\ref{eq:xi_GG}), and fitted to a power law function of
the form of $\xi_{\rm AA} = (r/r_{\rm AA})^{-\gamma}$.

The linear bias of AGNs $b$ is calculated as \citep{Koutoulidis+13}:
\begin{equation}
b = \frac{\sigma_{\rm 8, AGN}}{\sigma_{\rm 8,DM}},
\label{eq:b}
\end{equation}
where $\sigma_{8,\rm AGN}$ and $\sigma_{8,\rm DM}$ is the rms fluctuations 
of the AGN and dark matter density distribution within spheres of a comoving 
radius of 8 $h^{-1}$ Mpc, respectively.
$\sigma_{8,\rm AGN}$ is given by 
\begin{equation}
\sigma_{8,\rm AGN} = J_{2}(\gamma)^{1/2} \left( \frac{r_{\rm AA}}{8} \right) ^{\gamma/2},
\label{eq:sg8_AGN}
\end{equation}
\begin{equation}
J_{2}(\gamma) = \frac{72}{(3-\gamma)(4-\gamma)(6-\gamma)2^{\gamma}},
\end{equation}
and $\sigma_{8,\rm DM}$ is
\begin{equation}
\sigma_{8,\rm DM} = \sigma_{8} \frac{D(z)}{D(0)},
\label{eq:sg8_DMH}
\end{equation}
where $D(z)$ is the linear growth factor given as,
\begin{equation}
D(z) = \frac{5 \Omega_{\rm m} E(z)}{2} \int_{z}^{\infty} \frac{1+y}{E^{3}(y)}dy,
\end{equation}
\begin{equation}
E(z)^{2} = \Omega_{\rm m}(1 + z)^{3} + \Omega_{\Lambda}
\end{equation}

Using Equations~(\ref{eq:b}), (\ref{eq:sg8_AGN}) and (\ref{eq:sg8_DMH}),
bias is calculated from the AGN autocorrelation length as:
\begin{equation}
b = \left( \frac{r_{\rm AA}}{8} \right)^{\gamma/2} J_{2}(\gamma)^{1/2} 
       \left( \frac{\sigma_{8} D(z)}{D(0)} \right)^{-1}
\label{eq:b_r0}
\end{equation}

\subsection{Red galaxy fraction and luminosity distribution}
\label{sec:red_frac}

Using the UKIDSS/SDSS merged catalog obtained as described in 
section~\ref{sec:galaxies}, color of each galaxy was calculated
by performing SED fitting.
The SED fitting was performed by using EAZY software developed by
\citet{Brammer+08}.
As we are interested only in the galaxies associated with the AGNs,
the redshift was fixed to the AGN redshift in the SED fitting.
Photometric redshift was not used since the expected error can be large
and does not improve the statistics significantly.
It might also be possible that photo-z is not accurately
determined for blue galaxies since the photo-z is mainly determined by
the structure at 4000~{\AA}, so this can 
produce systematic bias for selecting red galaxies preferentially.
Instead of using the photometric redshift, we adapted 
filtering based on $\chi^{2}$ of the SED fitting to select galaxies 
located at  AGN redshifts inclusively.

A color estimator $D_{\rm opt-IR}$ is defined as 
\begin{equation}
   D_{\rm opt-IR} =  m_{\rm opt} - m_{\rm IR},
\end{equation}
where $m_{\rm opt}$ and $m_{\rm IR}$ are magnitudes at wavelength range 3,000--3,500~{\AA}
and 10,000--12,000~{\AA} in the rest frame, respectively.
In this analysis, we used only the galaxy samples which were detected both in UKIDSS 
and SDSS and for which the reduced $\chi^{2}$ of the SED fitting is less than
a given limit.

The $D_{\rm opt-IR}$ distribution for galaxies associated with the AGNs
can be derived by the subtraction method.
$D_{\rm opt-IR}$ distributions averaged over every AGNs in a group 
are calculated at a central and an offset region of AGN field.
We defined the central region as the region within 1~Mpc from the AGN,
and defined the offset region as an annulus region at a distance of 3 to 5~Mpc.
Since most of the galaxy samples at the offset region are located at redshifts
different from those of AGNs, it can be regarded 
as background galaxies.
By subtracting distribution at offset region from that at central region, we
can obtain the $D_{\rm opt-IR}$ distribution for galaxies associated with the
AGNs.

The reduced $\chi^{2}$ of the SED fitting for galaxies at the offset region
tends to be larger than that of galaxies at the central region,
since the fitting is performed by assuming incorrect redshifts.
We determined the limit of the reduced $\chi^{2}$ for selecting galaxies, by 
comparing the distributions for central and offset galaxies. 
We set the limit to 4.5 for $z<0.6$, and 2.0 for $z\ge0.6$.

\begin{figure*}
\includegraphics[width=0.8\textwidth]{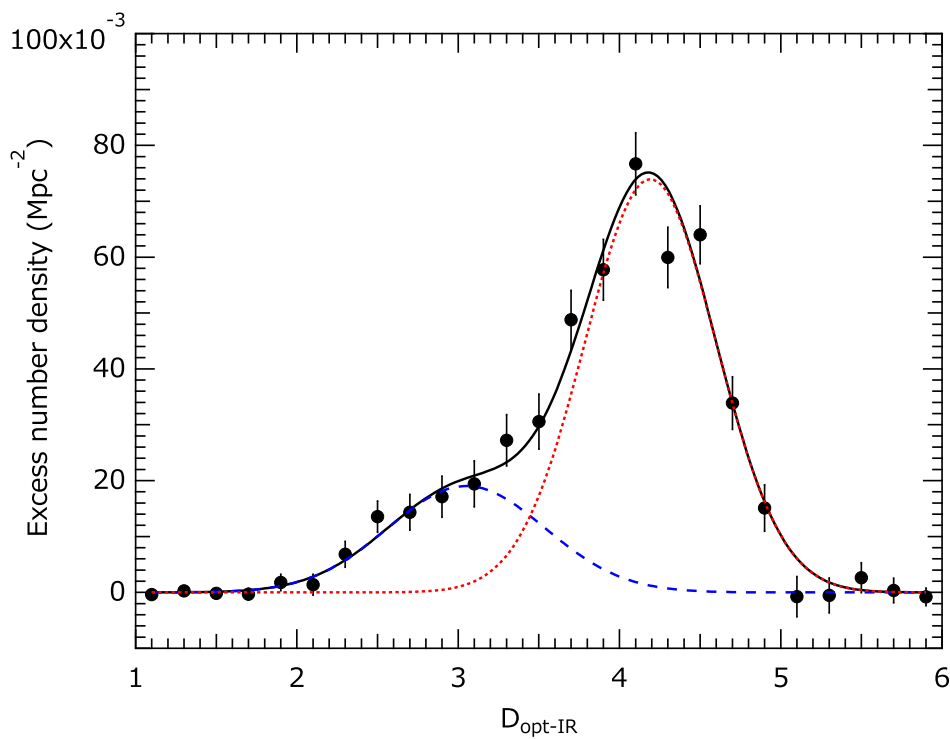}
\caption{$\Dx$ parameter distribution for all the AGN samples is shown with filled
black circles with statistical error bars. The distribution is fitted with a 
two Gaussian model (black solid line). Each component represented with a single 
Gaussian corresponds to blue cloud (peak with smaller $\Dx$) and red sequence 
galaxies (peak with larger $\Dx$).}
\label{fig:D_all}
\end{figure*}

In Figure~\ref{fig:D_all}, we show the distribution of $D_{\rm opt-IR}$ derived 
from all the AGN samples.
The bimodality of galaxy color distribution is clearly seen.
In the same figure, the color distributions for blue and red galaxies are deconvolved
by assuming normal distribution.
The peak values of the color distribution for blue and red galaxies are 3.1 and 4.2,
respectively.
We derive the red galaxy fraction for each AGN group by fitting the color distribution
to a model function given with a sum of two normal distributions.

\begin{figure*}
\includegraphics[width=0.8\textwidth]{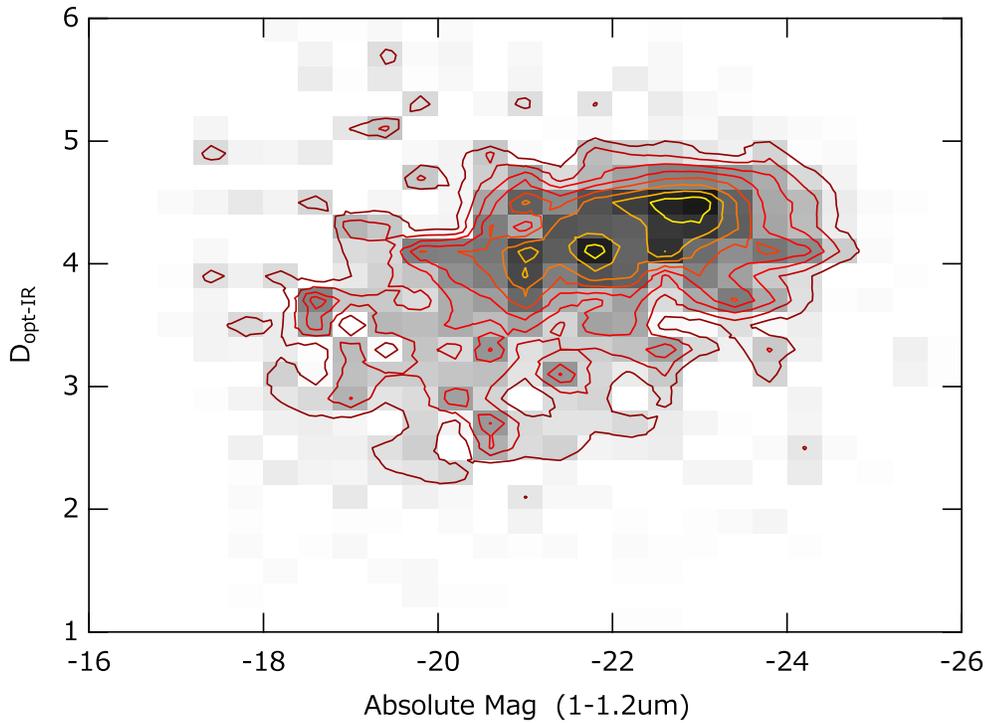}
\caption{Color magnitude distribution for all the AGN samples. Densities are 
represented with gray scales and contours of equi-density. Red sequence galaxies
are concentrated in a upper right region ($M < -20$ mag and $\Dx > 3.5$), while blue
cloud galaxies are widely spread in the dimmer part. }
\label{fig:CM_all}
\end{figure*}

The luminosity distribution can be derived with the same method as that used
for the color distribution described above.
The absolute magnitude was calculated by $m_{\rm IR} - DM(z)$, where $DM(z)$ is the 
distance modulus at redshift $z$.
In Figure~\ref{fig:CM_all}, we show the color-magnitude diagram obtained for all
the AGN samples by the subtraction method.
The red sequence and blue cloud are clearly seen.
Red sequence galaxies are concentrated in a upper right region ($M < -20$ and $\Dx > 3.5$), 
while blue cloud galaxies are widely spread in the dimmer part.

\section{Results}

\subsection{Cross-correlation}

\begin{figure*}
\includegraphics[width=0.8\textwidth]{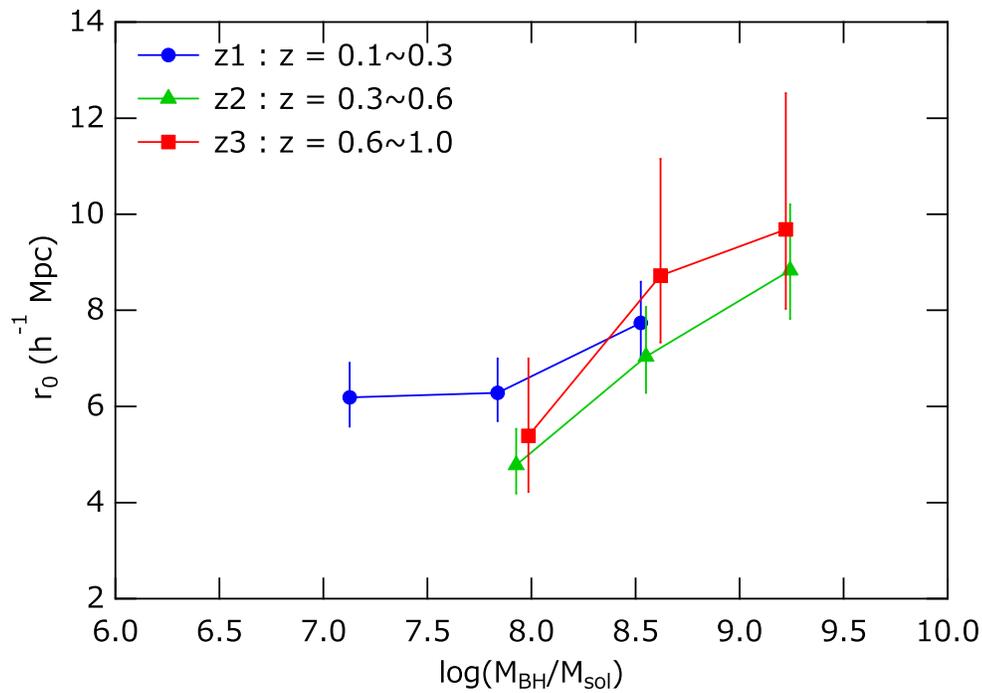}
\caption{Cross-correlation lengths derived for each redshift and mass group
are plotted as a function BH mass. The data point of the same redshift groups 
are connected with a line. Solid circles are for $z1$ redshift group, solid 
triangles and boxes are for $z2$ and $z3$ redshift groups, respectively.
}
\label{fig:r0_mass}
\end{figure*}

\begin{figure*}
\includegraphics[width=0.5\textwidth]{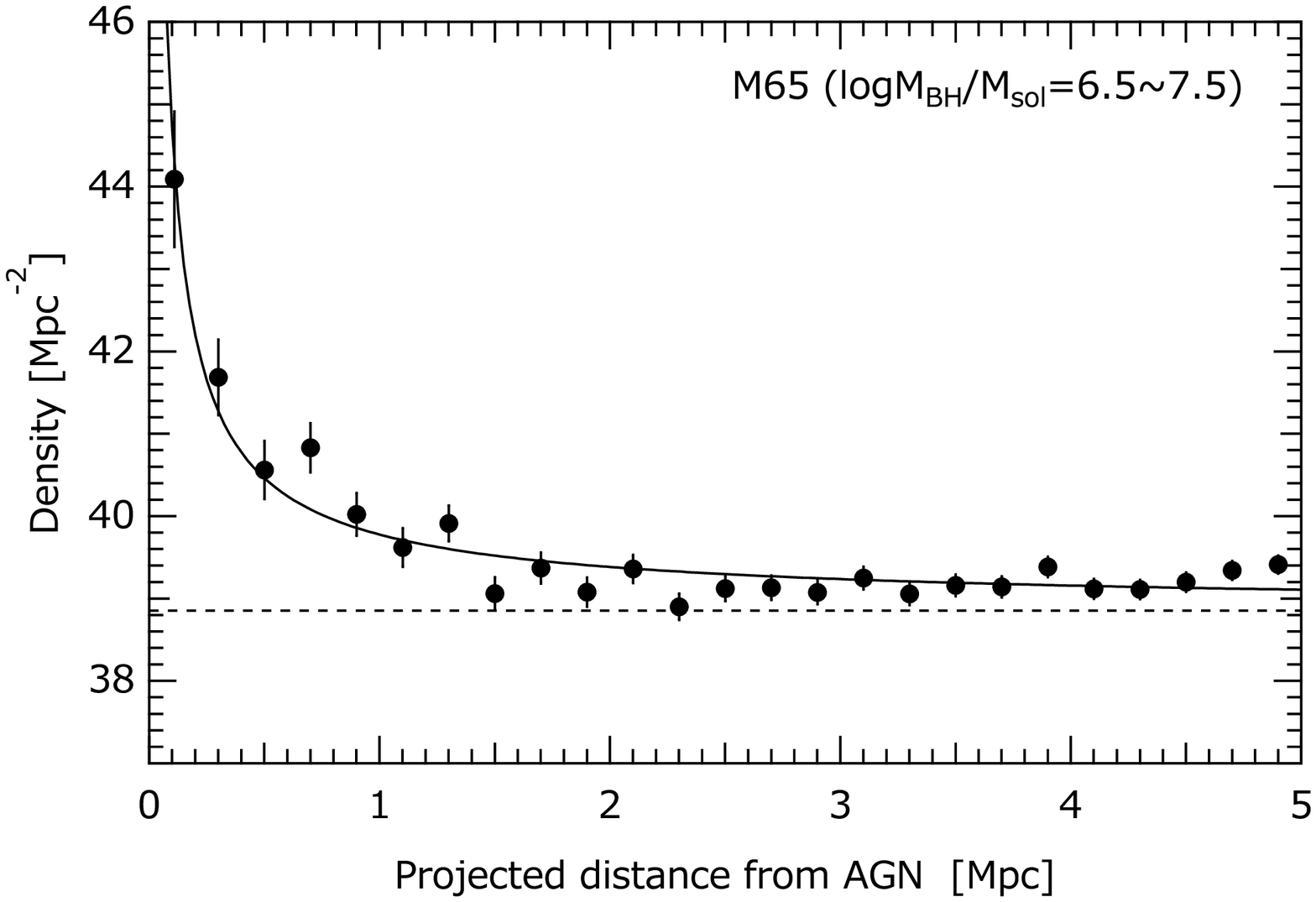}
\includegraphics[width=0.5\textwidth]{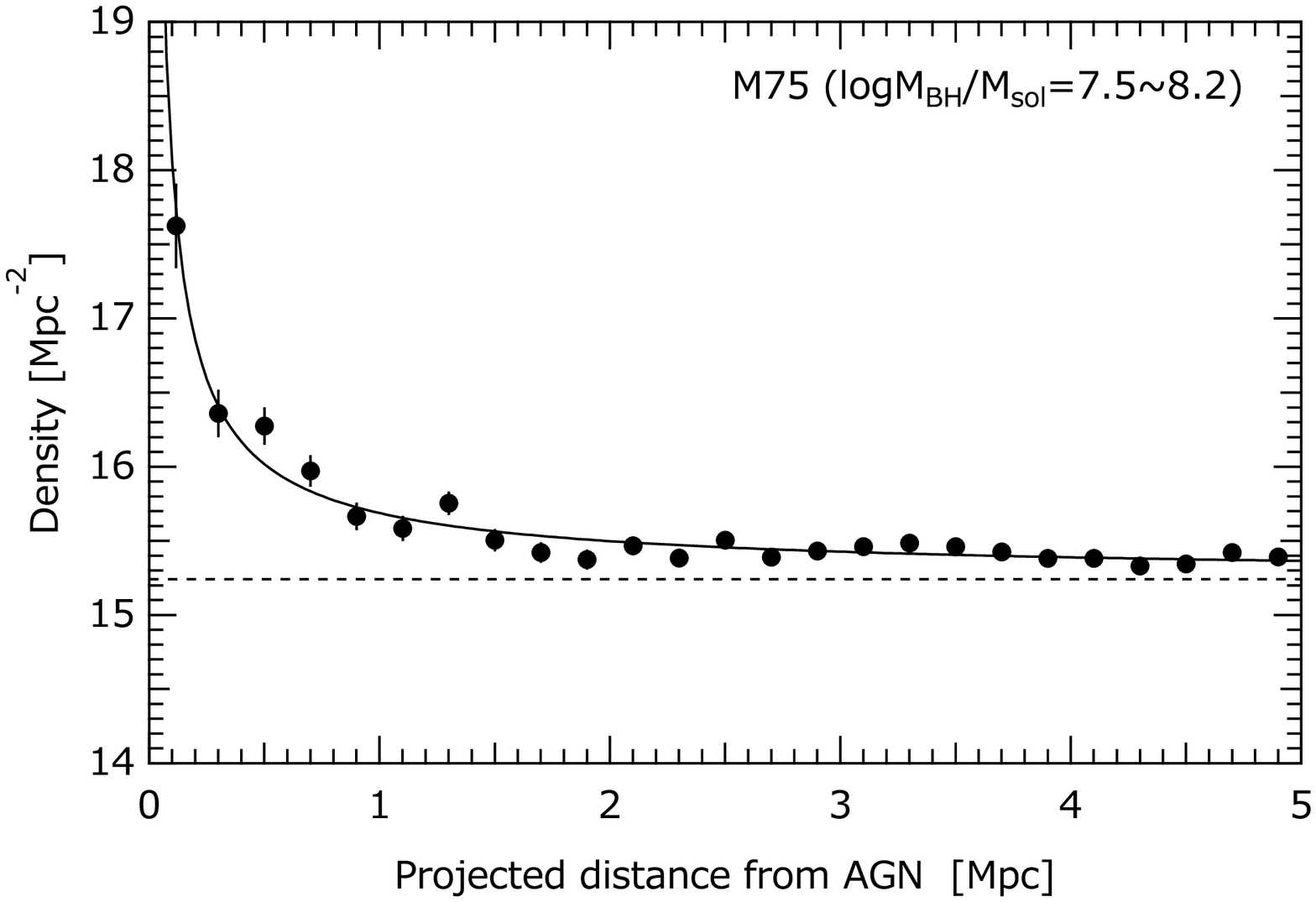}
\\
\includegraphics[width=0.5\textwidth]{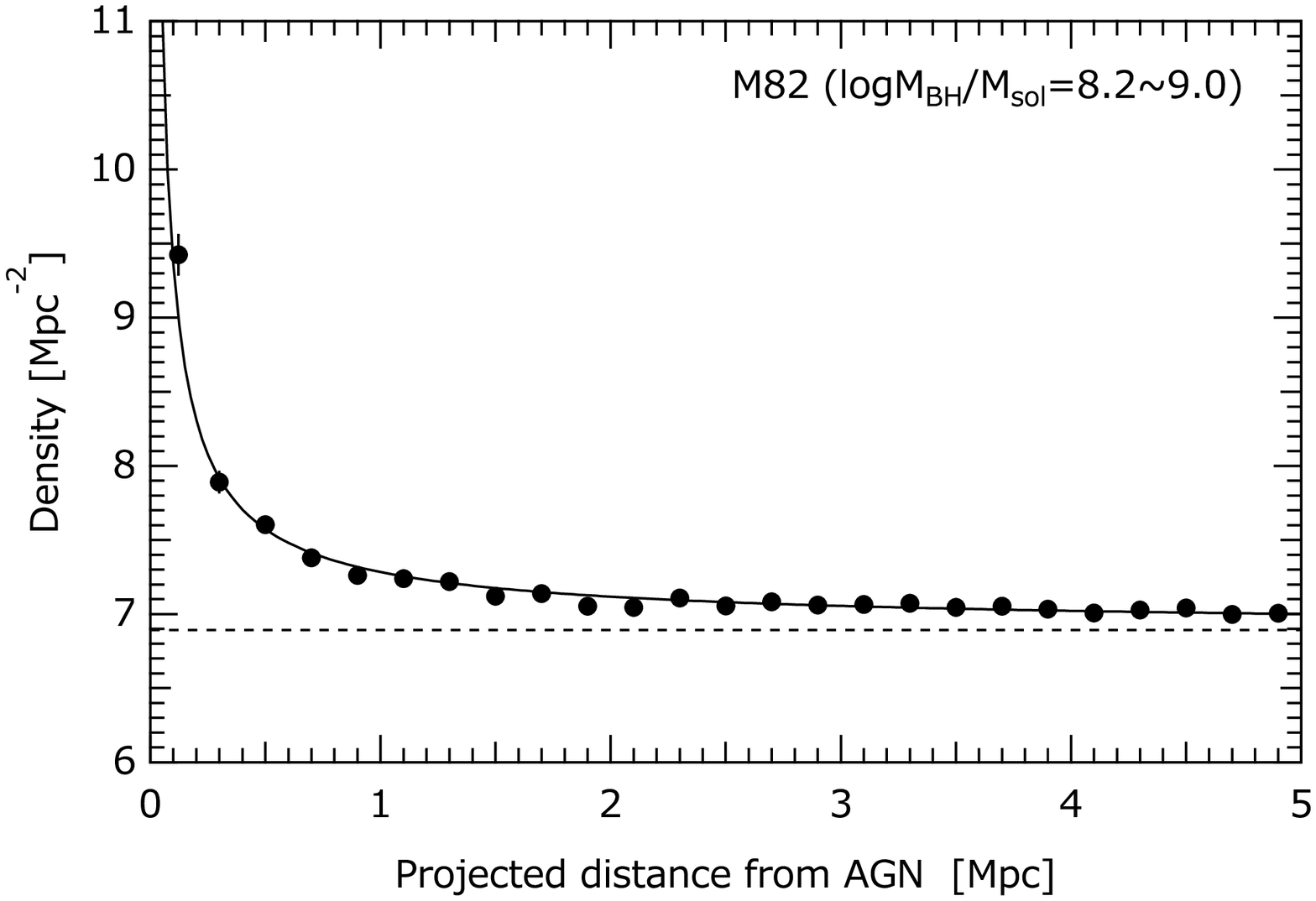}
\includegraphics[width=0.5\textwidth]{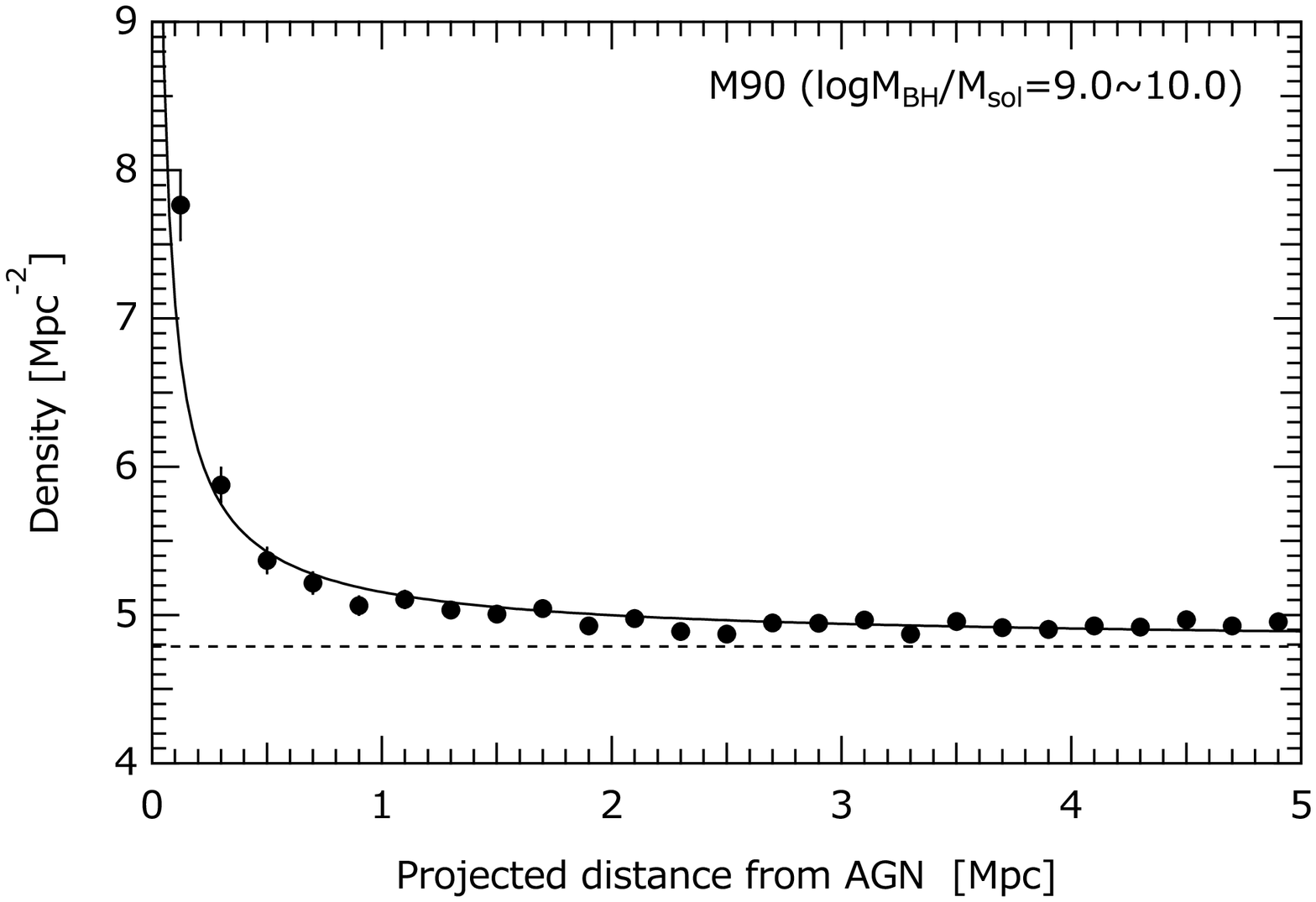}
\caption{Surface densities of galaxies plotted as a function of projected 
distance from AGN for mass group M65, M75, M82 and M90.}
\label{fig:density}
\end{figure*}

\begin{figure*}
\includegraphics[width=0.24\textwidth]{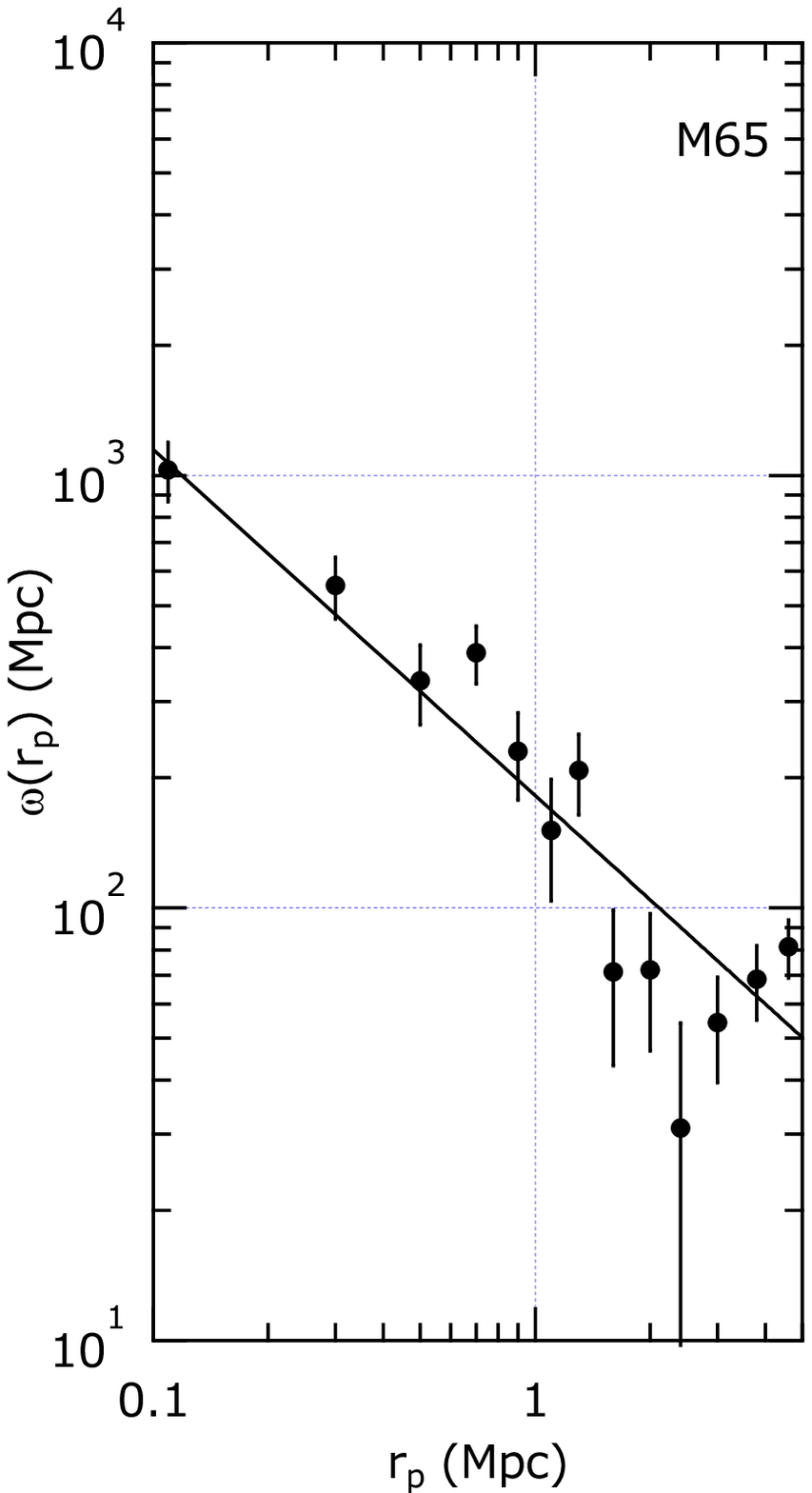}
\includegraphics[width=0.24\textwidth]{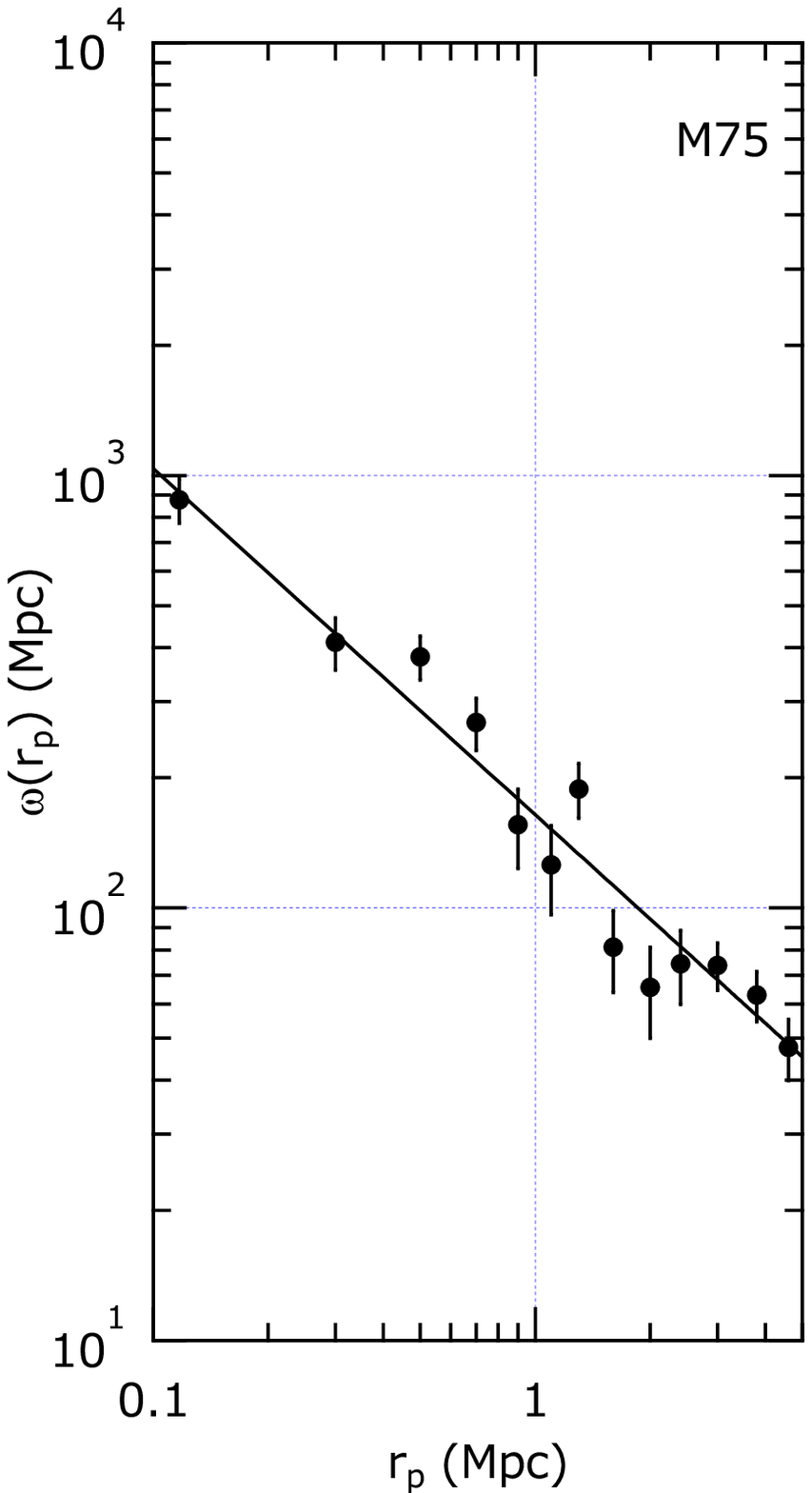}
\includegraphics[width=0.24\textwidth]{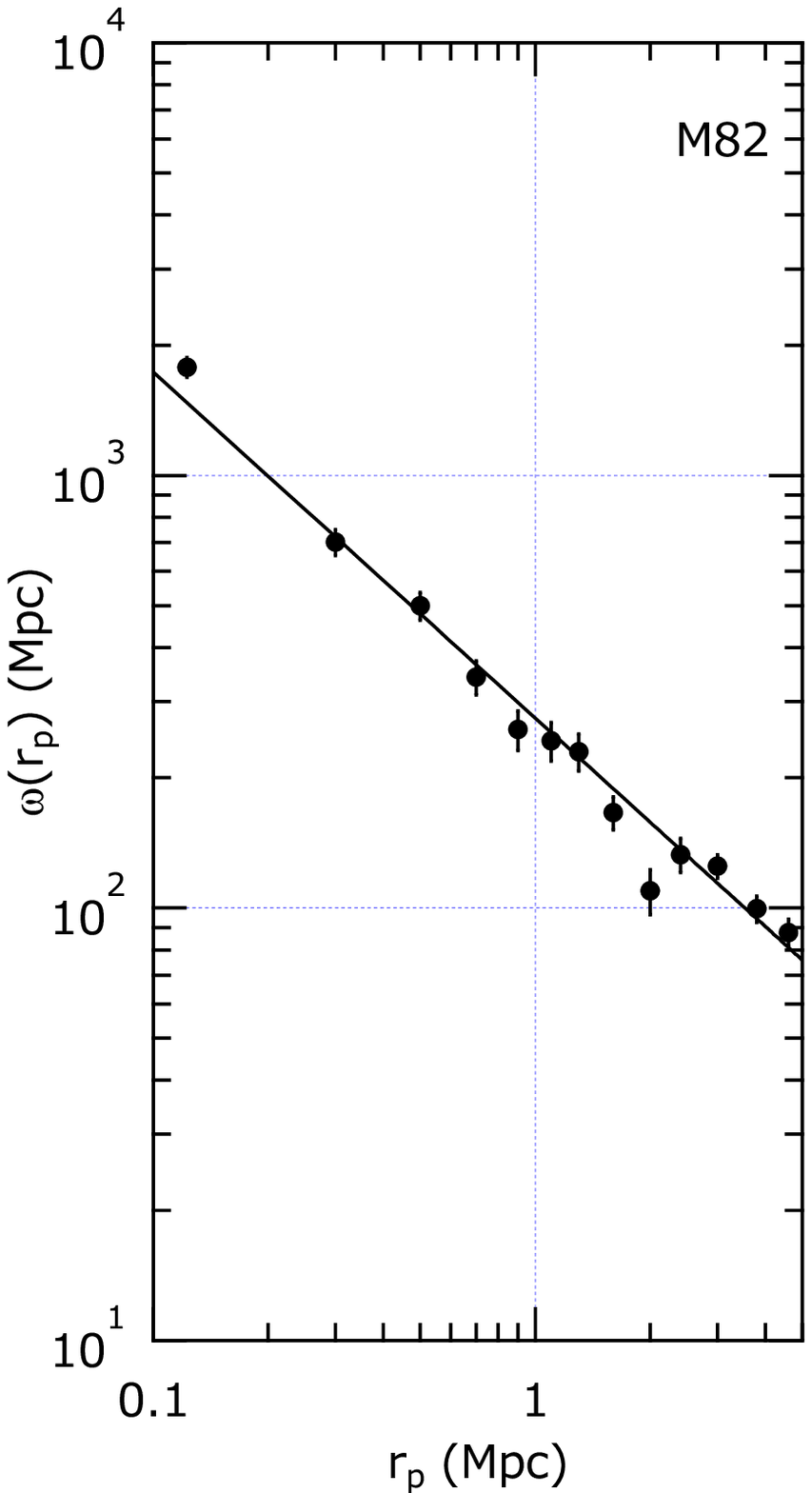}
\includegraphics[width=0.24\textwidth]{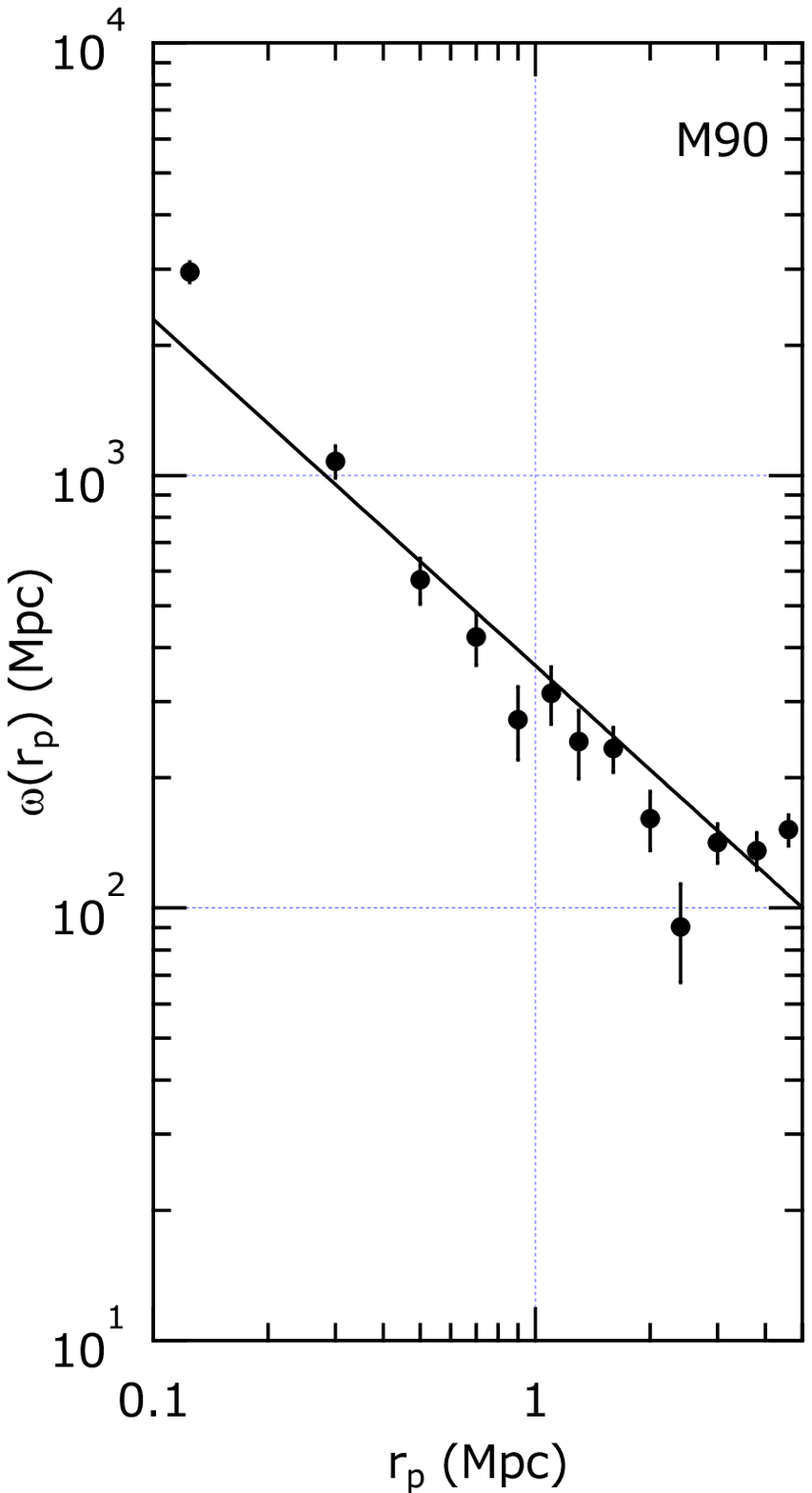}
\caption{Projected cross-correlation functions for mass group M65, M75, M82 and M90.}
\label{fig:omega}
\end{figure*}

\begin{figure*}
\includegraphics[width=0.8\textwidth]{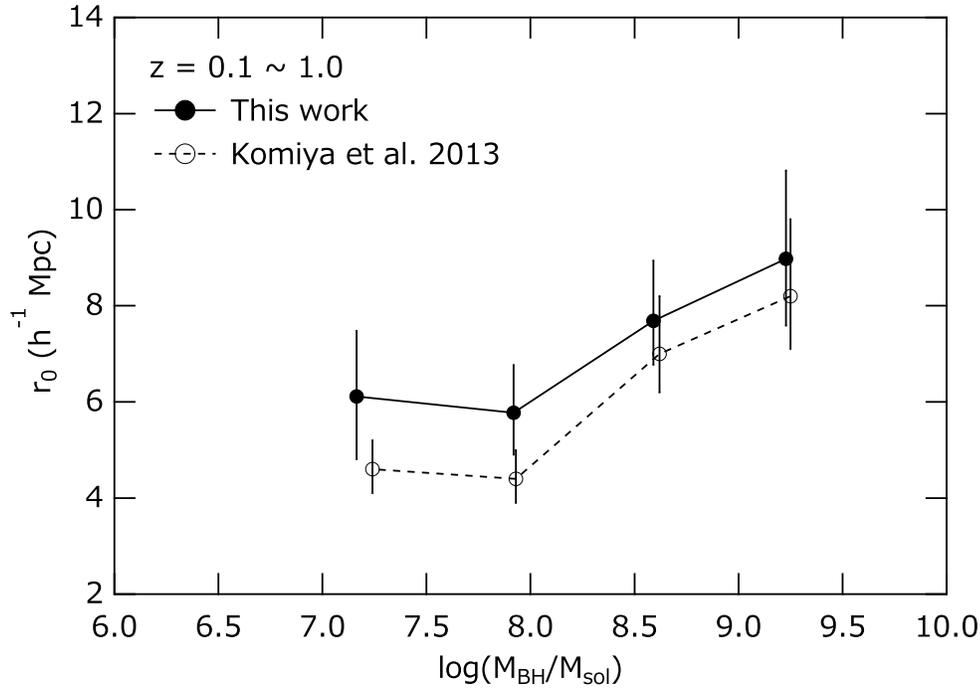}
\caption{Cross-correlation lengths derived for each mass groups are plotted
as a function of BH mass.
The redshift range of each mass groups are 
0.1$\sim$0.3 for M65 mass group,
0.1$\sim$1.0 for M75 and M82 mass groups,
and 0.3$\sim$1.0 for M90 mass group.
Results of this work are shown with filled circles and previous work by
\citet{Komiya+13} are shown with open circles.
The error bars of this work include statistical error, systematic error, and 
uncertainty related with the intrinsic variance among AGNs.
}
\label{fig:r0_mass_all}
\end{figure*}

All the AGN samples are divided into 12 redshift-mass groups as shown
in Figure~\ref{fig:M-z}.
Redshift is divided into three ranges 0.1--0.3, 0.3--0.6 and 0.6--1.0.
We designate them as z1, z2, and z3, respectively.
BH mass is divided into four ranges $\log{(\Mbh/\msun)} = $ 6.5--7.5,
7.5--8.2, 8.2--9.0, 9.0--10.0, and they are designated as M65, M75, M82, 
and M90, respectively.
Hereafter we call, for an example, a group of redshift range of 0.1--0.3 
and mass range of 6.5--7.5 as M65-z1.
The results obtained by fitting the model function of Equation~(\ref{eq:model}) to
the observed average number densities are summarized in the first part of
Table~\ref{tab:fit_r0}.
In Figure~\ref{fig:r0_mass}, the cross-correlation length $r_{0}$ obtained for 
each redshift-mass group is plotted as a function of $\Mbh$.
The result shows that there is a tendency that $r_{0}$ increases
with $\Mbh$, while there is no clear redshift dependence.
As no clear redshift dependence is seen,
the AGN samples are combined for the same mass and all the redshift 
groups to reduce the uncertainty resulting from intrinsic variance of 
clustering around each AGN, then $r_{0}$ is calculated for the combined mass groups.
The projected number densities for these samples and fitted model
functions are shown in Figure~\ref{fig:density}.
The horizontal dashed lines in these figures correspond to $n_{\rm bg}$
obtained from the fitting.
The projected cross-correlation functions calculated from
equation~(\ref{eq:omega_2}) are shown in Figure~\ref{fig:omega},
and are well expressed by a power law function.
The cross-correlation lengths derived from the fitting are 
plotted in Figure~\ref{fig:r0_mass_all}.
The error bars include statistical error, systematic error, and 
uncertainty related with the intrinsic variance among AGNs.
The uncertainty due to the intrinsic variance was estimated from a
difference between the cross correlation lengths calculated for two 
independent sub samples.
The sub samples were constructed by dividing the original AGN samples into 
two groups in random.
We generated 21 sets of two half-sized samples, calculated absolute
differences of cross correlation lengths between the two sub samples $\delta r_{0,1/2}$ 
for each set, and selected median of $\delta r_{0,1/2}$ as an estimate of the 
standard deviation for the half-sized sub sample.
The corresponding error on $r_{0}$ of the original sample was calculated 
as $\sigma_{r_{0}}=\delta r_{0,1/2}/\sqrt{2}$, which is summarized in 
Table~\ref{tab:halo_mass}.
The difference of $r_{0}$ between M75 and M90 groups is about
2.6 sigma excluding the systematic errors.
For comparison the results obtained in the previous work \citep{Komiya+13} 
are also plotted in the figure.
They are consistent with the results of this work.

The AGN linear bias
is estimated from the cross-correlation 
length assuming that the autocorrelation function of the galaxies 
is expressed by 
a linear combination of autocorrelation functions of three types of galaxies.

\citet{Zehavi+11} measured an autocorrelation function of galaxies up to
redshift of 0.25, 
and derived its color and luminosity dependence.
Their results show that the autocorrelation function significantly
different among blue, red, and the brightest galaxies.
Referring to their results, we used autocorrelation functions of the form
  $\xi_{\rm GG,blue}(r) = (r/3.6 h^{-1} {\rm Mpc})^{-1.7}$, 
  $\xi_{\rm GG,red}(r) = (r/6.6 h^{-1} {\rm Mpc})^{-1.9}$,  and
  $\xi_{\rm GG,bright}(r) = (r/10.7 h^{-1} {\rm Mpc})^{-1.9}$ 
for blue and red galaxies with absolute brightness in $r$ band  M$_{\rm r}\ge -22$,
and the brightest galaxies with M$_{\rm r}< -22$, respectively.
The parameter values adapted for blue and red galaxies are the ones obtained
for galaxies with $-20 < M_{\rm r} < -19$.
Although the absolute magnitude of our galaxy samples ranges over $-23 < M_{\rm r} < -17$,
we ignore the luminosity dependence in the range of $-22 < M_{\rm r} < -17$ as their
measured dependence is small.
For galaxies with brighter than $M_{\rm r} = -22$ significant increase in 
autocorrelation function was observed, so we used a separate function for those
galaxies.
We also assume that the autocorrelation functions are unchanged
at redshifts up to 1.0.

\begin{figure*}
\includegraphics[width=0.8\textwidth]{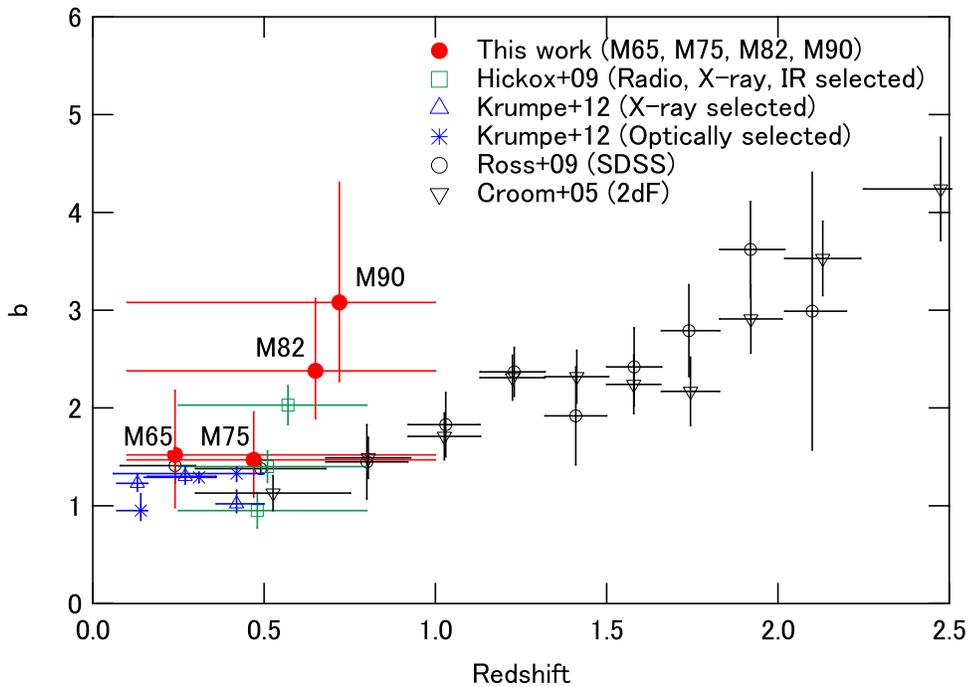}
\caption{AGN bias plotted as a function of redshift.
  The results of this work are plotted with filled circles for
  mass groups of M65, M75, M82 and M90.
  The results by Hickox et al. for radio, X-ray and IR selected AGNs
  are plotted with open squares at $b$ = 2.03, 1.40, and 0.95, respectively.
  The results by Krmpe et al. for X-ray and optically selected AGNs,
  Ross et al. for SDSS QSOs, and Croom et al for 2dF QSOs are plotted
  with markers specified in the legend in this figure.
  The horizontal error bars represent the redshift range of the samples.
}
\label{fig:b-z}
\end{figure*}

To derive weighting factors among these three components, fractions of blue and red
galaxies with $M_{\rm r} \ge -22$  and of galaxies with $M_{\rm r} < -22$ are measured
by the subtraction method described in section~\ref{sec:red_frac}.
The obtained fractions are summarized in Table~\ref{tab:halo_mass}.
Using the autocorrelation function of galaxies derived from a linear combination
of those for three galaxy types,
we obtained
$b = $
$1.52^{+0.66}_{-0.54}$,
$1.47^{+0.49}_{-0.38}$,
$2.38^{+0.74}_{-0.49}$,
$3.08^{+1.23}_{-0.81}$
for mass group M65, M75, M82 and M90, respectively.
The detailed method for deriving the
AGN bias 
is described in section~\ref{sec:halo_mass}.
These results are summarized in Table~\ref{tab:halo_mass} and plotted in
Figure~\ref{fig:b-z}.
The uncertainty of
$b$
includes statistical error, systematic error
corresponding to the uncertainty of $\rho_{0}$ estimation, and
the intrinsic variance of the clustering around each AGN.

As is inferred from the mass-redshift distribution shown in Figure~\ref{fig:M-z}, 
even for the same redshift group the higher mass groups tends to be biased 
to higher redshift.
To reduce the redshift bias, we constructed redshift matched samples for each 
redshift group. 
The redshift matched samples were constructed by selecting the same number of
AGN samples for each redshift bin with 0.02 width.

\begin{figure*}
\includegraphics[width=0.8\textwidth]{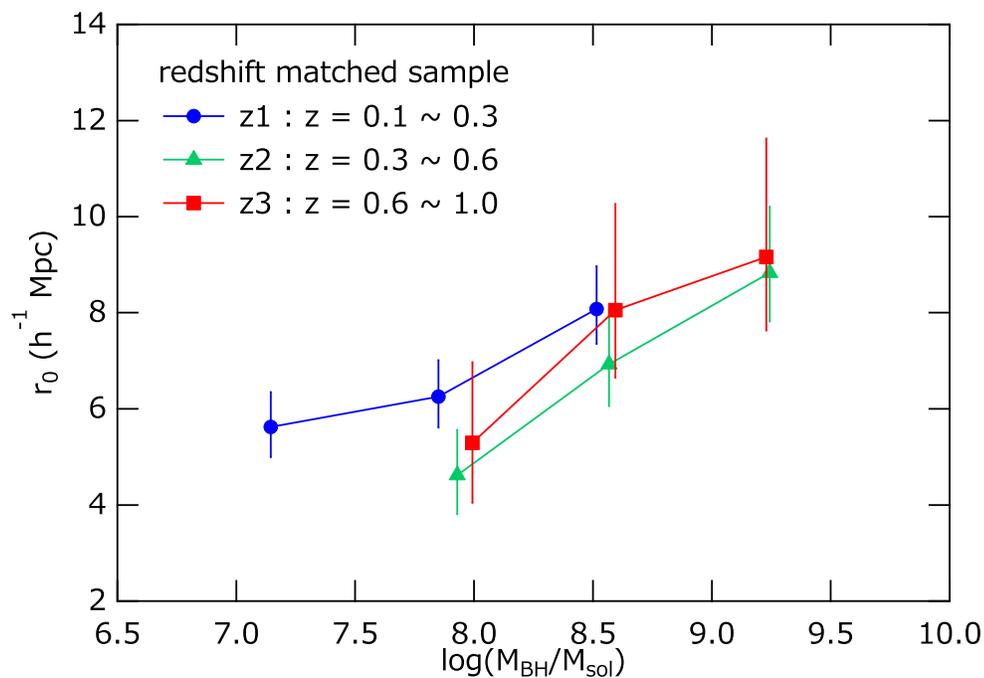}
\caption{Cross-correlation lengths derived for each redshift matched samples 
are plotted as a function BH mass. The data point of the same redshift groups
are connected with a line. Filled circles are for $z1$ redshift group, filled
triangles and boxes are for $z2$ and $z3$ redshift groups, respectively.}
\label{fig:r0_mass_zdist}
\end{figure*}

The number of samples selected in this way is reduced significantly from
the original samples, so the intrinsic fluctuation of clustering strength
of each sample can introduce large scatter to the cross-correlation length.
To obtain the most typical sample in term of clustering, we randomly
constructed 21 sets of redshift matched sample for each redshift group,
calculated cross-correlation length for each sample set, and selected a
sample set for which cross-correlation length was median among the sample sets.
The result is shown in Figure~\ref{fig:r0_mass_zdist} and summarized in
the middle part of Table~\ref{tab:fit_r0}.
The tendency of increase in $r_{0}$ as a function $\Mbh$ is seen in
the redshift matched samples as well.

\begin{figure*}
\includegraphics[width=0.8\textwidth]{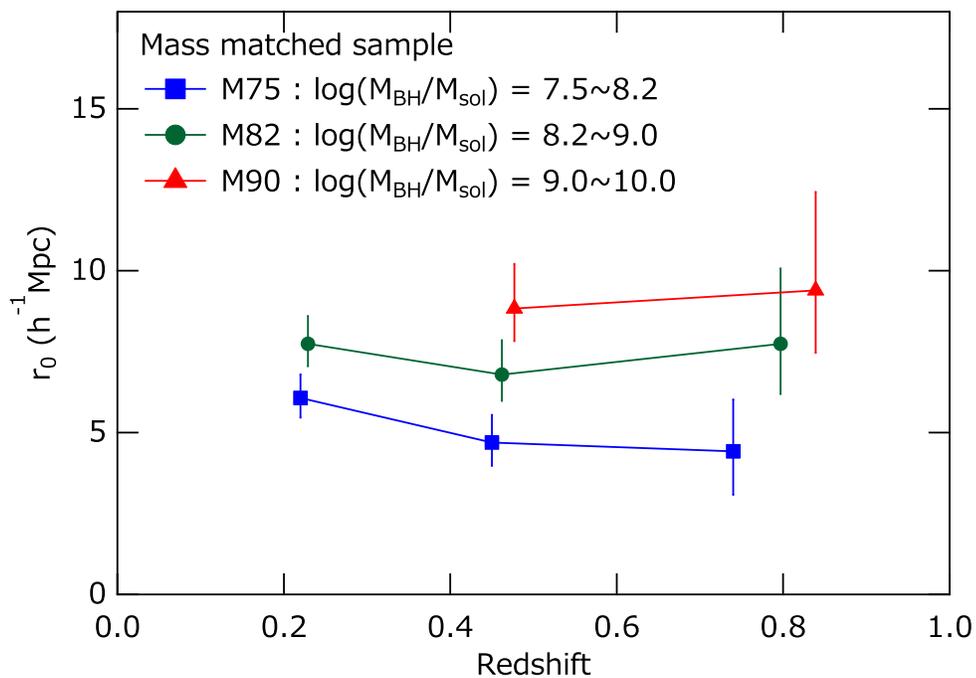}
\caption{Cross-correlation lengths derived for each BH mass matched samples
are plotted as a function of redshift. The data point of the same mass groups
are connected with a line. Filled boxes are for M75 mass group, filled
circles and triangles are for M82 and M90 mass groups, respectively.}
\label{fig:r0_z_mdist}
\end{figure*}

To see the redshift dependence of $r_{0}$ for the same $\Mbh$, we constructed mass 
matched samples in the same way as used to construct redshift matched samples
with 0.2 dex of bin width of mass histogram.
The results are shown in Figure~\ref{fig:r0_z_mdist} and summarized in the
last part of Table~\ref{tab:fit_r0}.
No clear redshift dependence is seen for any mass group.

These results are consistent with our previous result in \citet{Komiya+13}, although
we have used the most recent UKIDSS catalog and slightly different data selection
criteria and mass correction.

\subsection{Properties of Galaxies}

\begin{figure*}
\includegraphics[width=0.33\textwidth]{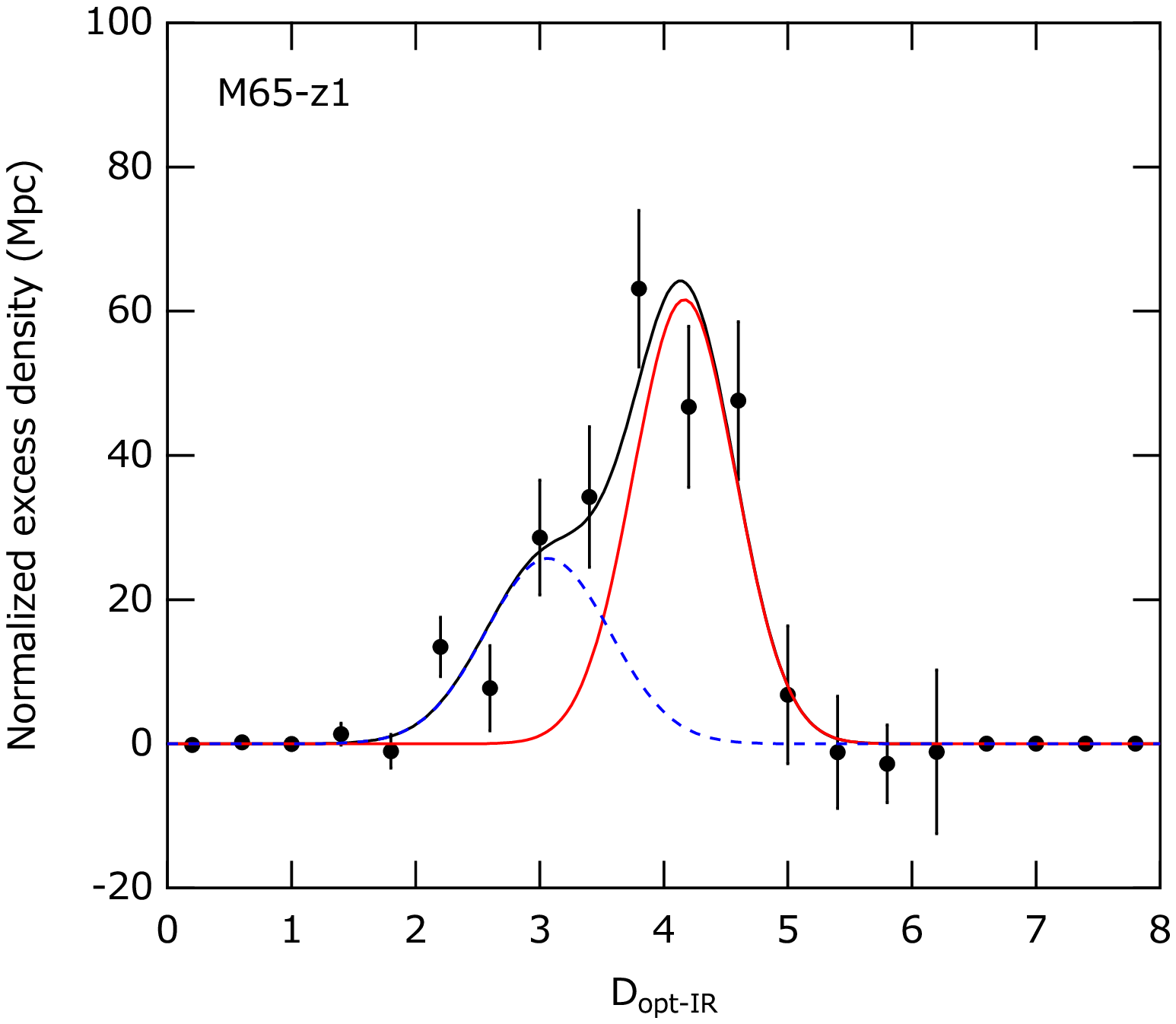}
\includegraphics[width=0.33\textwidth]{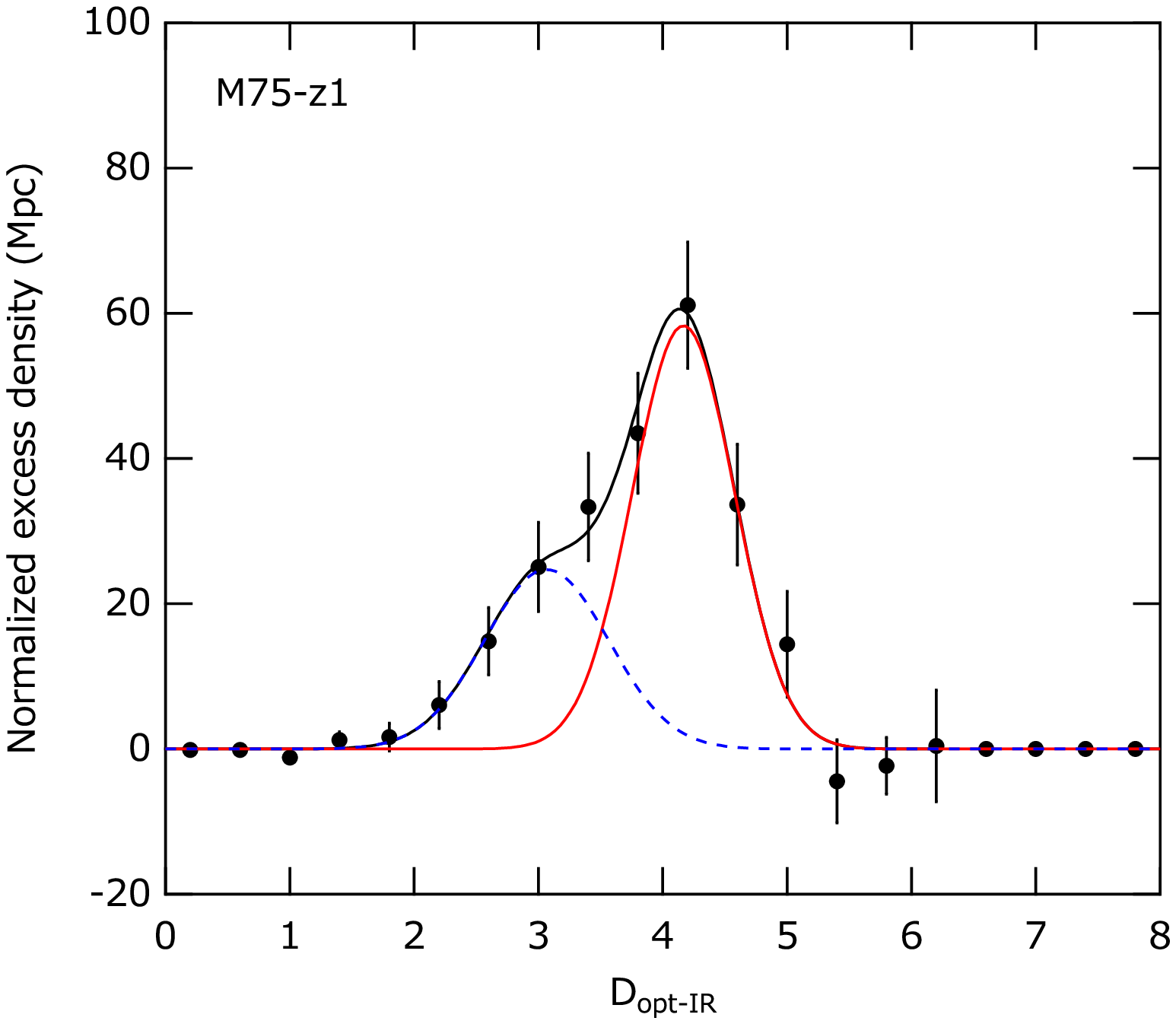}
\includegraphics[width=0.33\textwidth]{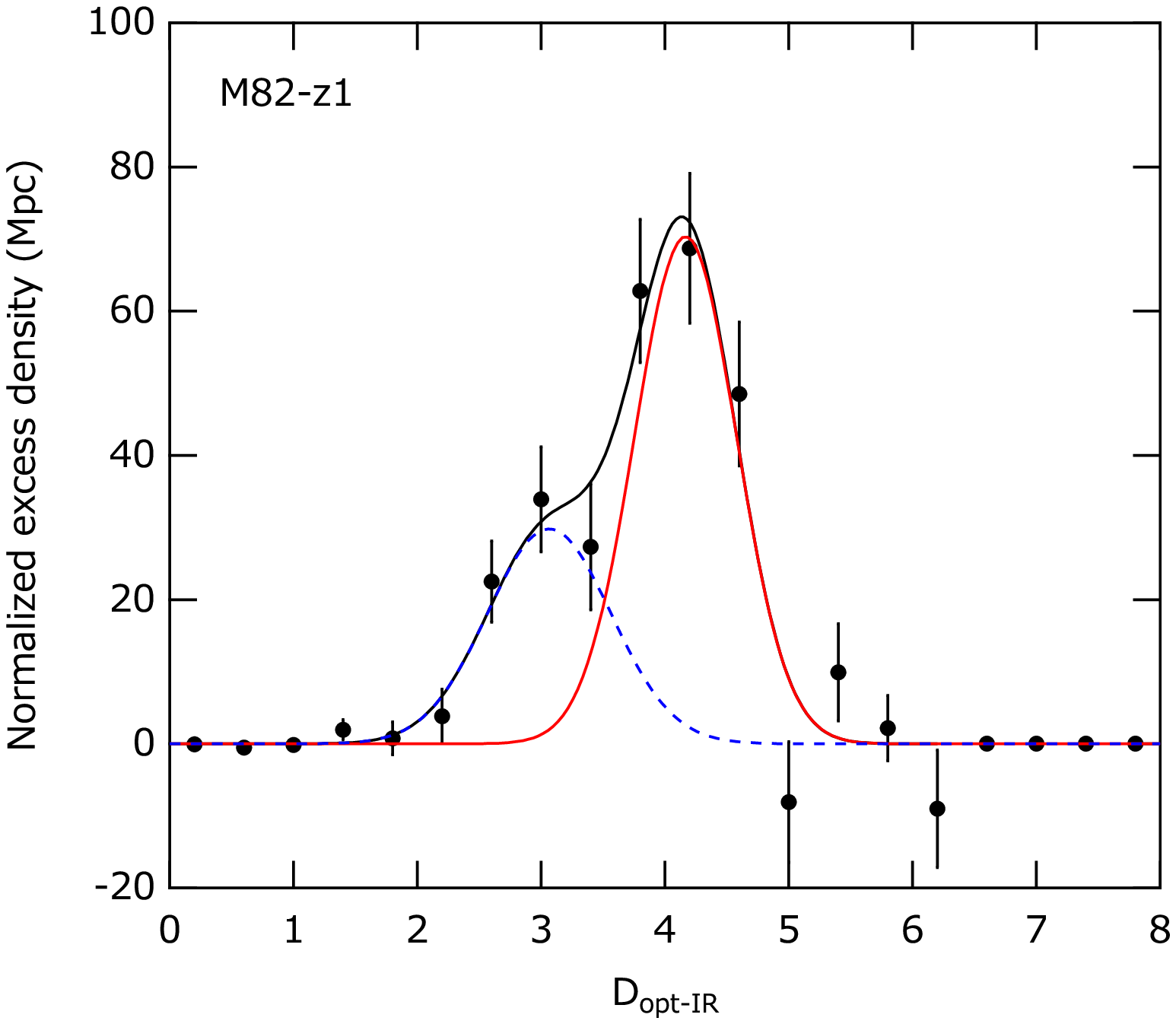}
\\
\includegraphics[width=0.33\textwidth]{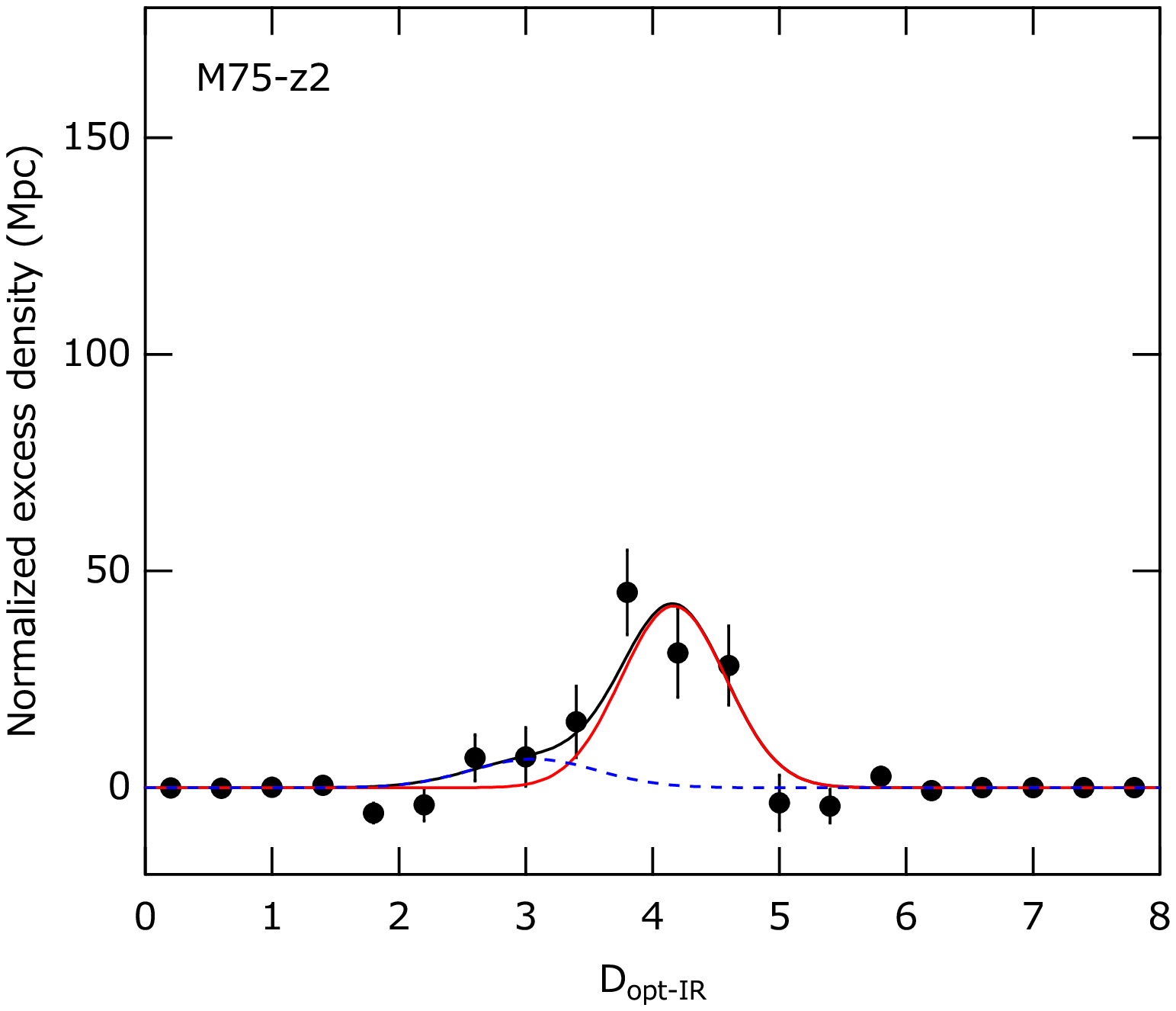}
\includegraphics[width=0.33\textwidth]{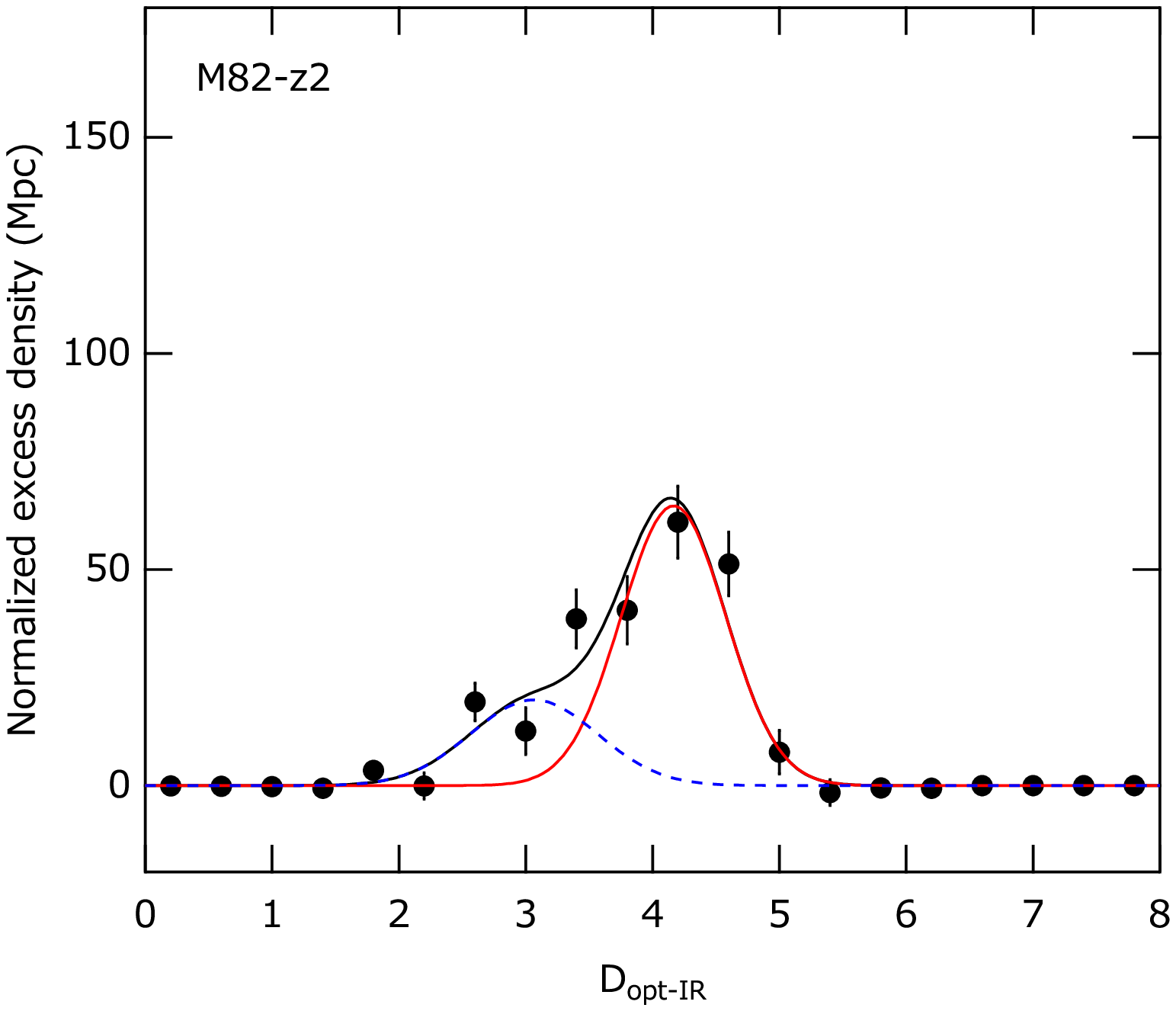}
\includegraphics[width=0.33\textwidth]{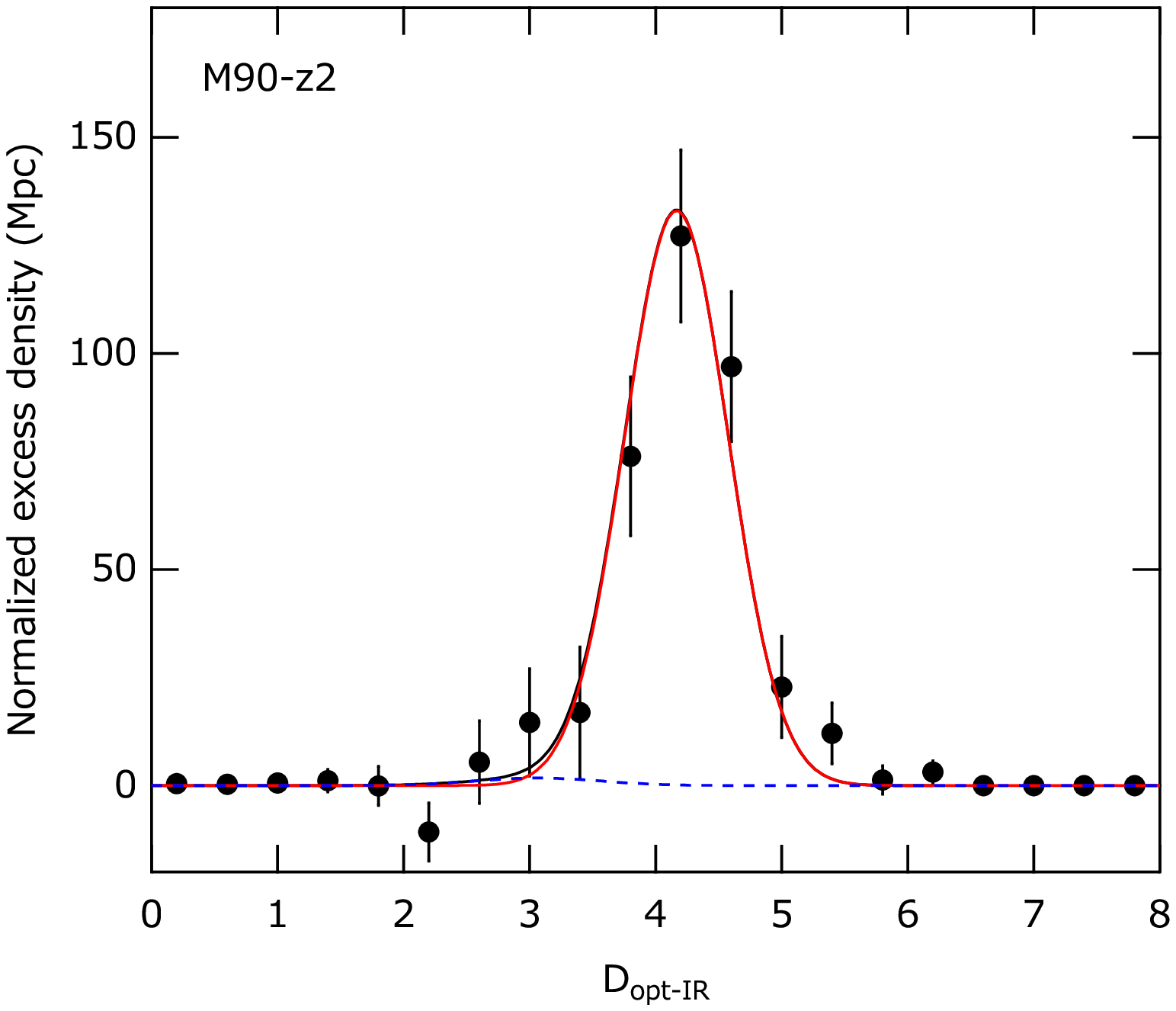}
\\
\includegraphics[width=0.33\textwidth]{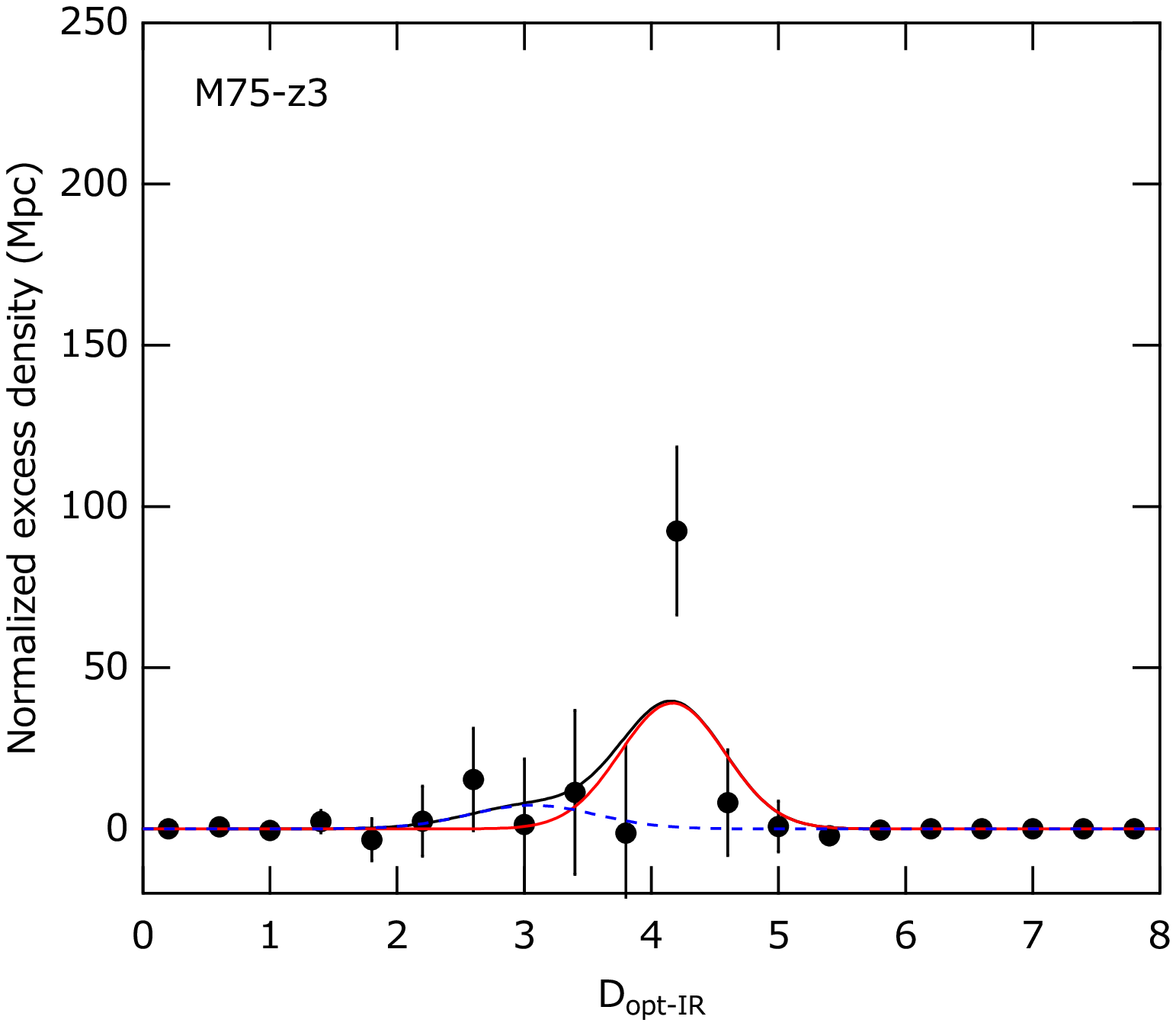}
\includegraphics[width=0.33\textwidth]{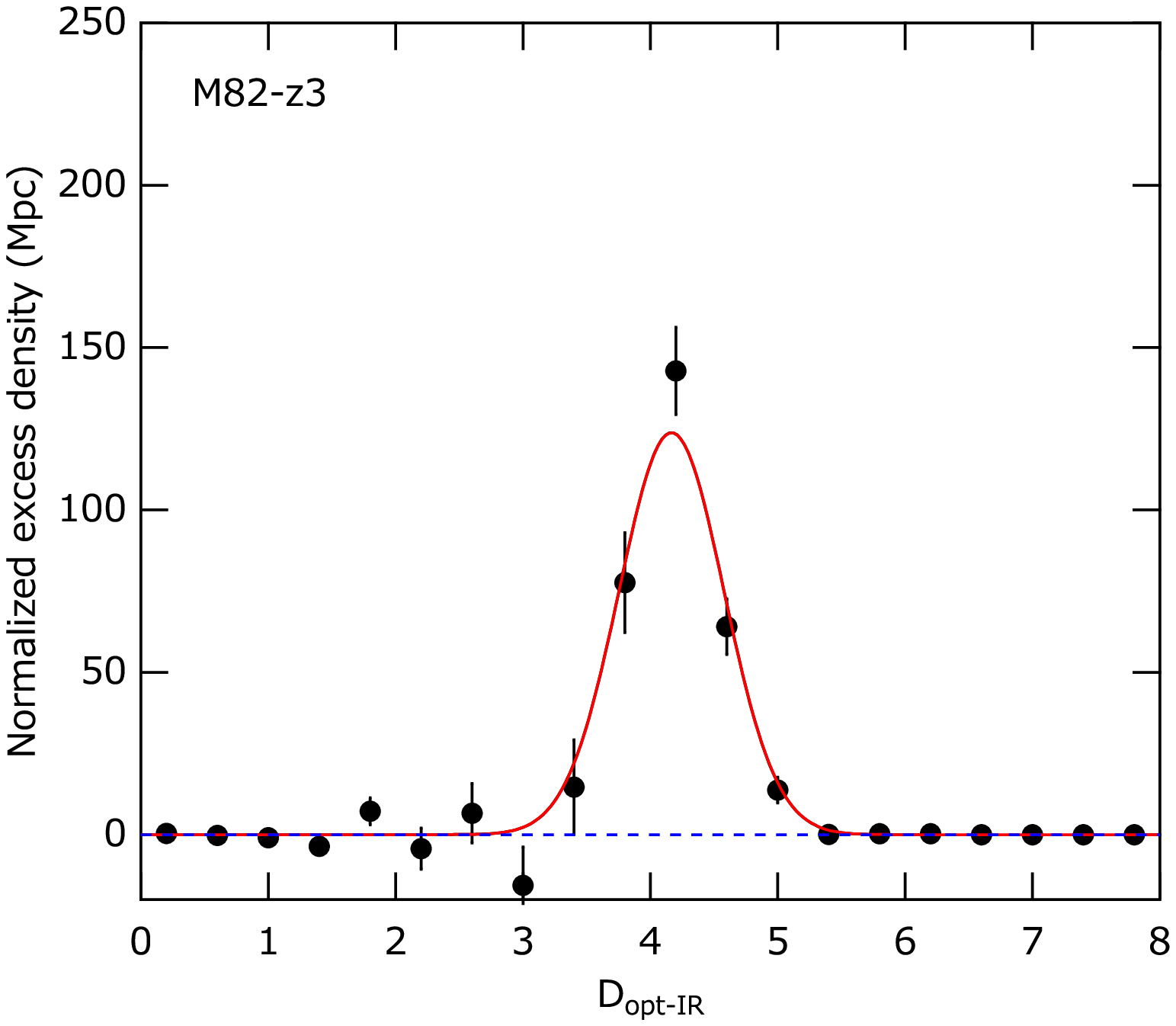}
\includegraphics[width=0.33\textwidth]{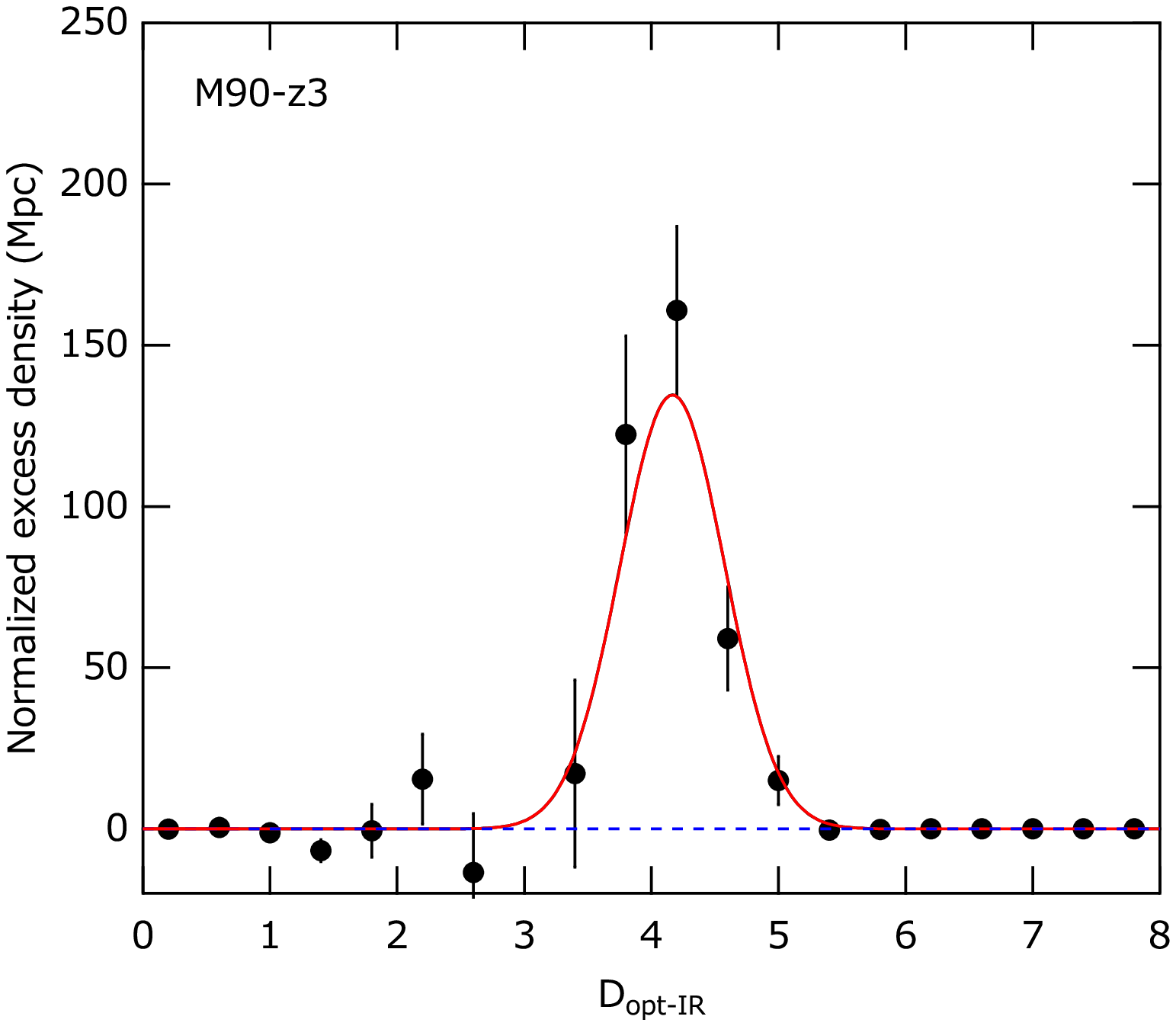}
\caption{Distributions of $\Dx$ parameter for each redshift and mass groups.
Filled circles represent observed data points, solid lines are fitting result
of the two component model, red solid and blue dashed lines are for red and blue
galaxy component, respectively.}
\label{fig:hist_D}
\end{figure*}

To investigate the properties of galaxies at environment of AGNs with the 
most massive SMBHs, where increase of clustering is found, we calculated 
the color parameter $\Dx$ for all the detected galaxies and derived its 
distributions for galaxies associated with the AGNs by using the 
subtraction method described in 
section~\ref{sec:red_frac}.
$\Dx$ parameter corresponds to the color defined as difference in 
brightness between the wavelength ranges of 3,000--3,500~{\AA} and 
10,000--12,000~{\AA} in the rest frame.
In Figure~\ref{fig:hist_D} we show $\Dx$ parameter distributions
for each redshift and mass sample.
The observed distributions are fitted with a two component model (red and 
blue galaxy components) assuming the normal distribution for both components.
The mean and standard deviation of the distribution of each component are
fixed to the predetermined values, and only the mixing ratio and the 
normalization constant are taken as free parameters.
The predetermined values of the mean and standard deviation are obtained
by fitting the model function to the observed distribution for all the
AGN samples
with taking all the parameters free.
The parameter values obtained from this fitting are summarized in 
Table~\ref{tab:fit_D}.

\begin{figure*}
\includegraphics[width=0.5\textwidth]{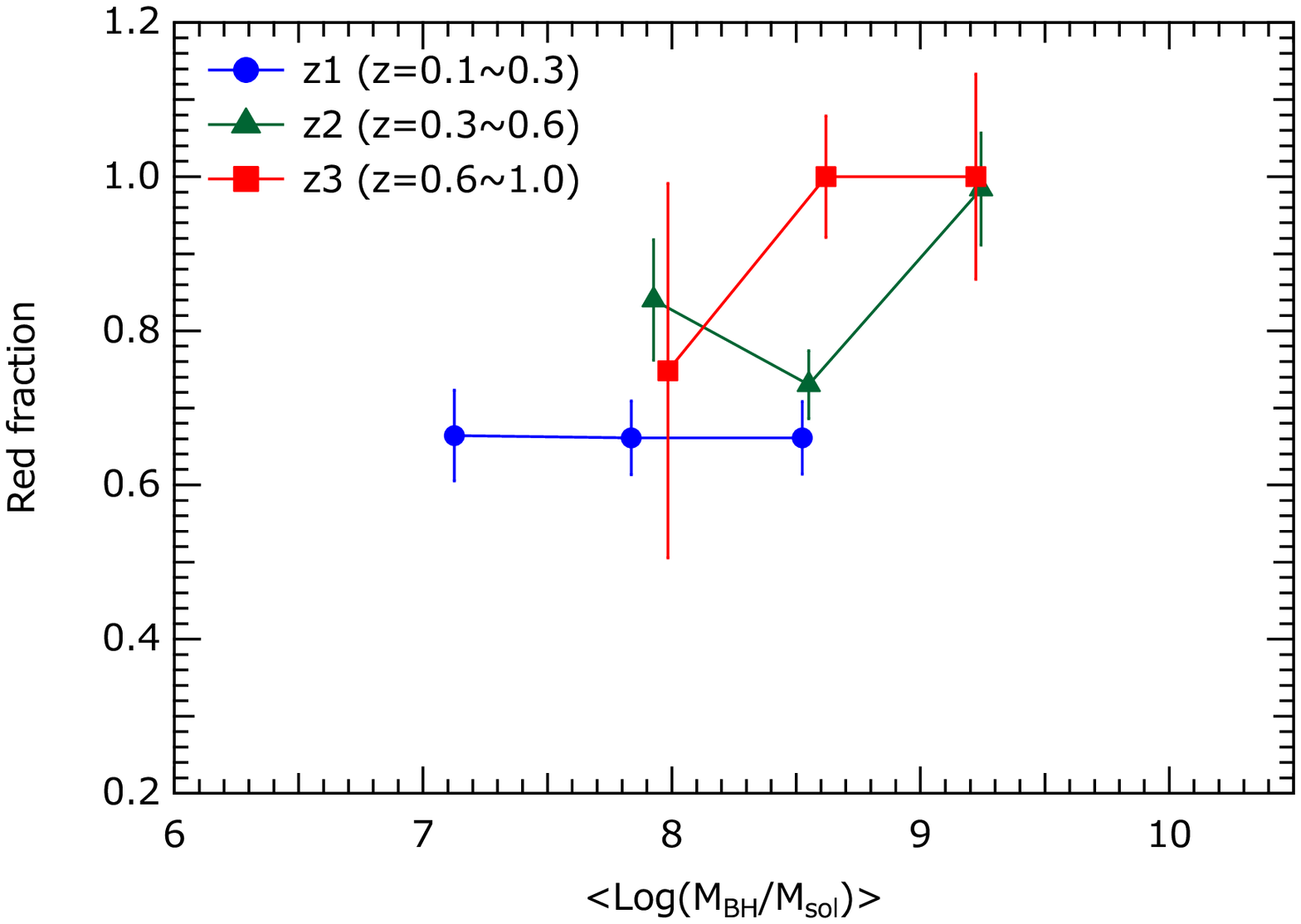}
\includegraphics[width=0.5\textwidth]{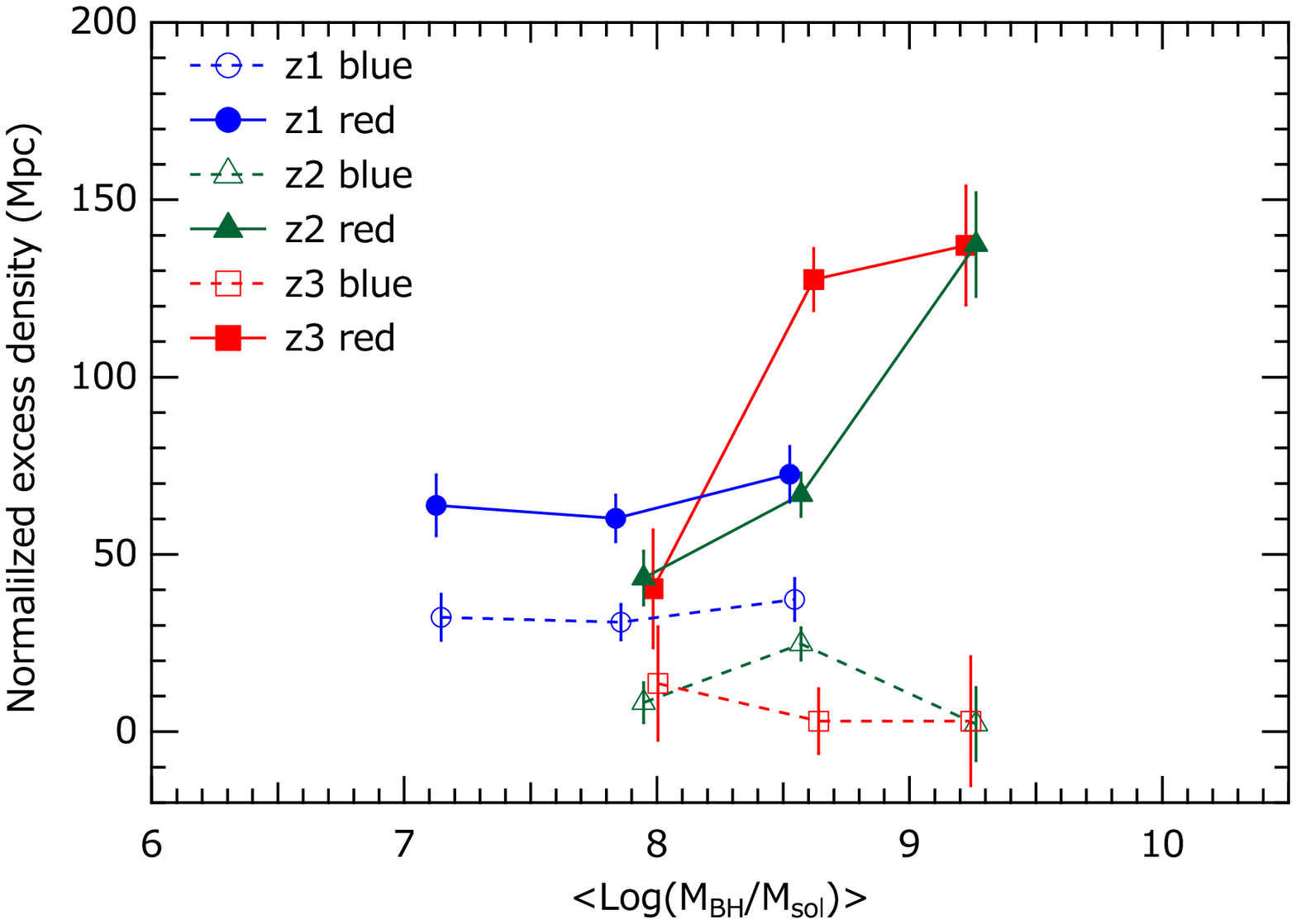}
\caption{Left panel: Red galaxy fractions derived from the two component fitting shown in 
Figure~\ref{fig:hist_D} are plotted as a function of BH mass. The results
of the same redshift groups are connected with a line. Circles are results for
redshift z1 group, and triangles and boxes are for redshift z2 and z3 groups,
respectively.
Right panel: Normalized excess density derived from the two component fitting.
Filled and open circles are results for blue and red galaxies respectively for z1 redshift 
group.
Filled and open triangles are results for z2 redshift groups with the same blue/red galaxy
assignment.
Filled and open boxes are results for z3 redshift groups with the same assignment,
and the arrows represents the upper limit for blue galaxies at redshift z3.
}
\label{fig:RedFraction}
\end{figure*}

The fractions of red galaxies are plotted in the left panel of 
Figure~\ref{fig:RedFraction} as a function of $\Mbh$.
The normalized excess densities $\omega'$, which are defined as
$\omega' = (n_{\rm on} - n_{\rm off}) / \rho_{0}$, are plotted
in the right panel of the figure.
$n_{\rm on}$ and $n_{\rm off}$ are the surface number density of 
galaxies at projected distance of 0--1Mpc (on region) and 3--5Mpc (off region)
from AGN, respectively.
The $\omega'$ can be considered as an approximate estimate of the projected
cross-correlation function.
The error bars represent one sigma Poisson statistical error for both
figures.
Since red galaxies are typically brighter than blue galaxies as seen in 
Figure~\ref{fig:CM_all}, the observed red galaxy fraction tends to be higher for the
higher redshift samples.
Thus, the comparison of the red galaxy fraction is meaningful only for the same 
redshift samples.
At redshift z1, both of the red galaxy fraction and the normalized excess 
density are almost unchanged among the the mass groups of M65, M75 and M82.
At redshift z2, significant increase of the normalized excess density is
seen for the red galaxies of the mass group M90.
At redshift z3, blue galaxies are hardly detected and red
fraction is $\sim$100\% for mass groups M82 and M90.

\begin{figure*}
\includegraphics[width=0.5\textwidth]{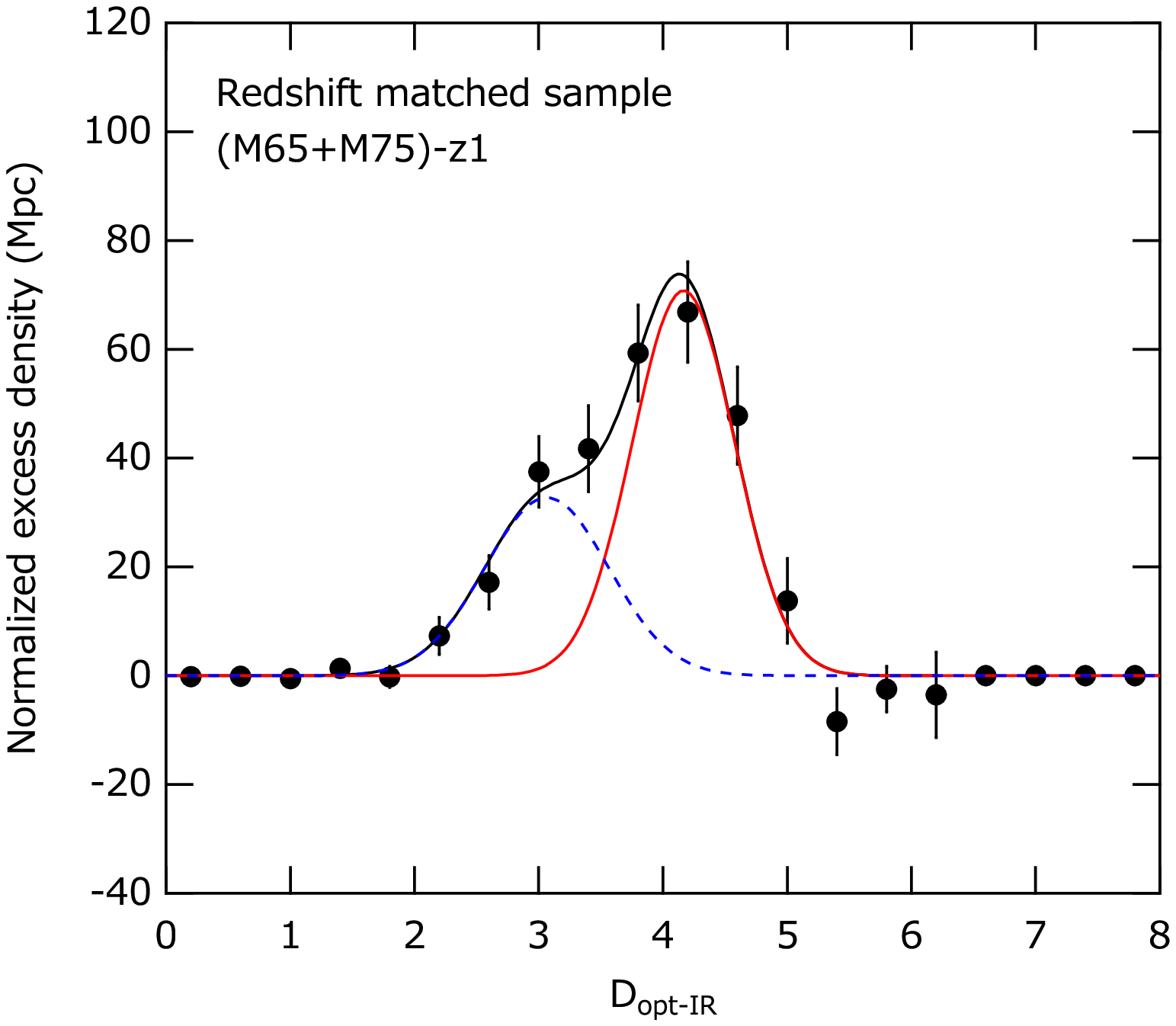}
\includegraphics[width=0.5\textwidth]{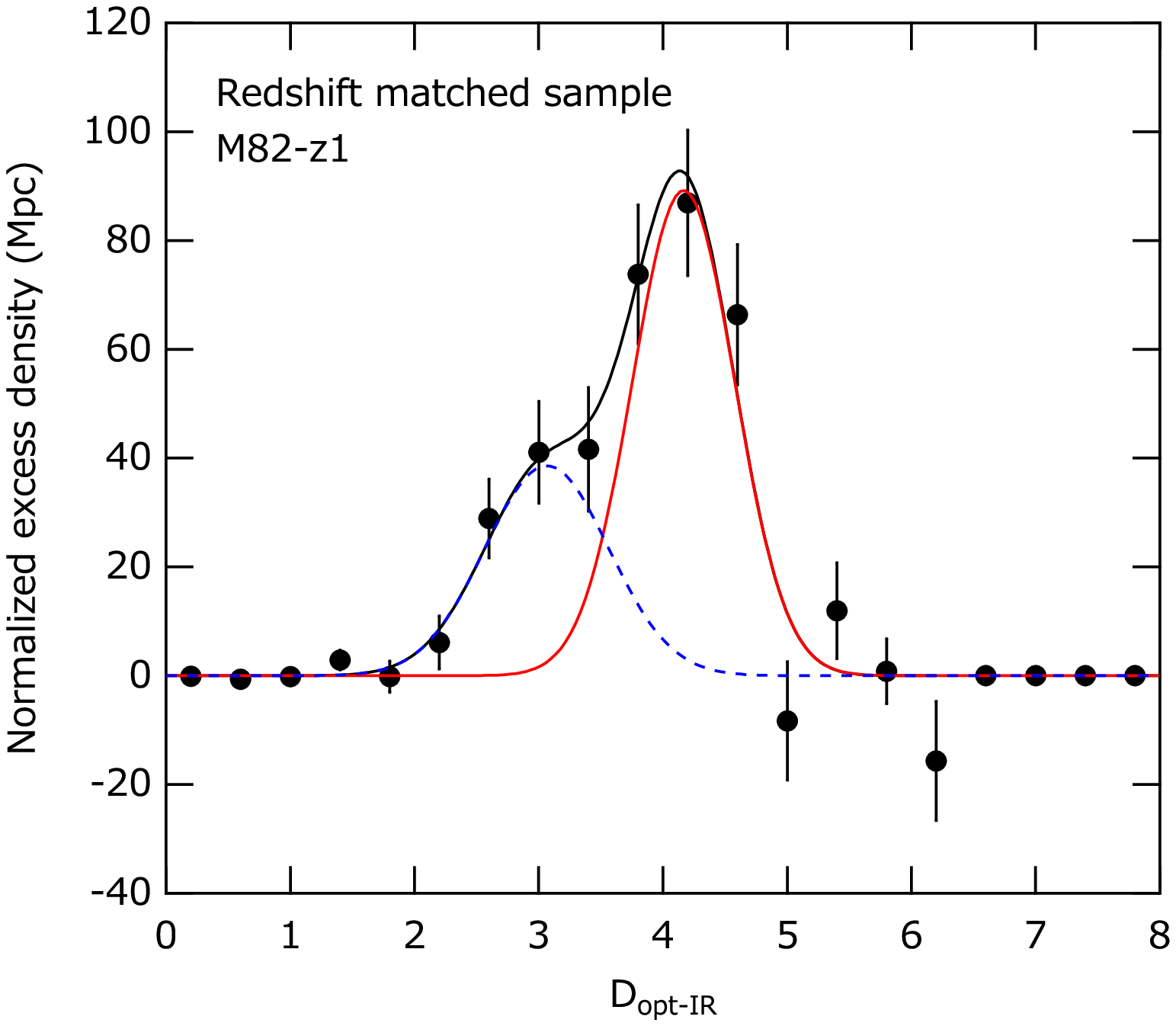}
\\
\includegraphics[width=0.5\textwidth]{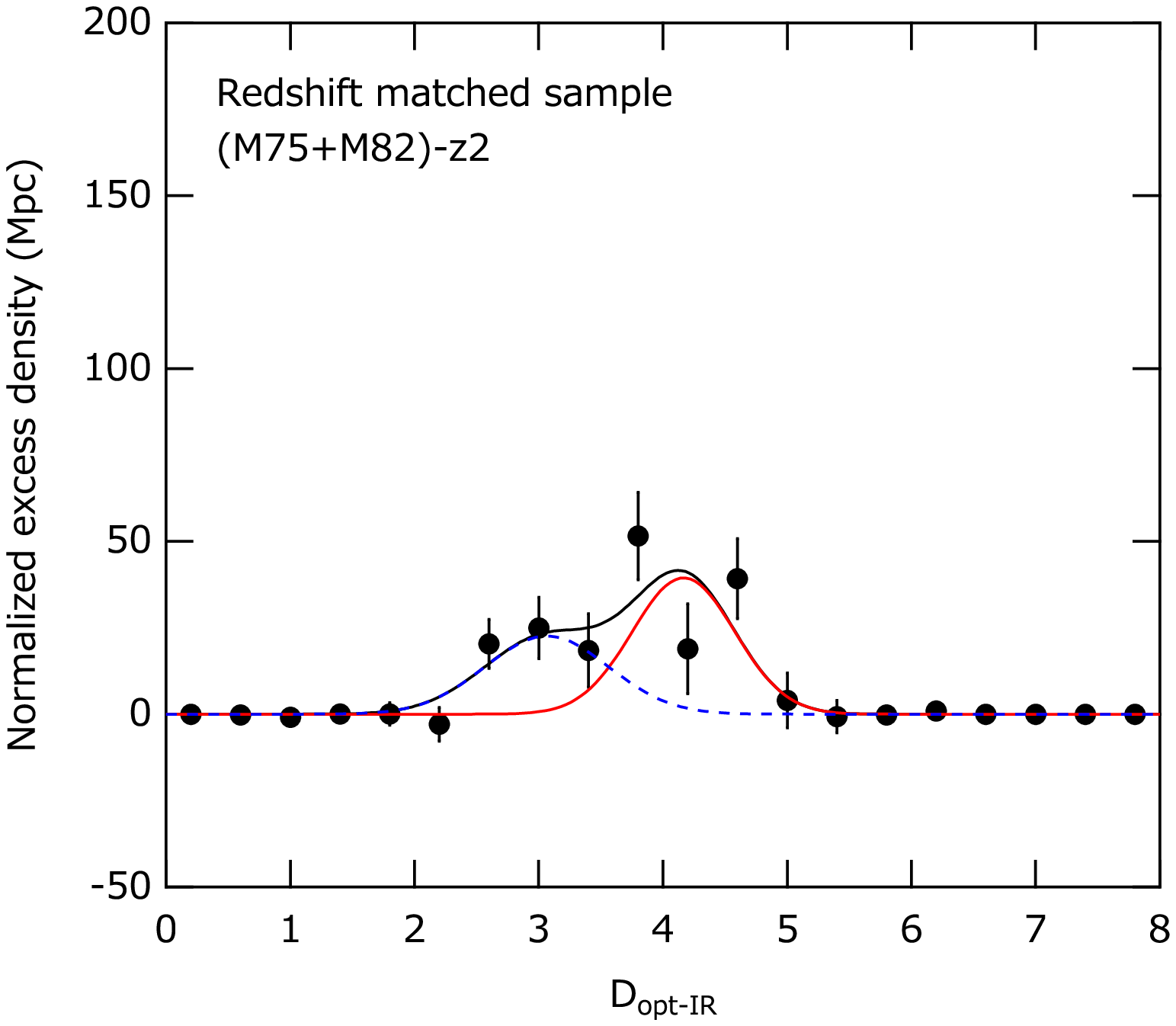}
\includegraphics[width=0.5\textwidth]{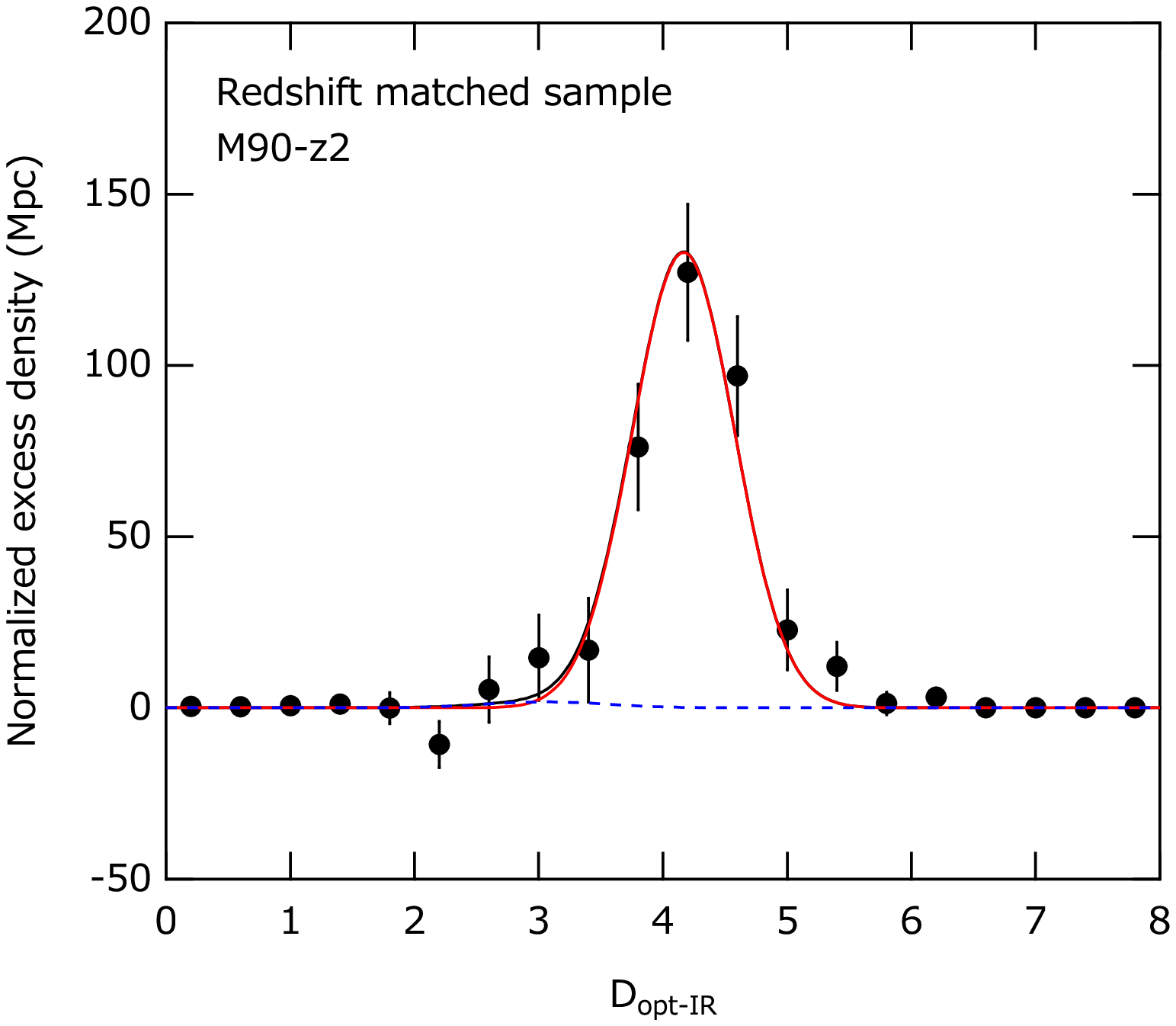}
\\
\includegraphics[width=0.5\textwidth]{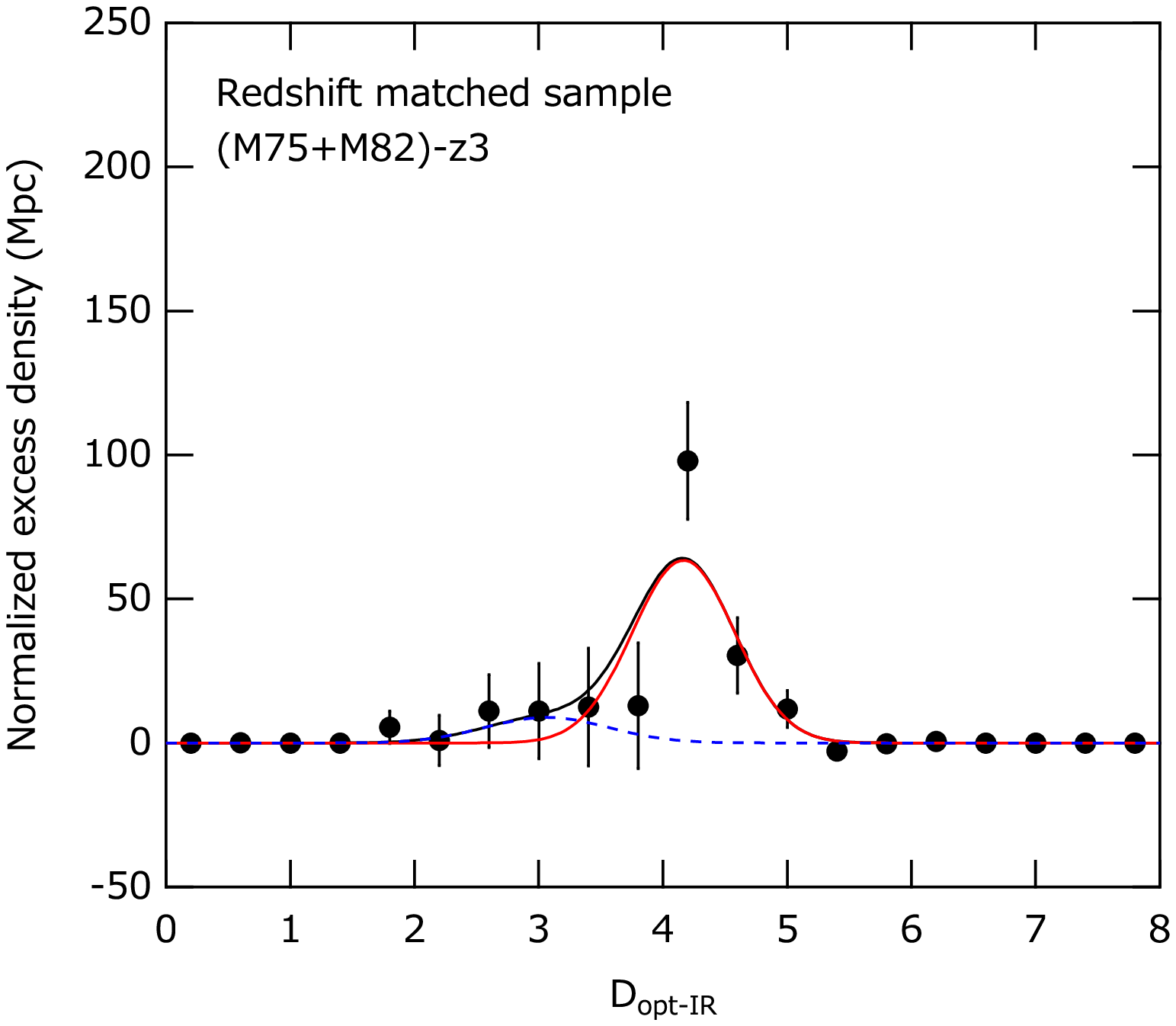}
\includegraphics[width=0.5\textwidth]{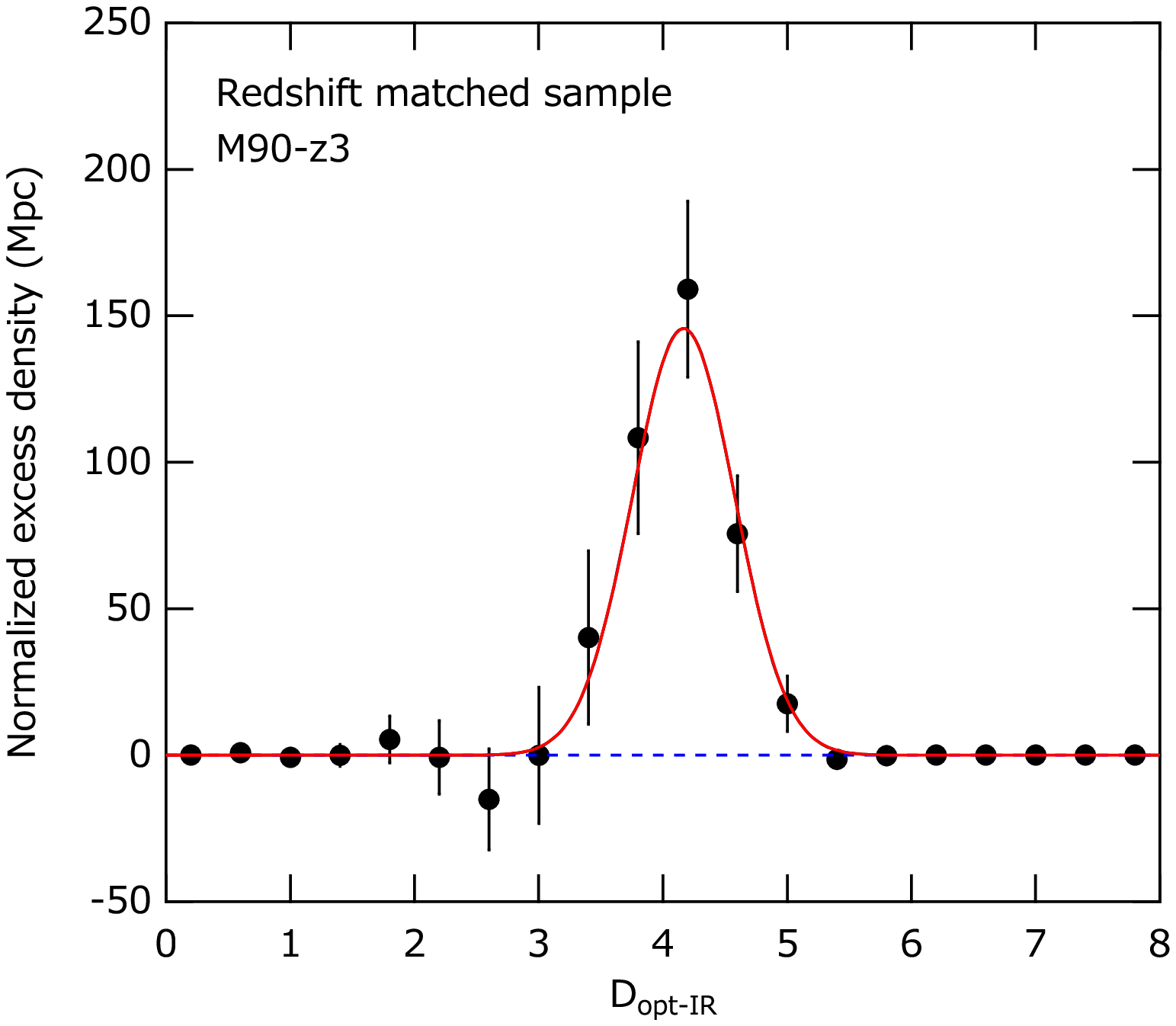}
\caption{Distributions of $\Dx$ parameter for redshift matched samples.
Filled circles represent observed data points, solid lines are fitting result
of the two component model, red solid and blue dashed lines are for red and blue
galaxy component, respectively.
}
\label{fig:hist_D_zdist}
\end{figure*}

\begin{figure*}
\includegraphics[width=0.5\textwidth]{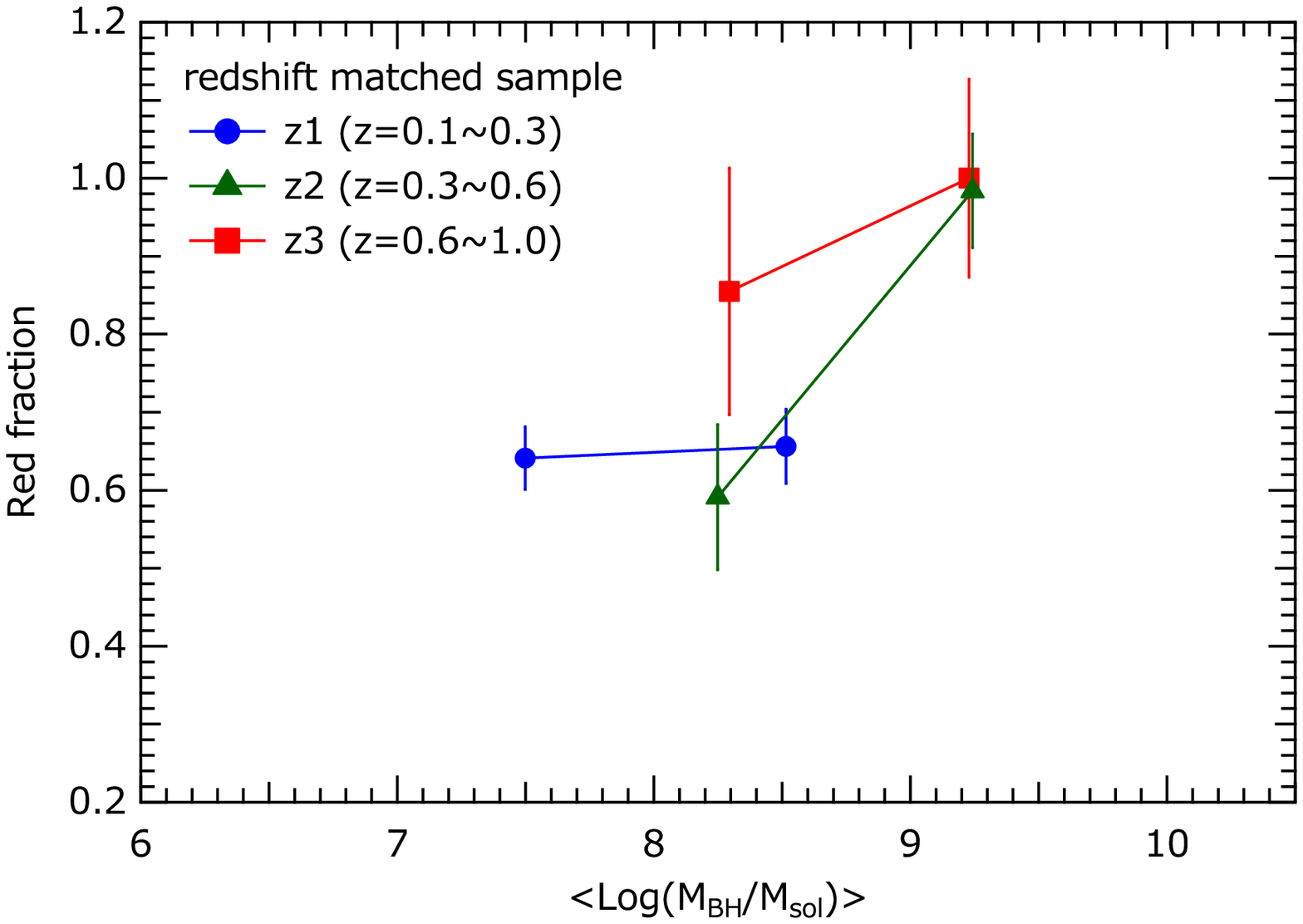}
\includegraphics[width=0.5\textwidth]{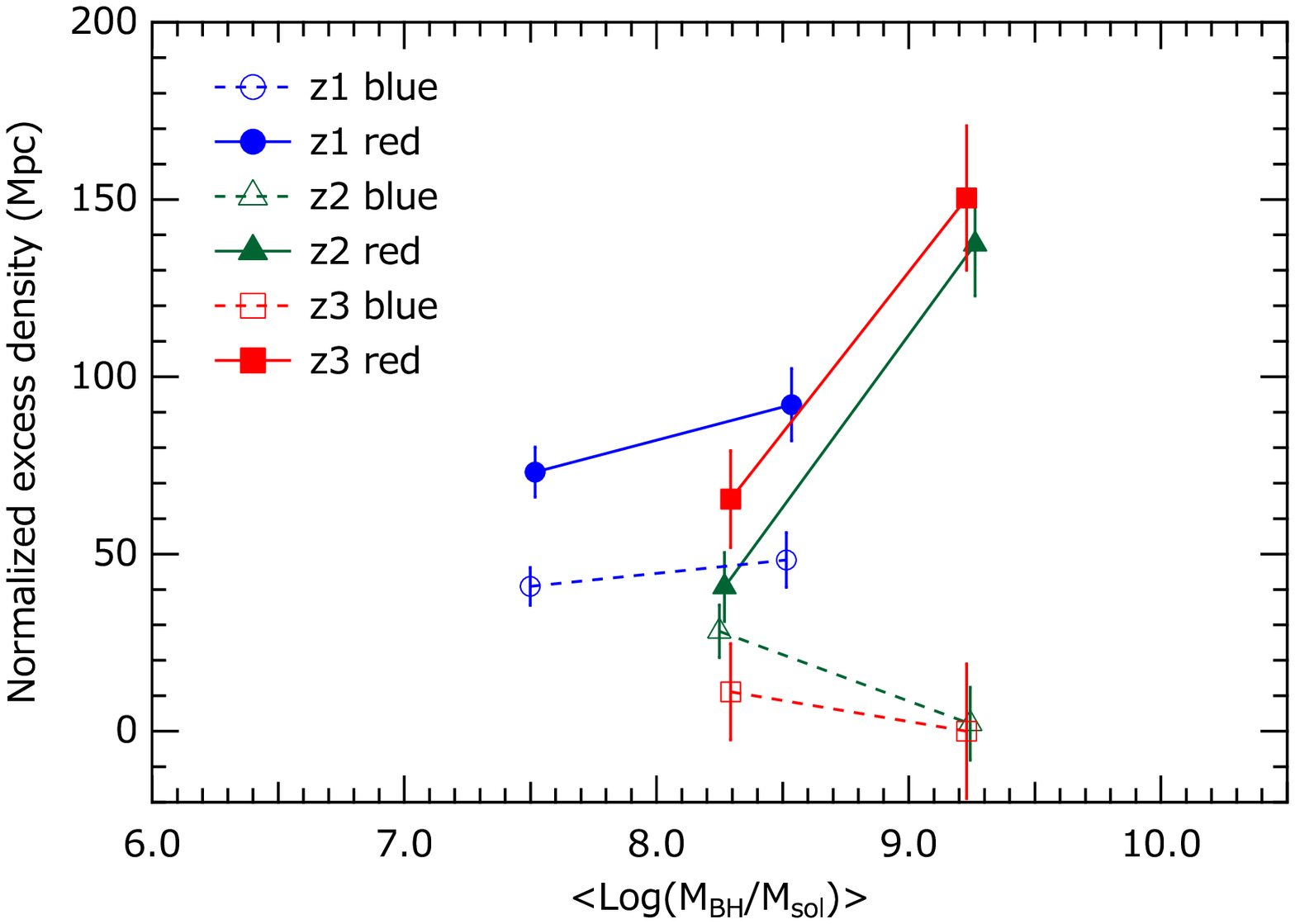}
\caption{
Left panel: Red galaxy fractions derived from the two component fitting shown in 
Figure~\ref{fig:hist_D_zdist} are plotted as a function of BH mass. The results
of the same redshift groups are connected with a line. Circles are results for
redshift z1 group, and triangles and boxes are for redshift z2 and z3 groups,
respectively.
Right panel: Normalized excess density derived from the two component fitting.
Filled and open circles are results for blue and red galaxies respectively for z1 redshift 
group.
Filled and open triangles are results for z2 redshift groups with the same blue/red galaxy
assignment.
Filled and open boxes are results for z3 redshift groups with the same assignment.
}
\label{fig:RedFraction_zdist}
\end{figure*}

The data samples used in Figures~\ref{fig:hist_D} and \ref{fig:RedFraction}
have a slight difference in the redshift distribution among the respective
mass groups even for the same redshift groups; the lower mass group has 
relatively smaller redshift than the higher mass group.
The red fraction of the higher mass group, therefore, can be biased to higher 
value.
To reduce the effect of redshift bias, the same analysis was performed also for
the redshift matched samples.
The distributions of $\Dx$ for the redshift matched samples are shown in 
Figure~\ref{fig:hist_D_zdist}.
Since the statistics of lower mass groups are poor, the two lower mass groups
are combined.
The obtained red galaxy fractions and normalized excess density are plotted as 
a function of $\Mbh$ in Figure~\ref{fig:RedFraction_zdist}.
These results also show that red galaxy becomes the dominant component in
the  highest mass group, M90.

\begin{figure*}
\includegraphics[width=0.5\textwidth]{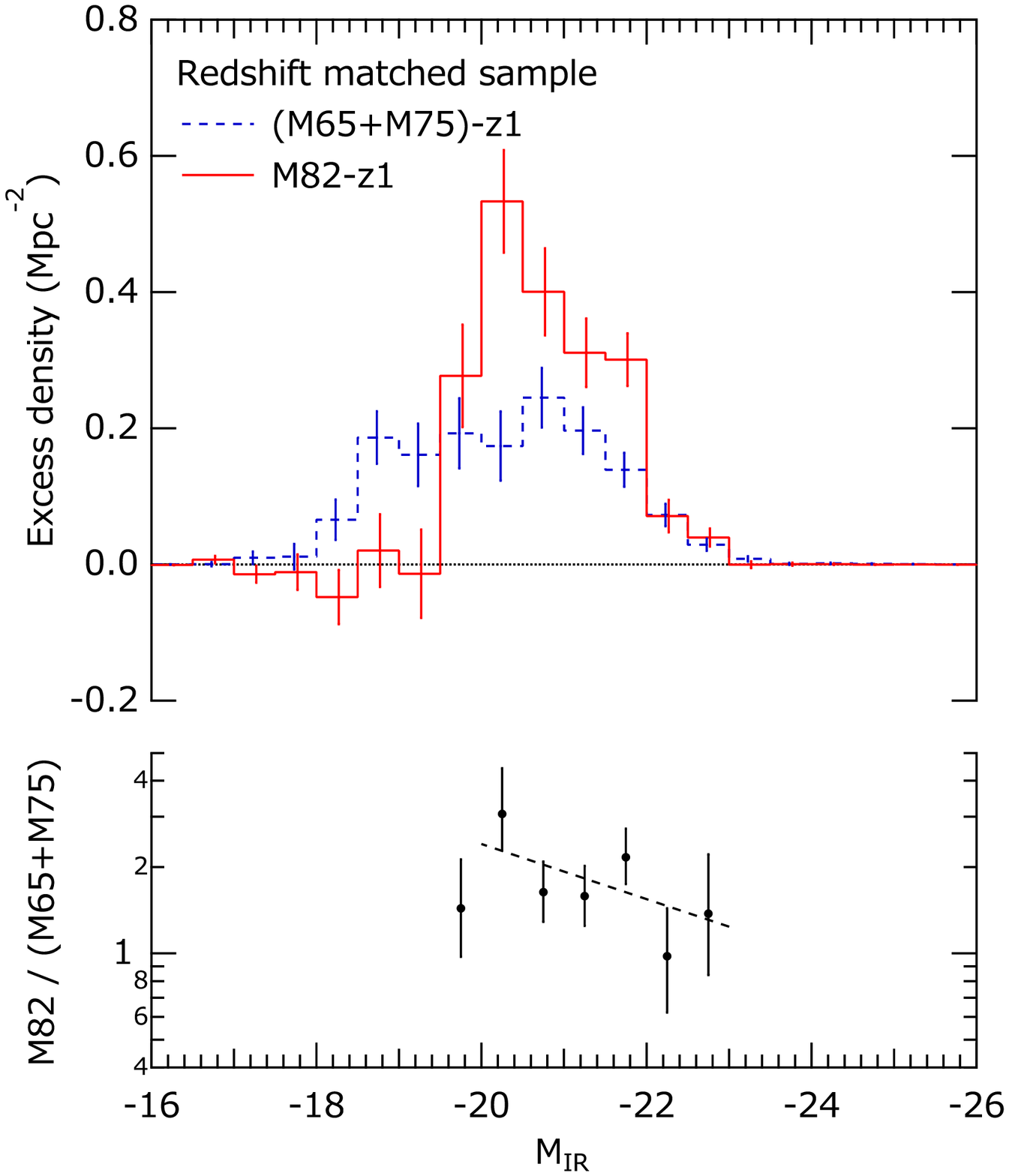}
\includegraphics[width=0.5\textwidth]{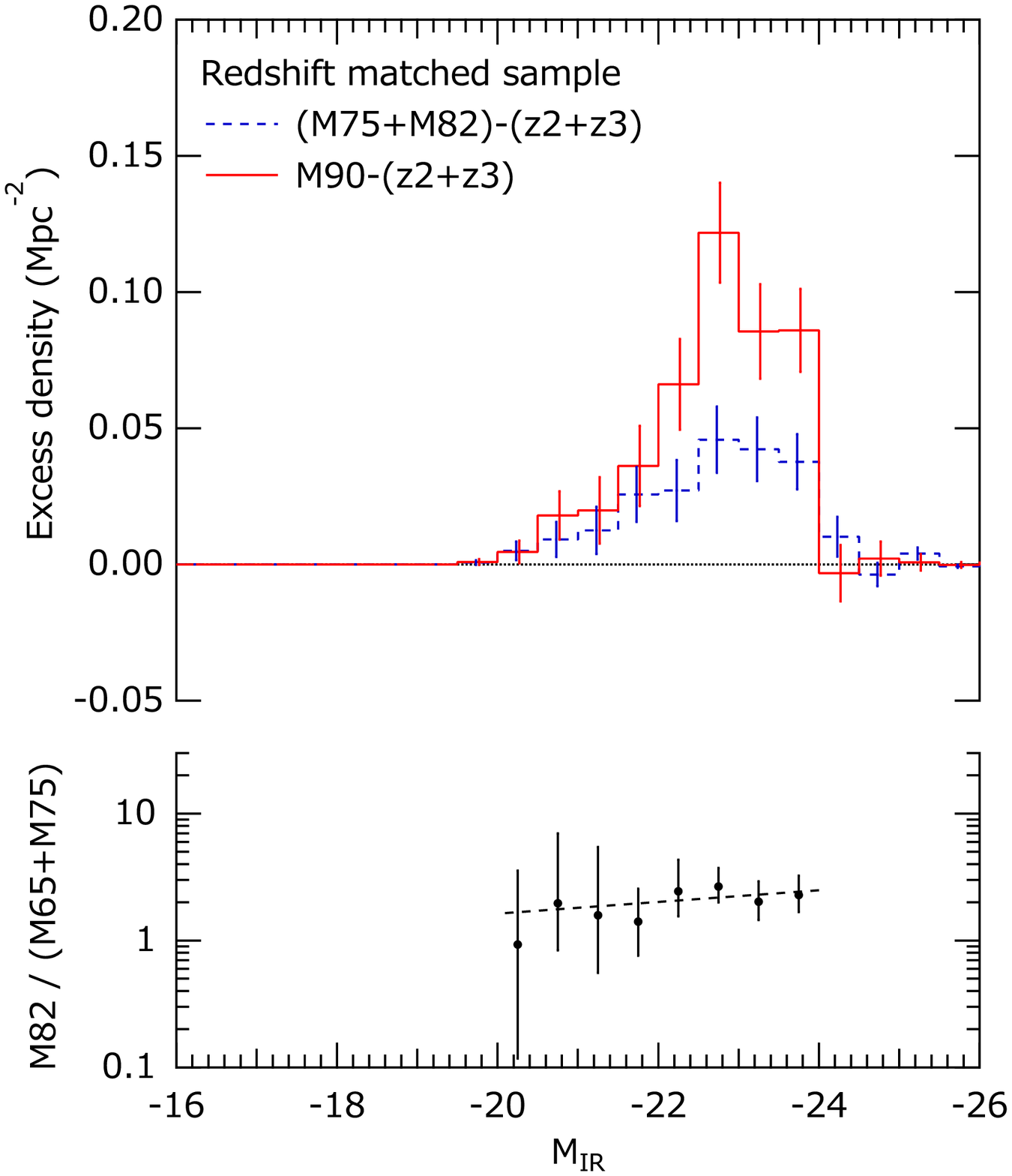}
\caption{Comparison of absolute magnitude distributions between higher (red histogram) 
and lower (blue histogram) mass groups. The ratios of high/low are plotted in the panel
below the histograms. Dashed lines shown in the ratio plot is a result of a linear fit 
in linear-log scale.}
\label{fig:absmag}
\end{figure*}

\begin{figure*}
\includegraphics[width=0.5\textwidth]{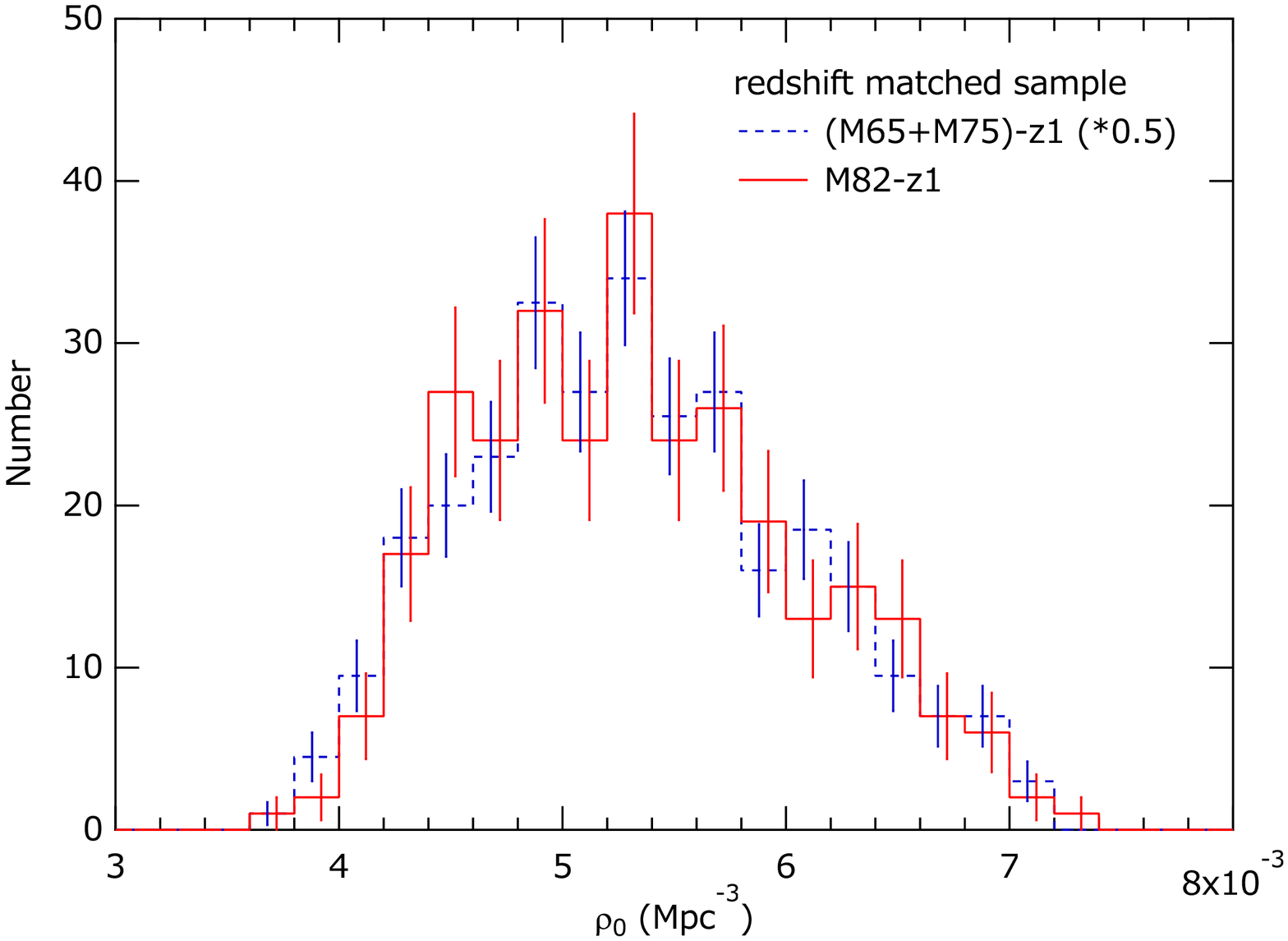}
\includegraphics[width=0.5\textwidth]{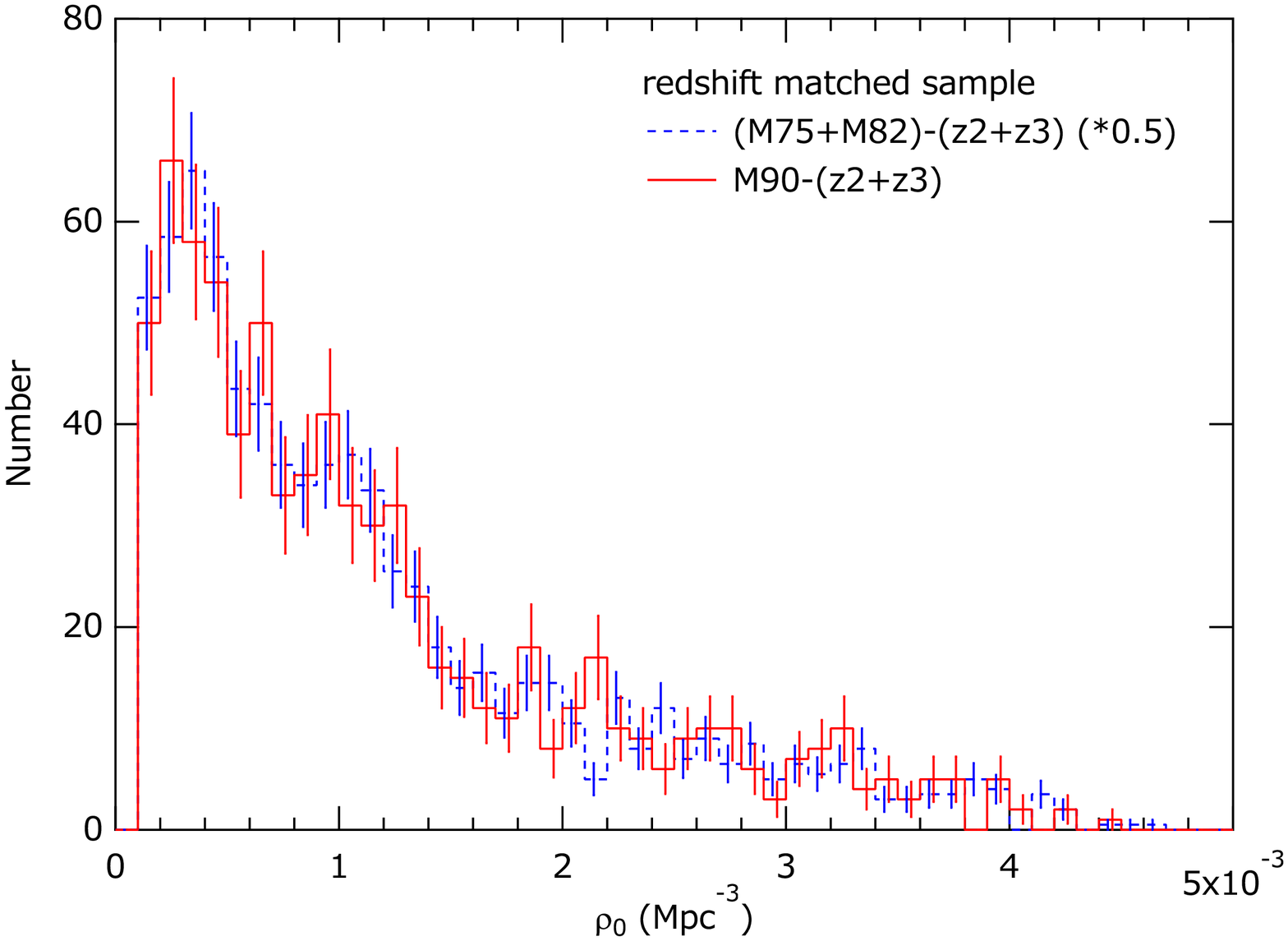}
\caption{Comparisons of the distributions of average number densities of galaxies 
at AGN redshift between higher (red histogram) and lower (blue histogram) mass 
groups. The left (right) panel is the comparison at redshift z1 (z2+z3).}
\label{fig:hist_rho0}
\end{figure*}

To see the relative difference of luminosity function of galaxies around SMBHs with 
lower and higher mass, absolute magnitude distributions are compared in 
Figure~\ref{fig:absmag}.
They are obtained by using the subtraction method described in section~\ref{sec:red_frac}.
In this comparison, redshift matched samples are used so that both of the samples
have the same sensitivity on the detection of galaxies.
To check the equivalence among the samples in term of the sensitivity of 
observations, the distributions of $\rho_{0}$ are compared in
Figure~\ref{fig:hist_rho0}.
We found that they are consistent with each other.

The left panel of Figure~\ref{fig:absmag} shows the comparison between the AGN 
groups of M65+M75 and M82 for redshift range of z1, and the ratio (high mass/low mass) 
of the magnitude distributions is shown in the bottom of the panel.
Since the lower mass samples has smaller excess density than the higher mass sample,
we combined two lower mass samples to increase the statistics.
The red histogram is the distribution of absolute magnitude for higher mass sample,
and the blue histogram is for lower mass sample.
The absolute magnitude is estimated at wavelength range of 1--1.2{\AA} in the rest 
frame by fitting model SED to the observed one using the EAZY code as described in
section~\ref{sec:red_frac}.
The ratio is peaked around $-20.25$ mag, and it shows a steep decline at dimmer side.
The significance of the depletion of galaxies in the M82-z1 group against (M65+M75)-z1
is $\sim$2.5 sigma at magnitude range from $-$19.5 to $-$18.0 mag.
The dashed line on the ratio plot is a power law function fitted
to the observations at brighter side below $-20.0$ mag.
The best fit value of the power law index is $-$0.096$\pm$0.075
for the ratio between M82-z1 and (M65+M75)-z1 samples.

\begin{figure*}
\includegraphics[width=0.8\textwidth]{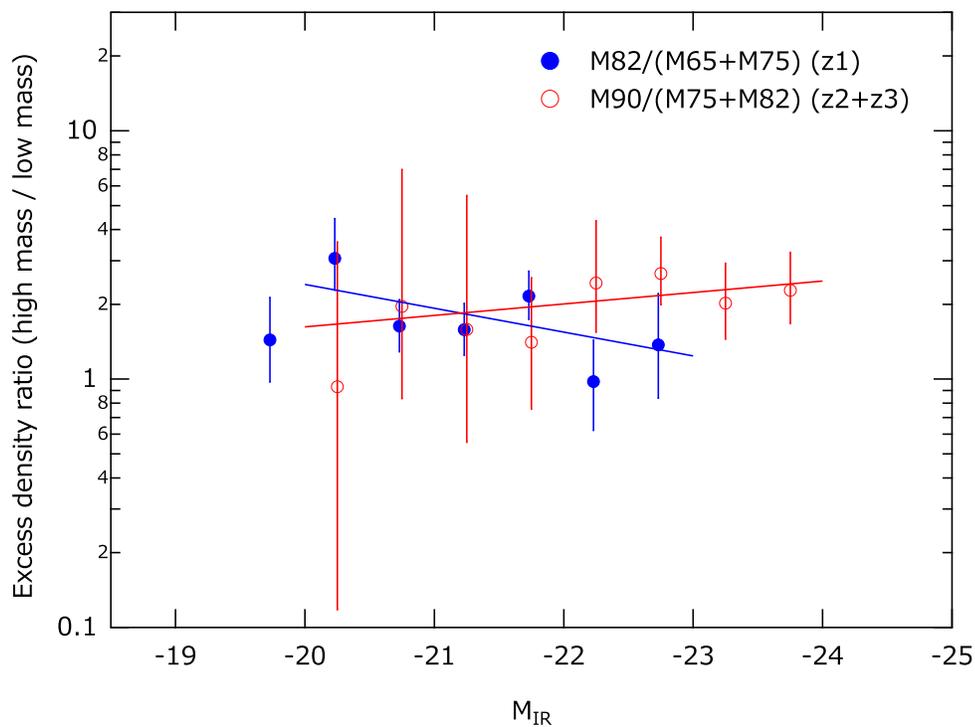}
\caption{Comparison of the ratio of absolute magnitude distributions. 
Filled blue circles represents the ratio of M82 and M65+M75 at redshift z1.
Open red circles represents the ratio of M90 and M75+M82 at redshift z2 and z3.
Fitting results with a power law function are also shown with the solid lines.}
\label{fig:absmag_ratio}
\end{figure*}

The same comparison between the mass groups M75+M82 and M90 at redshift z2 and z3
is shown in the right panel of Figure~\ref{fig:absmag}.
The observed ratio is fitted with the power law function and the power law
index is estimated to be +0.046 $\pm$ 0.093, which is consistent with 
a constant over the luminosity range.
The ratio of M82 and M65+M75 at redshift z1 and that of M90 and M75+M82 at redshift
z2 and z3 are compared in Figure~\ref{fig:absmag_ratio}.

\section{Discussion}

We have successfully reconfirmed the increase of cross-correlation length
above $\Mbh = 10^{8.2} \msun$ (Figure~\ref{fig:r0_mass_all}), that was already
reported in the previous paper by \citet{Komiya+13}, by adapting the improved 
data selection and using the most recent UKIDSS DR9 dataset.
We also confirmed that the trend is also seen in the redshift matched samples 
(Figure~\ref{fig:r0_mass_zdist}), and the cross-correlation length does not 
depend on the redshift (Figure~\ref{fig:r0_z_mdist}).
Although the AGN luminosity dependence was not investigated in this paper,
the previous work reported no significant dependence on the luminosity
\citep{Komiya+13}.

The derived
AGN bias $b$
(Figure~\ref{fig:b-z}) 
increase with $\Mbh$ above $10^{8.2} \msun$.
The AGN bias for M75 mass group is $b = 1.47^{+0.49}_{-0.38}$ at average
redshift of 0.47, which is consistent with those of SDSS QSOs by \citet{Ross+09},
2dF QSOs by \citet{Croom+05}, 
X-ray selected AGNs by \citet{Hickox+09}, and
optically selected AGNs by \citet{Krumpe+12}.

The AGN bias of mass group M90 is estimated to be $b = 3.08^{+1.23}_{-0.81}$,
which is larger than that of radio AGNs of \cite{Hickox+09} by one sigma
and indicates that those AGNs reside in haloes with mass larger than $10^{13.5}h^{-1}\msun$.

\citet{Shen+09} also studied virial mass dependence of clustering strength
using SDSS QSOs at redshift range from 0.4 to 2.5.
They found that the difference in the clustering strength for the
10\% most massive QSOs and the remaining 90\% is significant at 
the $\sim$ 2$\sigma$ level.
They also studied radio activity dependence of clustering by comparing between
the radio-loud and radio-quiet QSOs samples, and found that
the radio-loud QSOs are more strongly clustered 
than the radio-quiet QSOs at the $\sim$ 2.5$\sigma$ level.
Although these radio-loud QSOs tend to have systematically
larger BH masses than those radio-quiet QSOs by 0.12 dex,
the difference in their clustering remains when the comparison
is made for mass matched samples.
Thus they argue that more massive host halos, and denser environments 
may be related to the triggering of radio activity.

\citet{Mandelbaum+09} also found that the radio AGNs are hosted by
dark matter halo with $\sim$1.6$\times$10$^{13}h^{-1}\msun$.
There are observational evidences that radio-loud AGNs are associated
with the massive galaxies~\citep{Best+05} and have
SMBHs with masses typically larger than 10$^{9}\msun$~\citep{Laor-00}.
\citet{Ishibashi+14} also claimed that radio galaxies are preferentially 
associated with the more massive black holes.

According to the theoretical predictions by \citet{Fanidakis+11} for radio 
loudness and black holes mass  (see bottom panel of Figure~14 
in their paper), 
AGNs in the hot halo mode are distributed at a region of higher 
radio-loudness and their black hole mass extends from $10^{6} \msun$
up to $10^{10} \msun$, 
while the AGNs in the cold accretion mode are at lower part in 
radio-loudness and their black hole mass is limited below $\sim 10^{8.6} \msun$.
It is, therefore, expected that AGNs in the hot halo mode, i.e.
in dark matter haloes with higher mass, can preferentially
be selected by higher radio-loudness or higher black home mass.
Thus the increasing trend of AGN bias at higher black hole masses 
found in this work are consistent with those results obtained by the other 
authors, and they all are also consistent with the picture drawn from \citet{Fanidakis+11}.

The main purpose of this paper is to investigate what type of galaxy 
contributes to the increase of galaxy density around the most massive SMBH.
In this work we considered two kind of galaxy types which are classified
based on the distribution in a color-magnitude diagram.
One is a blue cloud galaxy (blue galaxy) which occupies a bluer 
and dimmer side of the distribution, and the other is a red sequence 
galaxy (red galaxy) which has a redder and narrower distribution 
in the color-magnitude diagram  (Figure~\ref{fig:CM_all}).

It should be noticed that detection efficiency for blue galaxy decreases
more rapidly 
with increasing redshift
than for red galaxy due to the difference in their brightness.
As is shown in the bottom panels of Figure~\ref{fig:hist_D},
at redshift range of z3 ($z=$0.6--1.0)
the observed fraction of blue galaxies 
is very low, this is 
largely
because most of the blue galaxies are below the detection limit.
At redshift range of z2 ($z=$0.3--0.6) the fraction of the detected blue galaxy 
constitutes about 20\% of all the detected galaxies, while at redshift range 
of z1 ($z=$0.1--0.3) the fraction increase to around 40\%.
As shown here, the observed red/blue fraction has strong redshift dependence
and should not be compared between different redshift groups.

The results of the two component analysis on the $\Dx$ distribution
(Figures~\ref{fig:hist_D},\ref{fig:RedFraction}) indicate that the increase 
of clustering strength found in the cross-correlation analysis 
for the most massive mass group M90 relative to
the lower mass groups are mainly due to the contribution of red galaxies.
This can be justified from the observational evidence that increase of 
the normalized excess density at the transition from M82 to M90 for redshift
z2 is significant for red galaxies, while it is almost unchanged or rather
decreasing for blue galaxies.
It is known that early type galaxies tend to be found at a high density region
such as a cluster/group core~\citep{Dressler-80,Postman+84,Balogh+04}.
It is, therefore, naturally expected that the high density around the high 
mass SMBH is coupled with early type galaxies.
The other indication from this analysis is that the increase of the
clustering strength found in the cross-correlation analysis for M82 
relative to the M75 is due to the 
contribution from both of the red and blue galaxies.
The red fraction is almost unchanged between M82 and M75, and slight 
increase in the normalized excess density from M75 to M82 is seen 
in both of the red and blue components for redshift groups z1 and z2.

The same analysis is performed also for the redshift matched samples
(Figure~\ref{fig:hist_D_zdist} and \ref{fig:RedFraction_zdist}).
The trend of the increase in the red fraction and normalized excess density
for red galaxies of the most massive group (M90) is also seen in this result.
The comparison between M82 and M65+M75 at redshift z1 shows that increase
in the normalized excess density are seen in both the red and blue galaxies.

Two physical mechanisms 
could
be relevant to this evolution.
One is related with the increase of brightness of blue dim galaxies 
induced by
starburst triggered in a secular and merger mode, which can result
in the increase of the detectable galaxy density.
Another one is related with the transition from blue to red galaxies,
which can be the result of a feedback
or gas removal
mechanism shutting off the star burst activity.
The latter process may follows the former process.
In case where there is an enough time delay between the both processes, 
we 
could
observe the increase of density both for red and blue 
galaxies as is the case of M82 group.
If the time delay is short, only the increase of red galaxy density
will be observed, and this 
could
be the case of M90 group.
The evolution of properties of galaxies around AGNs observed in this work
could
be explained in this manner.

According to the theoretical works by \citet{Fanidakis+13}, the hot halo
mode can be the most effective accretion mechanism in massive haloes with
$M_{\rm h} > 10^{12.5}\msun$.
In such a massive halo most of the galaxies are 
expected to be
in red sequence since AGN 
feedback shut off the gas cooling and star formation
in their model.
In galaxies turned into red sequence, i.e. early type galaxies, gas accretion
from the evolving stars to the center of SMBH could also occur as claimed by
\citet{Ciotti+01}.
Considering these theoretical predictions,
the dominance of red galaxies around AGNs with higher BH mass obtained
in this work is indicative of predominance of hot halo mode
and/or gas accretion from evolving stars
for the growth of the most massive SMBHs.

The Eddington ratios of our AGN samples in M90-z1 group are distributed
in the range $\lambda = $ 0.003 -- 0.1, and peaked around 
$\log(\lambda) \sim -1.8$. 
The lower bound of the distribution is limited by the detection limit, 
so the intrinsic peak can be expected to be much lower than 
$\log(\lambda) = -2$.
The bolometric luminosity of the same samples is above $10^{45}$ erg/s.
According to the model of \citet{Fanidakis+13b} 
for those bolometric luminosity, two branches exist in the 
$M_{\rm halo}$-$L_{\rm bol}$ diagram (see Figure~1 in their paper);
one is the branch of starburst mode for AGNs in dark matter halo of $10^{12}$ $\msun$,
the other is the branch of hot halo mode for AGNs in dark matter halo of 
$10^{15}$ $\msun$.
Considering that the Eddington ratios of our samples are lower than those
expected for the starburst mode ($\sim$0.1), the sample of M90-z1 
are consistent with the AGNs in the branch of hot halo mode.

We also investigated relative difference of luminosity function between the 
low and high mass groups.
The redshift matched samples are used for making them directly comparable
with each other. 
The equivalence among the samples in term of the sensitivity of 
observations was checked by comparing the distributions of $\rho_{0}$ 
(Figure~\ref{fig:hist_rho0}), and we found that they are consistent with 
each other.
Thus the distributions of completeness fraction are the same for the redshift 
matched samples, and the completeness fraction can be canceled by calculating 
the ratio.

For the redshift range z1, comparison between the M82 mass sample and the lower 
mass sample, which is a combined sample of M65$+$M75, is made in the left 
panel of Figure~\ref{fig:absmag}.
The figure indicates that an increase of galaxy density
for the higher mass sample would be caused by galaxies brighter 
than $M_{\rm IR} = -19.5$ mag.
At a dimmer side of $M_{\rm IR} > -19.5$ mag, on the other hand, the luminosity 
function of the higher mass sample is smaller than that of the smaller mass group.
The significance of the deficit of galaxies for the higher mass sample relative
to the smaller mass sample is 2.5 sigma at absolute brightness range from $-$19.5
to $-$18.0 mag.

Considering that blue galaxies dominate the dimmer side of the luminosity function,
the difference
of the luminosity functions 
could
be the result of brightening of the dim blue galaxies
in the transition from lower mass group (M65+M75) to higher mass group (M82).
Besides, to make the red/blue galaxy ratio unchanged some 
but not all of the blue galaxies are need to be transformed to red galaxies.

The plot of the ratio between luminosity function of M82 and M65+M75
shows that the increase fraction is larger at the dimmer side and peaked around 
$M_{\rm IR} = -20.5$ mag, and it is approximately fitted with a power law 
function with the power index of $-$0.096$\pm$0.075.
This may indicate that the dimmer galaxies have higher probability to increase
their brightness by interaction and/or major/minor merger.
For the redshift ranges z2 and z3, a comparison between the M90 and M75$+$M82 
groups is made in the right panel of Figure~\ref{fig:absmag}, which shows 
the ratio of the luminosity functions is almost constant at $M_{\rm IR} < -20$ mag.
If
the brightening of blue galaxies occurs at a same rate between the two groups,
the difference is only the rate of blue to red transformation, 
and the transformation occurs without much affecting the brightness in the 1-1.2~{\AA} band, 
the constant ratio of the luminosity functions is 
naturally expected.

\section{Conclusions}

In this paper, using the updated UKIDSS catalog we have successfully 
reconfirmed the previous findings of
\citet{Komiya+13} that the clustering of galaxies around AGNs with the 
most massive SMBH is larger than those with less massive SMBH.
The AGN bias was derived for each BH mass group.
The obtained AGN bias are
$b = 1.52^{+0.66}_{-0.54}$,
$1.47^{+0.49}_{-0.38}$,
$2.38^{+0.74}_{-0.49}$,
$3.08^{+1.23}_{-0.81}$
for mass group M65, M75, M82 and M90, respectively.
We further investigated what type of galaxies are associated in the
environment of the most massive SMBH to reveal the nature of evolution
mechanism of SMBHs and galaxies.
As a result, it is found that red galaxies are dominated around 
AGNs with the most massive SMBH. 
The red galaxy fraction increases from 0.73$\pm$0.04 for M82 mass group
to 0.98$\pm$0.07 for M90 at redshift z2.
This is the first observational result that revealed the nature of 
galaxies at such environments 
from more than thousands of AGN samples.
We also compared, for the first time, luminosity functions of galaxies 
at environments of AGNs with lower and higher mass SMBH, and found that
there is an indication that shows brightening of dim and blue galaxies
along with the evolution of SMBHs.
Summarizing the results obtained in this work, we can deduce the following 
scenario on the evolution of SMBH and its environment galaxies as a 
function of $\Mbh$.

Below the critical mass around $10^{8.2} \msun$, the environment of SMBH
does not depend on $\Mbh$ and is almost equivalent to that of a quiescent 
galaxy.
At that environment, the fraction of blue galaxy is more than 40\%.
Secular evolution can be the main driver for the evolution of these SMBHs.
Above the critical mass,
dim and blue galaxies increase its brightness, which is presumably caused via 
starburst.
Some of the brightened blue galaxies are transformed to red galaxies
after the AGN feedback or some other mechanisms which shut off
star formation in the galaxies operates. 

In the environment of AGNs of M82 mass group, increases in densities for 
both blue and red galaxies are observed, 
while in the environment of AGNs of M90 mass group, the 
increase is seen only for the red galaxies.
This difference may be due to the difference of time lag between the
starburst and some mechanism to shut off the starburst activity.

Considering that there is a correlation between the transition in properties 
of environmental galaxies and mass evolution of SMBHs above $\Mbh = 10^{8.2} \msun$,
they are governed by a sort of environmental effect rather than internal secular
evolution.
This transition makes galaxies observable by pushing up the brightness 
above the detection limit, and results in the increase of observed 
galaxy number density and the clustering length.
The hot halo mode
as well as gas accretion from evolving stars
may
be one of the most dominant mechanism 
in the growth of the most massive SMBHs,
as inferred from properties of environment galaxies.

\begin{ack}

We would like to thank the anonymous referee for his/her helpful comments.
Results are based on data obtained from the Japanese Virtual Observatory, 
which is operated by the Astronomy Data Center, National Astronomical 
Observatory of Japan. 
Funding for SDSS-III has been provided by the Alfred P. Sloan Foundation,
the Participating Institutions, the National Science Foundation, and the 
U.S. Department of Energy Office of Science. The SDSS-III web site is 
http://www.sdss3.org/.
SDSS-III is managed by the Astrophysical Research Consortium for the 
Participating Institutions of the SDSS-III Collaboration including the 
University of Arizona, the Brazilian Participation Group, Brookhaven 
National Laboratory, Carnegie Mellon University, University of Florida, 
the French Participation Group, the German Participation Group, Harvard 
University, the Instituto de Astrofisica de Canarias, the Michigan 
State/Notre Dame/JINA Participation Group, Johns Hopkins University, 
Lawrence Berkeley National Laboratory, Max Planck Institute for Astrophysics, 
Max Planck Institute for Extraterrestrial Physics, New Mexico State 
University, New York University, Ohio State University, Pennsylvania 
State University, University of Portsmouth, Princeton University, the 
Spanish Participation Group, University of Tokyo, University of Utah, 
Vanderbilt University, University of Virginia, University of Washington, 
and Yale University.
This work is based in part on data obtained as part of the UKIRT Infrared 
Deep Sky Survey.

\end{ack}


\begin{table*}
\begin{center}
\tbl{Parameters derived from the cross-correlation analysis for each BH mass and redshift range}{
\begin{tabular}{llrcccrrr}
\hline\hline
BH mass  
 & redshift & $n_{\rm AGN}^{\rm a}$
 & $\langle \log{M} \rangle^{\rm b}$
 & $\langle z \rangle^{\rm c}$
 & $r_0^{\rm d}$
 & $\sigma_{\rm stat}^{\rm e}$
 & $\langle n_{\rm bg} \rangle^{\rm f}$
 & $\langle \rho_{0} \rangle^{\rm g}$ \\
 $\log{M}$ & & & &  & $\hMpc$ & $\hMpc$ & $\Mpc^{-2}$ & $10^{-3}\Mpc^{-3} $ \\
\hline
\multicolumn{9}{c}{All samples}\\
\hline
6.5--7.5 & 0.1--1.0 & 524 & 7.2 & 0.24 & $6.1^{+0.7}_{-0.6}$ & 0.34 & 38.85$\pm$0.05 & 5.08$\pm$0.8 \\
         & 0.1--0.3 & 429 & 7.1 & 0.20 & $6.2^{+0.7}_{-0.6}$ & 0.36 & 45.17$\pm$0.06 & 5.61$\pm$0.9 \\
\hline
7.5--8.2 & 0.1--1.0 &1880 & 7.9 & 0.47 & $5.8^{+0.8}_{-0.6}$ & 0.23 & 15.24$\pm$0.02 & 2.71$\pm$0.5 \\
         & 0.1--0.3 & 520 & 7.8 & 0.22 & $6.3^{+0.7}_{-0.6}$ & 0.32 & 37.60$\pm$0.05 & 5.32$\pm$0.9 \\
         & 0.3--0.6 & 804 & 7.9 & 0.45 & $4.8^{+0.8}_{-0.6}$ & 0.35 &  9.41$\pm$0.02 & 2.48$\pm$0.5 \\
         & 0.6--1.0 & 556 & 8.0 & 0.75 & $5.4^{+1.6}_{-1.2}$ & 0.85 &  2.78$\pm$0.01 & 0.61$\pm$0.2 \\
\hline
8.2--9.0 & 0.1--1.0 &4485 & 8.6 & 0.65 & $7.7^{+1.2}_{-0.8}$ & 0.16 &  6.89$\pm$0.01 & 1.42$\pm$0.3 \\
         & 0.1--0.3 & 362 & 8.5 & 0.23 & $7.7^{+0.9}_{-0.7}$ & 0.32 & 33.10$\pm$0.06 & 5.18$\pm$0.8 \\
         & 0.3--0.6 &1380 & 8.6 & 0.46 & $7.0^{+1.0}_{-0.8}$ & 0.21 &  8.59$\pm$0.02 & 2.29$\pm$0.5 \\
         & 0.6--1.0 &2743 & 8.6 & 0.80 & $8.7^{+2.4}_{-1.4}$ & 0.33 &  2.58$\pm$0.01 & 0.49$\pm$0.2 \\
\hline
9.0--10.0& 0.1--1.0 &1170 & 9.2 & 0.72 & $9.0^{+1.6}_{-1.1}$ & 0.34 &  4.79$\pm$0.01 & 1.01$\pm$0.3 \\
         & 0.3--0.6 & 289 & 9.2 & 0.48 & $8.8^{+1.4}_{-1.0}$ & 0.42 &  8.19$\pm$0.04 & 2.17$\pm$0.5 \\
         & 0.6--1.0 & 845 & 9.2 & 0.83 & $9.7^{+2.8}_{-1.7}$ & 0.62 &  2.50$\pm$0.01 & 0.44$\pm$0.2 \\
\hline
\multicolumn{9}{c}{Redshift matched samples}\\
\hline
6.5--7.5 & 0.1--0.3
                    & 298 & 7.1 & 0.22 & $5.6^{+0.7}_{-0.6}$ & 0.44 &  35.04$\pm$0.07 & 5.29$\pm$0.8 \\
7.5--8.2 &  
                    & 298 & 7.8 & 0.22 & $6.3^{+0.8}_{-0.6}$ & 0.41 &  36.23$\pm$0.07 & 5.35$\pm$0.9 \\
8.2--9.0 & 
                    & 298 & 8.5 & 0.22 & $8.1^{+0.9}_{-0.7}$ & 0.34 &  35.96$\pm$0.07 & 5.33$\pm$0.9 \\
\hline
7.5--8.2 & 0.3--0.6
                    & 289 & 7.9 & 0.48 & $4.6^{+0.9}_{-0.8}$ & 0.65 &  8.47$\pm$0.03 & 2.18$\pm$0.5 \\
8.2--9.0 &  
                    & 289 & 8.6 & 0.48 & $6.9^{+1.1}_{-0.9}$ & 0.49 &  8.25$\pm$0.03 & 2.17$\pm$0.5 \\
9.0--10.0& 
                    & 289 & 9.2 & 0.48 & $8.8^{+1.4}_{-1.0}$ & 0.42 &  8.19$\pm$0.04 & 2.17$\pm$0.5 \\
\hline
7.5--8.2 & 0.6--1.0
                    & 483 & 8.0 & 0.76 & $5.3^{+1.7}_{-1.3}$ & 0.97 &  2.73$\pm$0.01 & 0.58$\pm$0.2 \\
8.2--9.0 &  
                    & 483 & 8.6 & 0.76 & $8.1^{+2.2}_{-1.4}$ & 0.72 &  2.67$\pm$0.02 & 0.58$\pm$0.2 \\
9.0--10.0& 
                    & 483 & 9.2 & 0.77 & $9.2^{+2.5}_{-1.5}$ & 0.68 &  2.76$\pm$0.02 & 0.58$\pm$0.2 \\
\hline
\multicolumn{9}{c}{Mass matched samples}\\
\hline
7.5--8.2
         & 0.1--0.3 & 371 & 7.9 & 0.22 & $6.1^{+0.7}_{-0.6}$ & 0.38 & 36.91$\pm$0.06 & 5.30$\pm$0.9 \\
         & 0.3--0.6 & 371 & 7.9 & 0.45 & $4.7^{+0.8}_{-0.7}$ & 0.54 &  9.53$\pm$0.03 & 2.45$\pm$0.5 \\
         & 0.6--1.0 & 371 & 7.9 & 0.74 & $4.4^{+1.6}_{-1.3}$ & 1.17 &  2.85$\pm$0.02 & 0.64$\pm$0.2 \\
\hline
8.2--9.0
         & 0.1--0.3 & 362 & 8.5 & 0.23 & $7.7^{+0.9}_{-0.7}$ & 0.32 & 33.10$\pm$0.06 & 5.18$\pm$0.8 \\
         & 0.3--0.6 & 362 & 8.5 & 0.46 & $6.8^{+1.1}_{-0.8}$ & 0.42 &  8.54$\pm$0.03 & 2.32$\pm$0.5 \\
         & 0.6--1.0 & 362 & 8.5 & 0.80 & $7.7^{+2.3}_{-1.5}$ & 0.99 &  2.61$\pm$0.02 & 0.50$\pm$0.2 \\
\hline
9.0--10.0
         & 0.3--0.6 & 289 & 9.2 & 0.48 & $8.8^{+1.4}_{-1.0}$ & 0.42 &  8.19$\pm$0.04 & 2.17$\pm$0.5 \\
         & 0.6--1.0 & 289 & 9.2 & 0.84 & $9.4^{+3.0}_{-1.9}$ & 1.18 &  2.49$\pm$0.02 & 0.40$\pm$0.2 \\
\hline                    
\end{tabular}}\label{tab:fit_r0}
\begin{tabnote}
$^{a}$number of sample AGNs.\\
$^{b}$average of logarithm of BH mass.\\
$^{c}$average redshift. \\
$^{d}$correlation length, the error contains systematic error due to uncertainty of $\rho_0$ and $1\sigma$ statistical error.\\
$^{e}$statistical error of $r_{0}$.\\
$^{f}$average of projected number density of background galaxies.\\
$^{g}$average of the averaged number density of galaxies at the AGN redshift.\\
\end{tabnote}
\end{center}
\end{table*}

\begin{table*}
\begin{center}
\tbl{Parameters derived from the bias estimate}{
\begin{tabular}{ccccccccc}
\hline\hline
mass & 
redshift &
$f_{\rm blue }^{\rm a}$ &
$f_{\rm red}^{\rm b}$ &
$f_{\rm bright}^{\rm c}$ &
$\sigma_{r_{0}}^{\rm d}$  &
$r_{\rm AA}^{\rm e}$ &
$\gamma_{\rm AA}^{\rm f}$ &
$b^{\rm g}$ \\ 
$\log{\msun}$ & 
   & 
  &
  &
  &
$h^{-1}$Mpc & 
$h^{-1}$Mpc & 
&  \\
\hline
6.5--7.5  & 0.1--1.0 & 0.321 & 0.669 & 0.010 & 1.16 &  $6.33^{+3.4}_{-2.5}$ & 1.69 & $1.52^{+0.66}_{-0.54}$ \\
7.5--8.2  & 0.1--1.0 & 0.296 & 0.650 & 0.054 & 0.65 &  $5.29^{+2.1}_{-1.6}$ & 1.69 & $1.47^{+0.49}_{-0.38}$ \\
8.2--9.0  & 0.1--1.0 & 0.230 & 0.581 & 0.189 & 0.35 &  $8.45^{+3.2}_{-2.0}$ & 1.69 & $2.38^{+0.74}_{-0.49}$ \\
9.0--10.0 & 0.1--1.0 & 0.112 & 0.668 & 0.220 & 0.82 & $11.09^{+5.4}_{-3.4}$ & 1.68 & $3.08^{+1.23}_{-0.81}$ \\
\hline
\end{tabular}}\label{tab:halo_mass}
\begin{tabnote}
$^{\rm a}$fraction of blue galaxies with brightness $M_{\rm r} \ge -22$\\
$^{\rm b}$fraction of red galaxies with brightness $M_{\rm r} \ge -22$\\
$^{\rm c}$fraction of galaxies with brightness $M_{\rm r} < -22$\\
$^{\rm d}$uncertainty of the cross-correlation length $r_{0}$ due to the intrinsic variance 
of clustering around each AGN.\\
$^{\rm e}$autocorrelation length of AGNs.\\
$^{\rm f}$power law index of AGN autocorrelation function.\\
$^{\rm g}$AGN bias.\\
\end{tabnote}
\end{center}
\end{table*}

\begin{table*}
\begin{center}
\tbl{Parameters derived from the two component analysis on $\Dx$ distributions}{
\begin{tabular}{ccccccccc}
\hline\hline
redshift 
 & mass
 & $\mu_{\rm red}^{\rm a}$
 & $\sigma_{\rm red}^{\rm b}$
 & $\mu_{\rm blue}^{\rm c}$ 
 & $\sigma_{\rm blue}^{\rm d}$
 & $F_{\rm red}^{\rm e}$
 & $\omega'_{\rm red}$$^{\rm f}$
 & $\omega'_{\rm blue}$$^{\rm g}$ \\
 & $\log{\msun}$ & & & & & & 10$^{2}$Mpc & 10$^{2}$Mpc \\
\hline
\multicolumn{9}{c}{\it All samples} \\
\hline
0.1--1.0 & 6.5--10.0& 4.17$\pm$0.05 & 0.41$\pm$0.03 & 3.06$\pm$0.23 & 0.50$\pm$0.13 & 0.77$\pm$0.09 &          &          \\
\hline
0.1--0.3 & 6.5--7.5 & 4.17          & 0.41          & 3.06          & 0.50          & 0.66$\pm$0.06 & 0.64$\pm$0.09 & 0.32$\pm$0.06 \\
         & 7.5--8.2 & 4.17          & 0.41          & 3.06          & 0.50          & 0.66$\pm$0.05 & 0.60$\pm$0.07 & 0.31$\pm$0.05 \\
         & 8.2--9.0 & 4.17          & 0.41          & 3.06          & 0.50          & 0.66$\pm$0.05 & 0.73$\pm$0.08 & 0.37$\pm$0.06 \\
\hline
0.3--0.6 & 7.5--8.2 & 4.17          & 0.41          & 3.06          & 0.50          & 0.84$\pm$0.10 & 0.43$\pm$0.08 & 0.08$\pm$0.06 \\
         & 8.2--9.0 & 4.17          & 0.41          & 3.06          & 0.50          & 0.73$\pm$0.04 & 0.67$\pm$0.06 & 0.25$\pm$0.05 \\
         & 9.0--10.0& 4.17          & 0.41          & 3.06          & 0.50          & 0.98$\pm$0.07 & 1.37$\pm$0.15 & 0.22$\pm$0.10 \\
\hline
0.6--1.0 & 7.5--8.2 & 4.17          & 0.41          & 3.06         & 0.50           & 0.81$\pm$0.30 & 0.40$\pm$0.17 & 0.09$\pm$0.17 \\
         & 8.2--9.0 & 4.17          & 0.41          & 3.06         & 0.50           & 1.00$\pm$0.08 & 1.28$\pm$0.09 & 0.00$\pm$0.10 \\
         & 9.0--10.0& 4.17          & 0.41          & 3.06         & 0.50           & 1.00$\pm$0.14 & 1.39$\pm$0.17 & 0.00$\pm$0.19 \\
\hline
\multicolumn{9}{c}{\it Redshift matched samples} \\
\hline
0.1--0.3
         & 6.5--8.2 & 4.17          & 0.41          & 3.06         & 0.50           & 0.64$\pm$0.04 & 0.73$\pm$0.07 & 0.41$\pm$0.05 \\
         & 8.2--10.0& 4.17          & 0.41          & 3.06         & 0.50           & 0.66$\pm$0.05 & 0.92$\pm$0.10 & 0.48$\pm$0.08 \\
\hline
0.3--0.6
         & 7.5--9.0 & 4.17          & 0.41          & 3.06         & 0.50           & 0.59$\pm$0.09 & 0.41$\pm$0.10 & 0.28$\pm$0.07 \\
         & 9.0--10.0& 4.17          & 0.41          & 3.06         & 0.50           & 0.98$\pm$0.07 & 1.37$\pm$0.15 & 0.02$\pm$0.10 \\
\hline
0.6--1.0
         & 7.5--9.0 & 4.17          & 0.41          & 3.06         & 0.50           & 0.86$\pm$0.16 & 0.65$\pm$0.14 & 0.11$\pm$0.14 \\
         & 9.0--10.0& 4.17          & 0.41          & 3.06         & 0.50           & 1.00$\pm$0.13 & 1.50$\pm$0.20 & 0.00$\pm$0.19 \\
\hline
\end{tabular}}\label{tab:fit_D}
\begin{tabnote}
$^{\rm a}$mean of the $\Dx$ distribution for red component.\\
$^{\rm b}$standard deviation of the $\Dx$ distribution for red component.\\
$^{\rm c}$mean of the $\Dx$ distribution for blue component.\\
$^{\rm d}$standard deviation of the $\Dx$ distribution for blue component.\\
$^{\rm e}$fraction of red component.\\
$^{\rm f}$normalized excess density for red component.\\
$^{\rm g}$normalized excess density for blue component.\\
\end{tabnote}

\end{center}
\end{table*}


\end{document}